\documentclass{amsart}

\usepackage{amsfonts,amsmath,amsthm,amssymb,color,filecontents,float,graphicx,url,enumitem,lineno,mathtools}
\usepackage{stackengine,xcolor}
\usepackage{booktabs}
\usepackage{chngcntr}
\usepackage[toc,page]{appendix}
\usepackage{hyperref}
\hypersetup{colorlinks=true,linkcolor=blue, citecolor=blue, urlcolor=blue}
\usepackage[foot]{amsaddr}
\usepackage[mathscr]{eucal} \usepackage{mathrsfs}
\usepackage[all]{xy}
\usepackage{placeins}
\usepackage[superscript,biblabel]{cite}
\usepackage{bbm}
\let\Oldsection\section
\renewcommand{\section}{\FloatBarrier\Oldsection}

\let\Oldsubsection\subsection
\renewcommand{\subsection}{\FloatBarrier\Oldsubsection}

\let\Oldsubsubsection\subsubsection
\renewcommand{\subsubsection}{\FloatBarrier\Oldsubsubsection}

\newcount\Comments  
\definecolor{green}{rgb}{0.1,0.1,0.7}
\definecolor{red}{rgb}{1,0,0}
\definecolor{blue}{rgb}{0,0,1}
\definecolor{bluegreen}{rgb}{0,0.5,1}
\definecolor{redblue}{rgb}{0.5,0,1}
\definecolor{greenish}{rgb}{0.4,0.7,0.5}
\newcommand{\kibitz}[2]{\ifnum\Comments=0\textcolor{#1}{#2}\fi}

\DeclareGraphicsExtensions{.png,.jpeg,.jpg,.pdf}
\graphicspath{{./Figures/}}

\theoremstyle{definition}
\newtheorem{theorem}{Theorem}[section]
\newtheorem{definition}[theorem]{Definition}

\newcommand{\mc}{\mathcal}
\newcommand{\eps}{\varepsilon}

\newcommand{\bbm}{\mathbbm}

\DeclareMathOperator{\Nb}{Nb}
\DeclareMathOperator{\Emails}{Emails}

\begin{document}

\title{The effect of co-location on human communication networks}

\author{Daniel Carmody}\email{dcarmody@mit.edu}
\author{Martina Mazzarello}

\author{Paolo Santi}

\author{Trevor Harris}

\author{Sune Lehmann}

\author{Timur Abbiasov}

\author{Robin Dunbar}

\author{Carlo Ratti}

\address[A1,A2,A3,A6,A8]{Senseable City Lab, Massachusetts Institute of Technology, Cambridge, MA, USA}

\address[A4]{Department of Statistics, Texas A\&M University, College Station, TX, USA}

\address[A5]{Technical University of Denmark, Kgs Lyngby, Denmark}

\address[A7]{Department of Experimental Psychology, Oxford University, Oxford OX2 6GG, UK}

\address[A3]{Instituto di Informatica e Telematica del CNR, Pisa, Italy}

\keywords{Communication networks, email networks, weak ties, co-location, working environments, COVID-19}

\begin{abstract}
The ability to rewire ties in communication networks is vital for large-scale human cooperation and the spread of new ideas. We show that lack of researcher co-location during the COVID-19 lockdown caused the loss of more than 4,800 weak ties -- ties between distant parts of the social system that enable the flow of novel information -- over 18 months in the email network of a large North American university. Furthermore, we find that the re-introduction of partial co-location through a hybrid work mode led to a partial regeneration of weak ties. We quantify the effect of co-location in forming ties through a model based on physical proximity, which is able to reproduce all empirical observations. Results indicate that employees who are not co-located are less likely to form ties, weakening the spread of information in the workplace. Such findings could contribute to a better understanding of the spatio-temporal dynamics of human communication networks, and help organizations that are moving towards the implementation of hybrid work policies evaluate the minimum amount of in-person interaction necessary for a productive work environment.
\end{abstract}

\maketitle

\section{Introduction}

The ability to establish and leverage communication networks to share information and collaboratively accomplish sophisticated tasks is a distinguishing feature of humans\cite{ScienceCooperation}. While the ability to form social ties with others was originally developed when individuals were in close proximity, technological improvements have allowed increasingly remote  forms of communication and collaboration, that now include tele- and video-conferencing, email, and chats\cite{Chen2012TheIO}. 
These changes reinvigorated longstanding debates about the extent to which social relationships are predicated upon physical proximity\cite{McPherson2001BirdsOA}. While earlier studies in sociology and organizational science
discuss the role of spatial \textit{propinquity} in producing interpersonal ties \cite{NbhdTieSimul,reagans2011close}, the causal mechanisms through which co-location affects social networks remain understudied\cite{socialinteractfine,reynolds1993interactionism,BarabasiSong,Simini2012AUM}.

Addressing this question is all the more relevant today. First, in planning the transition towards the post-COVID-19-pandemic ``new normal", institutions and policy-makers world-wide are wondering about the best way to reshape work environments following COVID-19\cite{BenvenisteCNN,McLeanCNN,McKinsey800Execs,NBERw27344}. Second, the massive shift to remote work during the past two years has produced a trove of Big Data that promises to elucidate what happens when we remove physical presence as a main conduit of communication\cite{curbCreate}. While recent work has started to highlight changes in the communication networks of information workers due to mandatory remote work, the data was collected across several campuses in the continental US -- making it difficult to link these network changes to a lack of physical proximity \cite{yang2021the}. Hence, the following question remains open: what is the effect of co-location on human communication networks?

In this study, we explore the mechanism via which the complete removal and subsequent partial re-introduction of physical co-location at a large North American university -- the MIT campus -- affects the structure of its digital communication network. We find that, despite the robustness of many network measures to the shift to remote work, physical co-location plays a crucial role in the formation of weak ties. 

Since Mark Granovetter’s seminal work in 1973\cite{GranWeakTies}, weak ties have been identified as fundamental microscopic structures that enable the spread of ideas and opportunities in social networks. Our hypothesis is that weak ties form due to chance encounters
in and around the office, so that removing the possibility for chance encounters should affect the formation of weak ties. By hindering new weak tie formation, the removal of physical co-location leads to increased redundancy in email networks -- more information is spread between fewer people. Said differently, physical proximity is vital for updating the people one communicates with over time. This is in line with earlier work on the notion of propinquity, used to denote the tendency to form connections among proximate individuals\cite{McPherson2001BirdsOA}. Propinquity can be  modeled with a modification of the link-central preferential attachment model which includes a co-location factor $\tau$, accurately reproducing the dynamics of formation and stability of weak ties caused by long-term removal and partial re-introduction of physical co-location on the MIT campus.

\section{Results}

\subsection{Forming the daily email network}

We build and analyze a large email network of research workers at MIT in Cambridge, Massachusetts. Despite the wide-scale adoption of synchronous video-conferencing technology, email remains a universal mode of digital communication used by researchers to exchange information and organize meetings. In order to study changes in communication behavior due to fully remote work, we study the email habits of 2,834 MIT faculty and postdocs over 18 months starting on December 26, 2019. Each researcher belongs to a ``research unit" which describes their campus affiliation (see Supplementary Information for a complete list -- the partition of the MIT research community into research units is in general finer than the partition into departments). During March 2020 MIT started implementing COVID-19 contingency plans, which led to a progressive decrease in campus attendance and culminated on Monday, March 23, 2020 with the halting of in-person research activities. For each day, the number of emails sent between each pair of (anonymized) individuals is determined exactly for $> 66\%$ pairs of individuals from randomized, aggregated data and used to form the edge weights of an undirected network. For the remaining pairs we estimate the number of emails sent using non-negative matrix factorization (see Methods).

\subsection{Robustness of network metrics to the remote work transition}
\begin{figure}[h!]
    \centering
    \includegraphics[width=\linewidth]{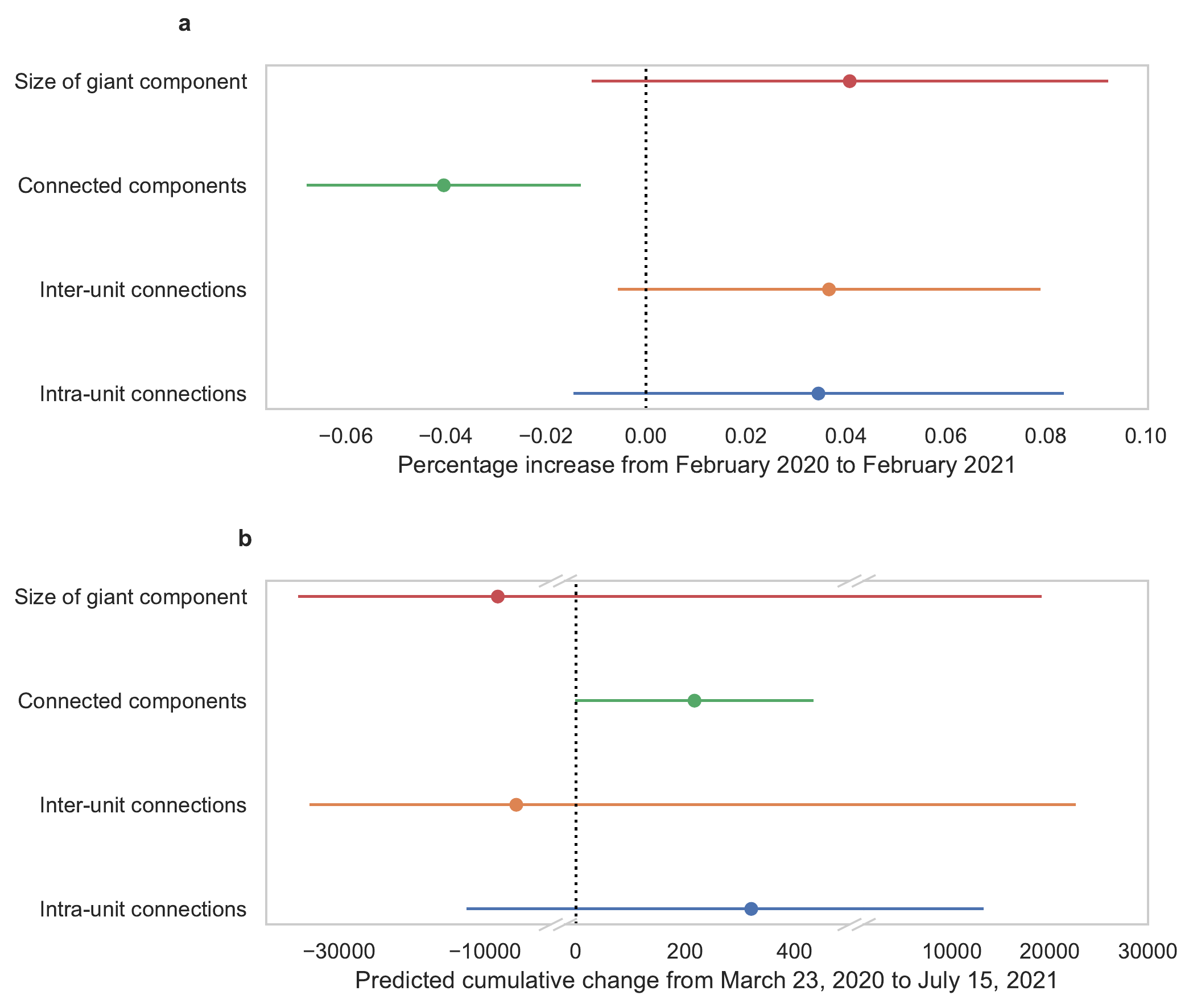}
    \caption{{\bf Robustness of global network topology in daily reciprocated email networks.} {\bf a}, The change from the beginning of the Spring 2020 semester (weekdays between February 5 and March 5, 2020) to the Spring 2021 semester (weekdays between February 17 and March 19, 2021) in the number of users in the largest component, the number of components, the number of connections (edges) between users who are in the same research unit, the number of connections (edges) between users in distinct research units. 95\% confidence intervals from two-sided $z$ test on the coefficients of a generalized least squares model fit to the log difference in means ($n = 16$ days for all variables). Full results are in Supplementary Tables 1-4. {\bf b}, Predicted cumulative change in network metrics compared to a synthetic counterfactual from March 23, 2020 to July 15, 2021. Posterior predictive intervals with 95\% coverage computed using Bayesian structural time series ($n_{\mathrm{pre}} = 8$ weeks, $n_{\mathrm{post}} = 72$ weeks). Fitted values/intervals use the mean as the measure of central tendency.}
    \label{fig:basic_stats}
\end{figure}

Our initial analysis did not show many changes in the network: directly comparing connected components and the number of intra/inter-research unit connections in the email network in February 2020 and February 2021 using a paired test on the logarithms of these network metrics (see Methods) highlights few significant differences (Fig \ref{fig:basic_stats} panel a). However, due to the seasonal nature of academic work, paired testing comparing only February is not sufficient to estimate the short- and cumulative long-term effects of fully remote working on the communication network.

To provide a statistically robust estimation of long-term effects, we design a methodology based on the Bayesian Structural Time Series (BSTS) approach. The approach is based on the construction of a synthetic counterfactual time series from a covariate unaffected by the treatment. A well-constructed synthetic counterfactual can capture fluctuations in the time series of interest due to seasonality and confounding factors other than the treatment. If the predictive power of the covariate wanes over time (which can happen because e.g. the model is only fit on data before the treatment is applied), then the width of the posterior predictive interval around the synthetic counterfactual will increase over time, representing our increased uncertainty about the distant future. As the treatment in this case corresponds to the termination of campus access for researchers, we use email data from weekends when most researchers were not entering the office to construct our counterfactual (see Methods). When studying these cumulative effects, there are also no significant differences in the number of intra/inter-research unit connections, number of connected components, and the size of the giant component due to remote work (Figure \ref{fig:basic_stats} panel b). A detailed plot highlighting the evolution of the uncertainty of the BSTS model over time can be found in Supplementary Information. The fact that the sign (loss vs gain) of the change in network metrics like connected components, inter-unit connections, and size of giant component differs in panels a and b is indicative of the fact that choosing a single month in 2020 and 2021 for comparison does not accurately capture cumulative effects.

\subsection{Weak tie formation is impeded by remote work}
To study the effect of remote work on the sorts of connections which might arise from serendipitous encounters on campus, we investigate the structure of weak ties in the email network. Because local bridges can be identified knowing only the topology of a social network, Granovetter suggested using the notion of a local bridge as an accessible proxy for the notion of a weak tie\cite{GranWeakTies} (see Supplementary Information for detailed definitions). To check that our results are robust to alternative definitions of weak ties, we repeat our analysis (obtaining very similar results) using low contact frequency ties in Supplementary Information. Figure \ref{fig:num_weak_bridges} panel a provides an illustration of the way in which chance encounters led to the formation of new local bridges.

The removal of physical co-location (as a consequence of mandatory fully remote work) caused an immediate and persistent drop in the number of weak ties formed in the MIT email network. Figure \ref{fig:num_weak_bridges} panel b shows the causal effect of a lack of co-location estimated with a piecewise polynomial regression discontinuity design on both the number of weak ties and new (not previously seen) weak ties. There is a statistically significant 6.2\% drop in the number of weak ties and a 38.7\% drop in the number of new weak ties coinciding with the sudden absence of co-location. Because we study a fixed population of users, as we see more  ties the number of new weak ties will naturally decrease -- thus the downward trend of new weak ties is expected. However, the significant jump discontinuity on March 23, 2020 indicates that the absence of physical co-location is negatively associated with the ability to form new weak ties. The decrease in the addition of new weak ties hints at a stagnation effect -- researchers are not updating their pool of weak ties as often as would be expected.

Not only is the drop in weak ties sudden and statistically significant at the onset of the transition to full remote working, but it is also cumulatively significant over the course of more than one year. Using BSTS to estimate the cumulative effect of a lack of co-location, we see in Figure \ref{fig:num_weak_bridges} panel c a significant predicted loss of more than 5,100 weak ties from March 23, 2020 until July 15, 2021 due to remote work -- approximately 1.8 ties per person in the 2,834 researchers we study. Thus we find that remote work leads to a long-lasting, statistically significant drop in weak ties. Panel c also shows a non-significant cumulative drop in the number of new weak ties formed in the network -- we explain this phenomenon in detail in the following section. 

Finally, we identify a striking difference in the mechanism via which local bridges disappear due to remote work: in the short-term through the end of the Spring 2020 semester, local bridges become embedded in triangles, while in the long-term they are dropped from the network (Figure \ref{fig:num_weak_bridges} panel d). We also confirm in the Supplementary Information that ego networks become more stagnant is the absence of co-location -- the social contacts of researchers become more similar from week to week after remote work. Specifically, by computing the intersection over union of the edges in the daily reciprocated email networks on day $d$ and day $d+7$, we see a lasting, significant increase in the stability of network edges from week to week after remote work (Supplementary Figure 1 panel h).

\begin{figure}[h!]
    \centering
    \includegraphics[width=\linewidth]{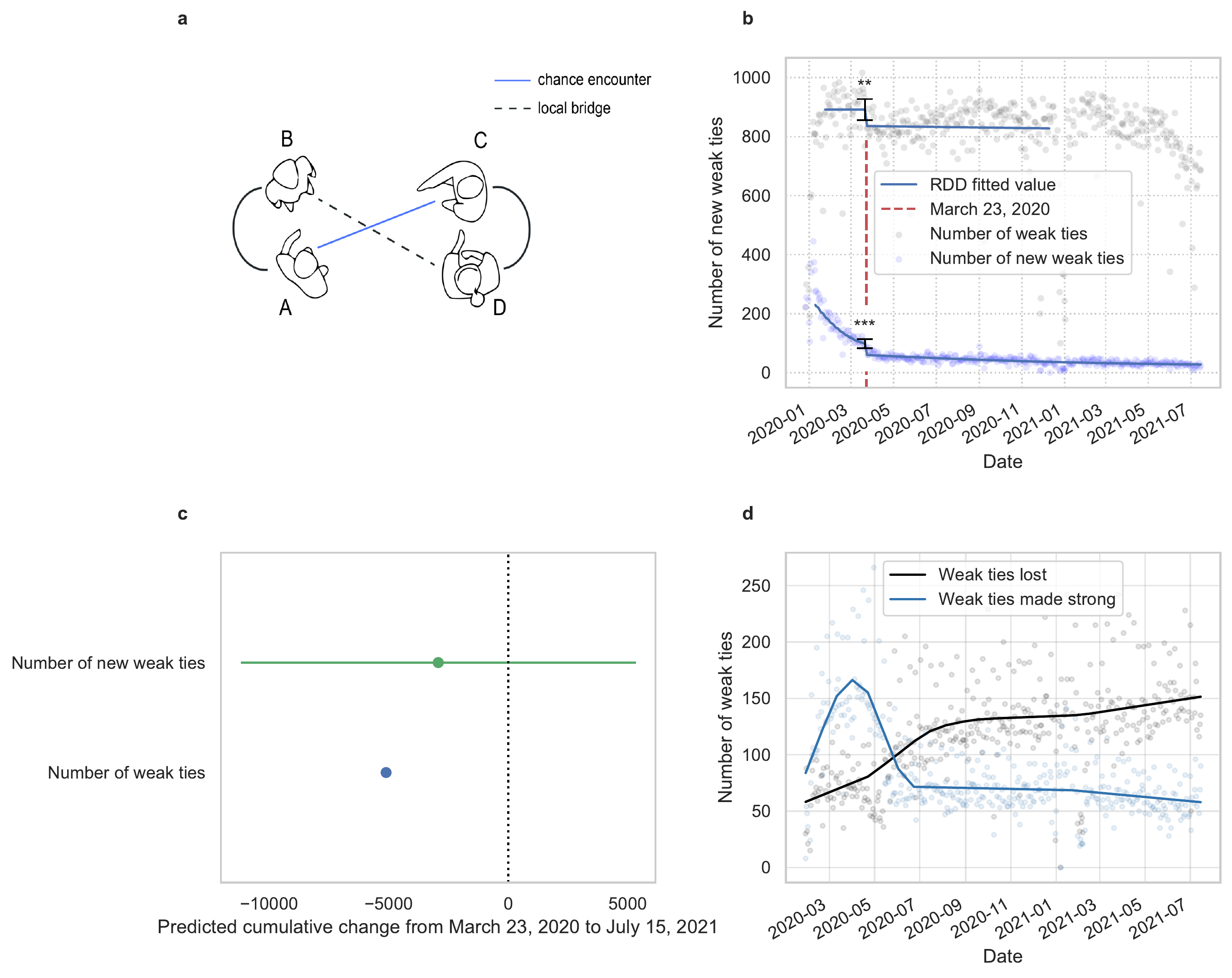}
    \caption{{\bf Changes in weak ties in the MIT email network after the shift to remote work}. ***$: p < .001$, **$: .001 \leq p <.01$, *$: .01 \leq p < .05$. {\bf a}, A candidate mechanism for local bridge formation in a social network which requires co-location. {\bf b}, A drop of 55.70 (6.2 \%) in the number of local bridges (weak ties) after March 23, 2020 ($p = .002$, 95\% CI: [-91.627, -19.774]). There is a drop of -38.03 (38.7\%) in the mean number of new (not previously seen) weak ties appearing each weekday after March 23, 2020 ($p < .001$, 95\% CI: [-53.38, -22.68]). Statistics represent the results of a two-sided $z$ test corresponding to a local polynomial regression discontinuity design ($n_{\mathrm{pre}} = 42$ days, $n_{\mathrm{post}} = 188$ days). {\bf c}, There is a cumulative loss of 5110 weak ties throughout an entire year ($p < .001$, 95\% PPI: [-4957,-5267]) and a non-significant loss of 2930 new weak ties ($p = .241$, 95\% PPI: [-11588, 5730]). Posterior predictive intervals computed using Bayesian structural time series ($n_{\mathrm{pre}} = 8$ weeks, $n_{\mathrm{post}} = 72$ weeks). {\bf d}, The number of weak ties which become strong (become embedded in triangles) or are churned (dropped from the network) in a 30 day rolling window. Fitted values/intervals use the mean as the measure of central tendency.}
    \label{fig:num_weak_bridges}
\end{figure}

\subsection{Weak tie formation and physical proximity}

As people are more likely to meet by chance on campus if their offices are nearby \cite{Simini2012AUM}, we expect to observe relatively more consistent changes in the formation of new weak ties for co-located MIT personnel. For each week we predict the mean value of the dependent variable (weak ties between researchers in a fixed distance range) during business days. We use the mean of weekend values as our covariate for the BSTS approach with treatment on March 23, 2020 outlined previously. To have a consistent measure of distance across all days in the data, we use distance between the campus offices of researchers rather than the distance between their active work environments -- during the shift to remote work the distance between researchers' campus offices does not change. We use four distance thresholds: $0$m (researchers working in the same lab), $0-150$m (close/nearby researchers in distinct labs), $150-650$m (researchers at medium distance), $> 650$m (far away researchers). The distribution of distances between researcher offices can be found in Supplementary Information.

\begin{figure}[h!]
    \centering
    
    \includegraphics[width=\linewidth]{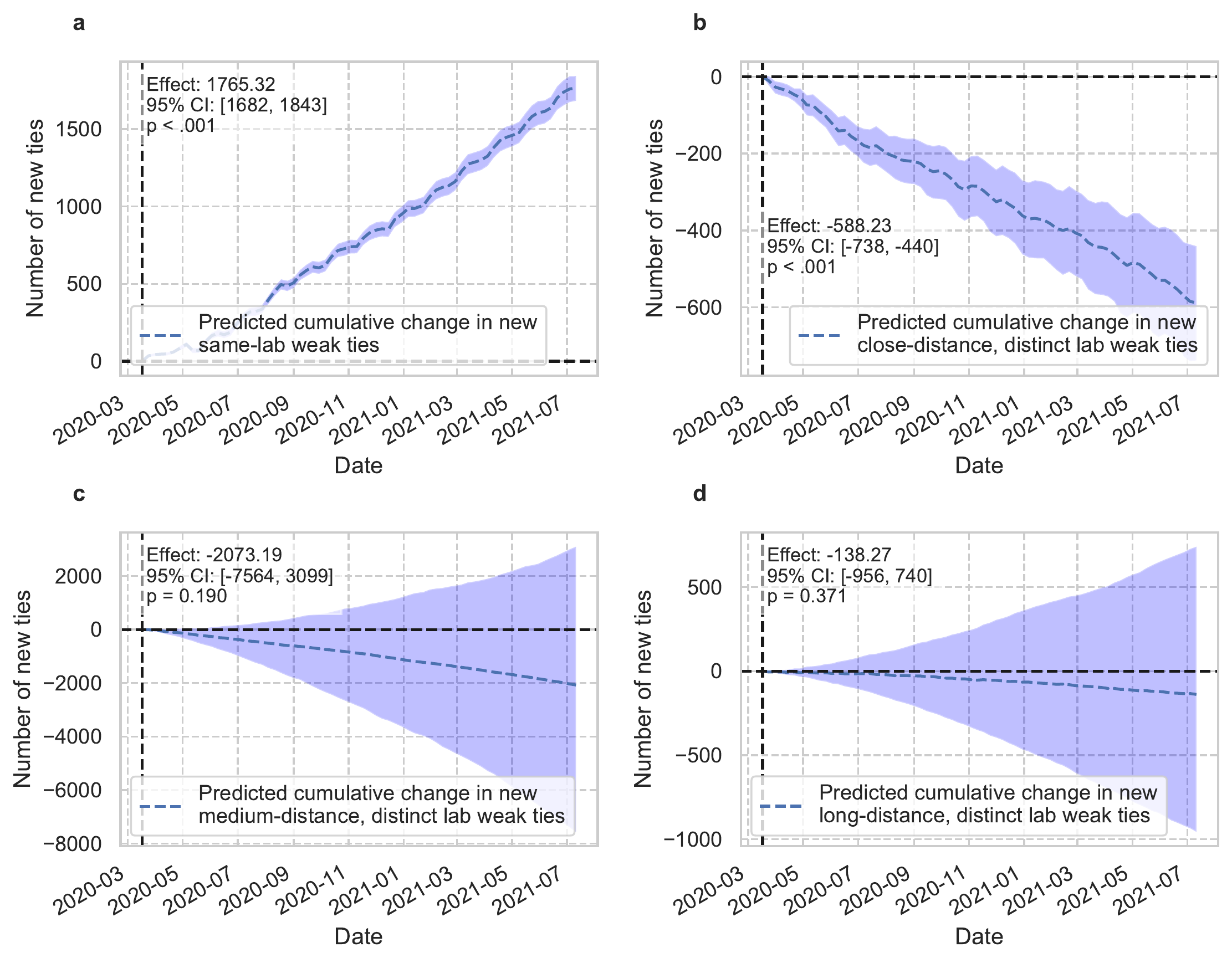}
    
    \caption{{\bf Formation of new weak ties stratified by distance}. {\bf a} The estimated effect  between researchers in the same lab. {\bf b} The estimated effect between researchers in distinct labs within 150 meters. {\bf c} The estimated effect between researchers in distinct labs between 150 and 650 meters. {\bf d} The estimated effect between researchers in distinct labs further than 650 meters. Shaded regions represent 95\% posterior predictive intervals computed using Bayesian structural time series with a synthetic counterfactual constructed from weekend data ($n_{\mathrm{pre}} = 8$ weeks, $n_{\mathrm{post}} = 72$ weeks for all panels). Fitted values/intervals use the mean as the measure of central tendency.}
    \label{fig:causal_spatial}
\end{figure}

 Using BSTS with a synthetic counterfactual constructed from weekend email network data, we find an immediate and lasting drop in the number of new weak ties between researchers in distinct but nearby research labs (Figure \ref{fig:causal_spatial} panel b). This is in line with our expectation that propinquity contributes to weak tie formation. Given this decrease, it may seem surprising that there is an increase in the number of new weak ties between researchers in the same lab (Figure \ref{fig:causal_spatial} panel b). However, Yang et al.\cite{yang2021the} discovered an increase in the use of asynchronous communication (e.g. email) after the COVID-19 pandemic. A plausible explanation for the increase in new weak ties between researchers in the same lab is that after the shift to remote work, email was used to schedule one-on-one meetings or ask small questions between same-lab researchers formerly scheduled or asked in person. Figure \ref{fig:causal_spatial} panels c and d show a non-significant decrease in the number of new weak ties formed between researchers in distinct labs at medium and far distances. This is also compatible with our intuition, as we don't expect researchers who work far away from one another to have many chance encounters even when working in-person.

\subsection{The effect of hybrid work on the formation of weak ties}

MIT re-opened its campus for the Fall 2021 semester starting on September 8, 2021. However, following MIT recommendation many research labs adopted a hybrid mode of work with researchers only physically present for (at most) 3 out of 5 business days each week, implying that the chance of serendipitous encounters was still lower than before the COVID-19 pandemic. Furthermore, limitations on the number of people allowed to eat together at a time and ongoing restrictions to international travel prevented departments from hosting large-scale events where researchers might typically mix. To estimate the causal effect of the end of remote work, we calculate a synthetic counterfactual for 2021 weekday email data from 2020 weekday email data (see Methods).

\begin{figure}[h!]
    \centering
   
    \includegraphics[width=\linewidth]{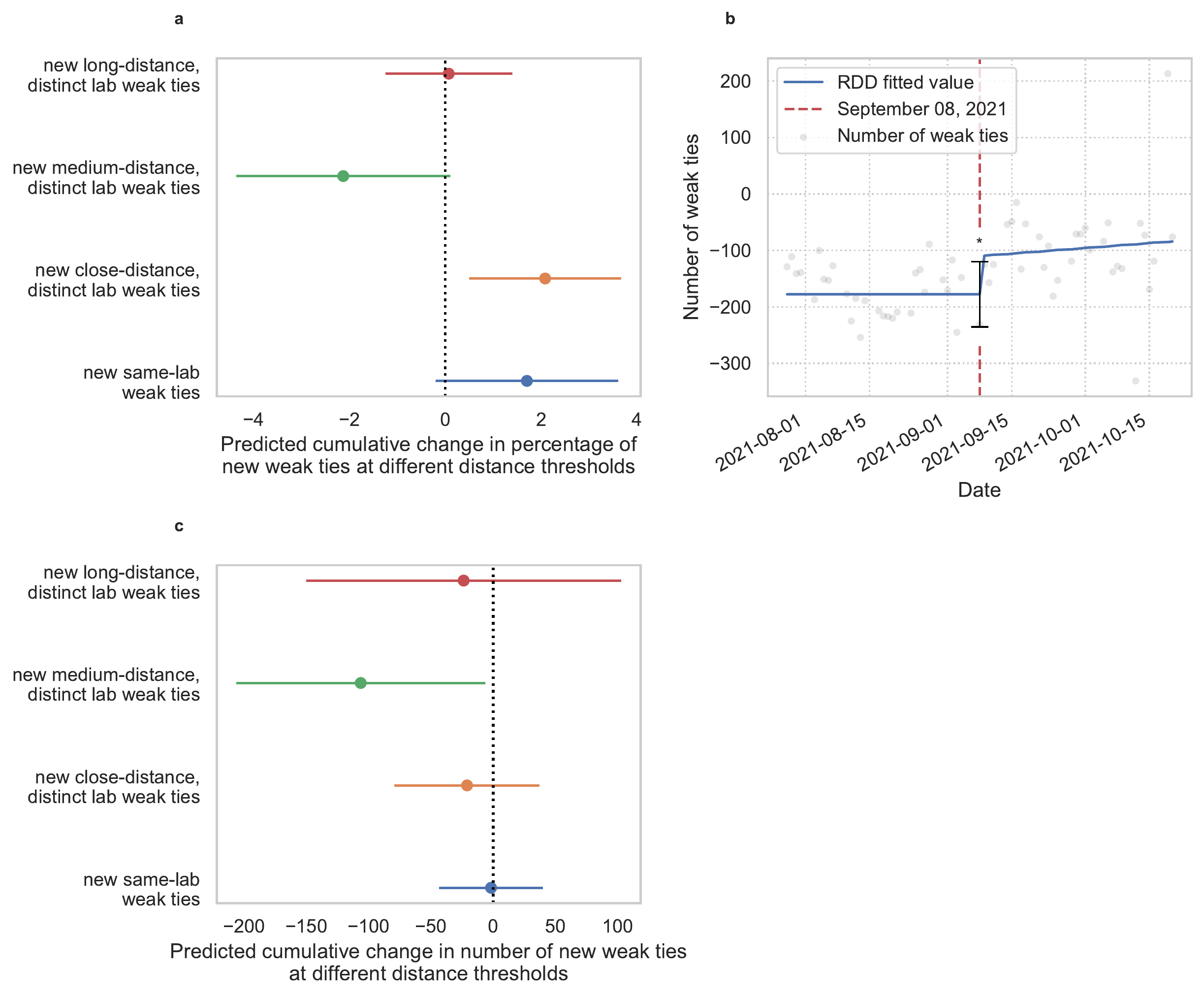}
    
    \caption{{\bf The effect of re-introducing co-location between researchers.} ***$: p < .001$, **$: .001 \leq p <.01$, *$: .01 \leq p < .05$. {\bf a}, The predicted cumulative change due to hybrid work in the \emph{percentage} of new weak ties between researchers in the same lab, researchers in distinct labs at distance less than 150m, researchers in distinct labs at distance between 150m and 650m, and researchers in distinct labs at distance larger than 650m. 95\% posterior predictive intervals computed using Bayesian structural time series ($n_{\mathrm{pre}} = 29$ days, $n_{\mathrm{post}} = 35$ days). {\bf b}, A RDD on the \emph{difference} of the number of weak ties in September 2020 and September 2021 using a two-sided $z$ test $(\mathrm{effect} = 67.5, p = 0.022, 95\%$ CI: $[9.740,	125.266], n_{\mathrm{pre}} = 34$ days, $n_{\mathrm{post}} = 26$ days). {\bf c}, The changes in the \emph{total number} of new ties between researchers at the distance thresholds detailed above with 95\% posterior predictive intervals computed using Bayesian structural time series. Fitted values/intervals use the mean as the measure of central tendency.}
    \label{fig:combined_ties_fall2021}
\end{figure}

The \emph{percentage} of new weak ties at close distances is higher than expected in Fall 2021 given the percentage at close distances in Fall 2020 (Figure \ref{fig:combined_ties_fall2021} panel a), with no significant differences observed at other distance thresholds. The number of weak ties rises more sharply than expected on September 8, 2020 (Figure \ref{fig:combined_ties_fall2021} panel b) given the rise at the beginning of the Fall 2020 semester. Despite this, the total number of new weak ties is lower or similar to the predicted value after hybrid work (panel b). Taken together, the results of Figure \ref{fig:combined_ties_fall2021} hint at the partial but incomplete success of the hybrid work model at allowing researchers to once more form new weak ties with other proximal researchers.

\subsection{Modeling the effect of distance on tie formation}

Our empirical results are consistent with the existence of some kind of mechanism via which co-location promotes weak tie formation. Here we seek to address the following question: given a collection of potential communicating pairs of researchers, how can we choose which pairs of researchers communicate each weekday in order to accurately capture the topological structure and temporal dynamics of real world networks?

\begin{figure}[h!]
    \centering
    \includegraphics[width=\linewidth]{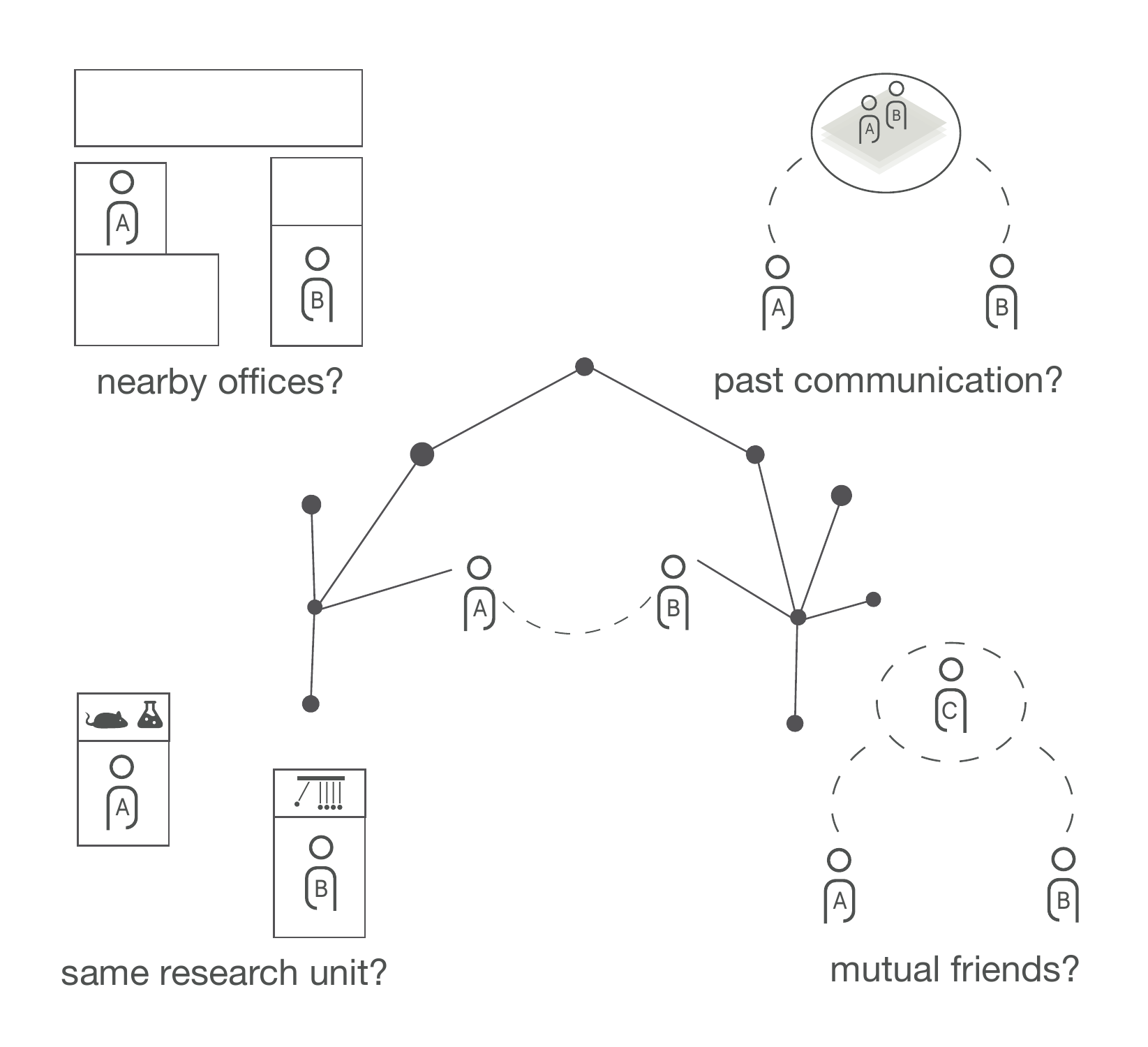}
    \caption{{\bf Illustration of model mechanisms.} In our tie choice model, illustrated above, the probability of two people interacting depends on four key mechanisms: Co-location, whether the pair belongs to the same research unit, whether the pair have a mutual connection, and their pattern of past communication.}
    \label{fig:simu}
\end{figure}

Previous work has identified at least four factors relevant to tie formation\cite{KossWattsEmpirical,linkcentPhys,propmult}: focal closure, triadic closure, link-centric preferential attachment, and physical co-location. Figure \ref{fig:simu} gives a diagramatic depiction of these four factors. However, determining a simple functional form via which these factors combine in order to govern the evolution of a dynamic communication network is still an open problem. Here we describe a simple network evolution model via which co-location multiplicatively scales the effect of homophily in order to determine which pairs of people communicate on a given day (see Methods for details). We simulate the formation of email networks on weekdays by creating an edge memory dictionary from the last two weeks of February 2020, then generating new graphs each day using our model. Previous work on the effect of distance on tie formation has primarily focused on distance and homophily as separate, additive factors which both contribute positively to tie formation\cite{McPherson2001BirdsOA,WimmerLewis,Burt2002BridgeD}. However, bringing people with clashing personalities close together is more likely to make them enemies than friends, and hence unlikely to make them contact one another via email. For this reason we view co-location as a multiplicative factor which scales the effects of homophily in order to form new ties. See Supplementary Information for an ablation study of our model together with a discussion of other random graph models.

\begin{figure}[h!]
    \centering
    \includegraphics[width=\linewidth]{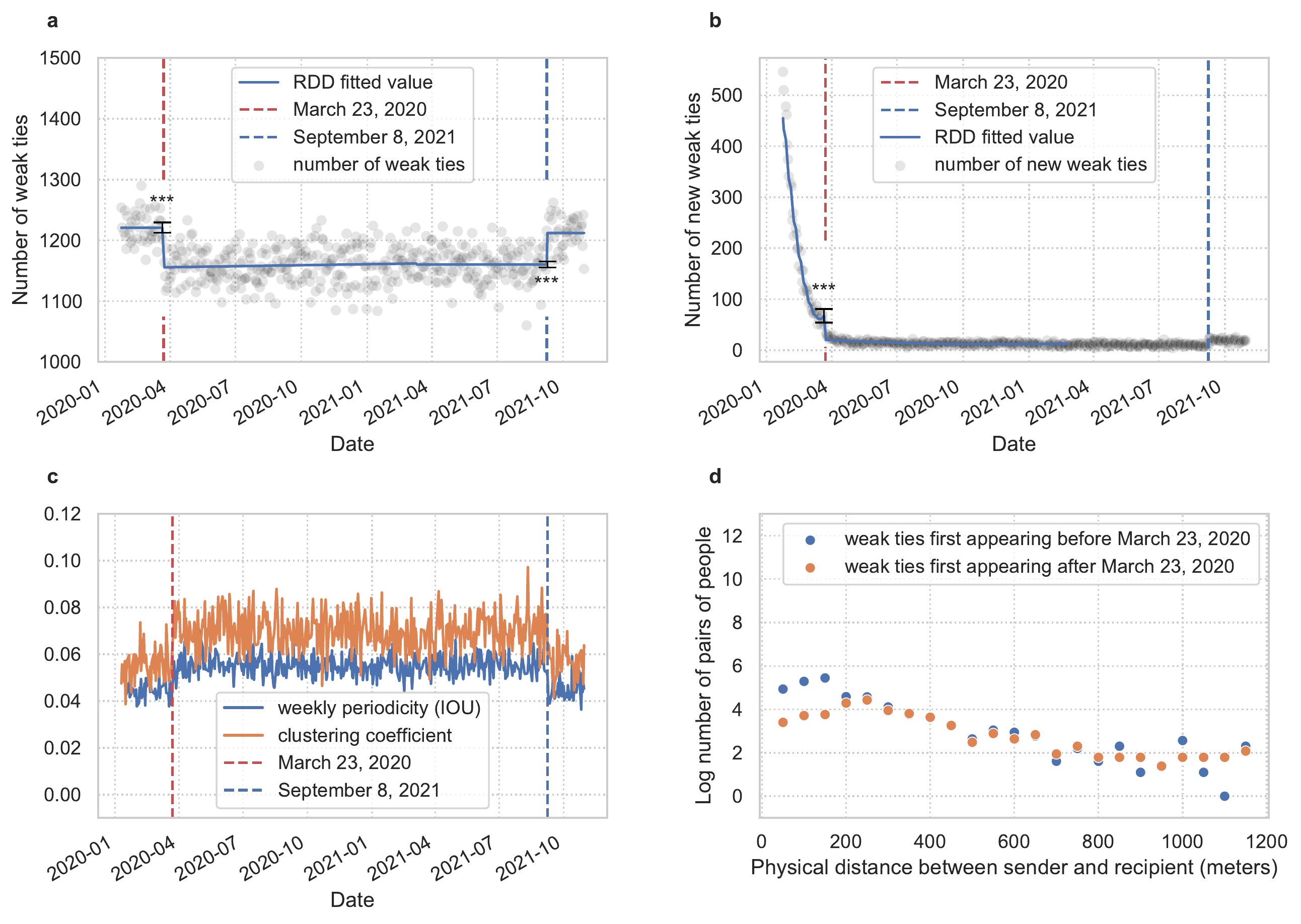}
    \caption{{\bf A tie choice model in which distance multiplicatively scales the effects of homophily.}  The output of our model when artificially setting office distances to a large fixed constant after March 23, 2020, then returning researcher offices to their initial positions on September 8, 2021. ***$: p < .001$, **$: .001 \leq p <.01$, *$: .01 \leq p < .05$. {\bf a}, A simulated drop in number of local bridges on March 23, 2020 $(\mathrm{effect} = -65.2, p < .001, 95\%$ CI: $[-76.109, -54.316], n_{\mathrm{pre}} = 42, n_{\mathrm{post}} = 238))$ followed by a rise in the fall $(\mathrm{effect} = 51.9, p < .001, 95\%$ CI: $[42.091, 61.693], n_{\mathrm{pre}} = 129, n_{\mathrm{post}} = 38)$. {\bf b}, The number of new weak ties entering the network $(\mathrm{effect} = -47.0, p < .001, 95\%$ CI: $[-60.566, -33.510], n_{\mathrm{pre}} = 42, n_{\mathrm{post}} = 238))$. All statistics are two-sided $z$ tests. {\bf c}, The weekly periodicity and daily clustering coefficient, {\bf d}, The number of simulated weak ties between users in distinct research units first appearing between February 4 and March 23, 2020 versus between March 23, 2020 and May 22, 2020. Fitted values/intervals use the mean as the measure of central tendency.}
    \label{fig:combined_simu}
\end{figure}

Our goal is to reproduce the dynamics of weak tie formation in networks with a simple network evolution model. To reproduce the qualitative features observed in the data, we set the co-location variable $\tau$ for each pair of individuals to zero (corresponding to no physical co-location) starting from March 23, 2020 then back to 1 on September 8, 2021. Upon the removal of co-location, our model produces a drop in the number of weak ties (Figure \ref{fig:combined_simu} panel a) and new weak ties (Figure \ref{fig:combined_simu} panel b) which is qualitatively similar to what we observe in the empirical data. It also reproduces the increase in edge stability (Figure \ref{fig:combined_simu} panel c), as well as the robustness of long-distance ties to a sudden absence of co-location (Figure \ref{fig:combined_simu} panel d). Here the weekly periodicity is measured as the intersection over union of the edge sets of the networks on days $d$ and $d-7$. By looking at the number of weak ties, the clustering coefficient, and the week to week periodicity (panels a,b) produced by our model after September 8, 2021, we see that our model  predicts that complete reintroduction of co-location results in a complete recovery of weak ties. The signs of the log of the distance interaction coefficients allows us to identify the following potential mechanism for the drop in weak ties: two researchers are more likely to form new weak ties when they are co-located. To further confirm this hypothesis, we have simulated a scenario without the sudden transition to fully remote work modeled through the change in the value of the physical co-location variable on March 23, 2020. The results, reported in the Supplementary Information, show no observable drop in weak ties, providing further evidence in support of our explanatory hypothesis.

\section{Discussion}

Several sociologists have argued that the lack of connections during the COVID-19 pandemic negatively impacted mental and physical well-being as well as innovation, collaboration, and creativity \cite{ForbesBrower, MITSloanPentland}. However, the mechanism via which such effects have occurred has yet to be explicitly identified. 
As businesses and universities make crucial decisions about the amount of in-person work after the COVID-19 pandemic, understanding the lasting effects of remote work on research communities is of paramount importance. 

Our study shows that the transition to fully remote work on the MIT campus -- with consequent complete removal of physical co-location between co-workers -- had notable effects on the email communication network: while some common topological features were preserved, the formation of weak ties was hindered, causing weak tie deterioration and network stagnation in the long term. Employees who are not co-located are less likely to form ties, weakening the spread of information in the network \cite{Hansen1999TheSP,Argote2000KNOWLEDGETA,Reagans2003NetworkSA}. The mechanism of weak tie formation can be successfully reproduced using a link formation model through which co-location multiplicatively scales the effects of homophily between researchers. 


Our findings have implications for the design of future research campuses and work environments, as well as for the development of new virtual technologies which seek to recreate interactions that happen in physical offices.  
Today it is of the utmost importance to identify what is the ``minimum amount" of in-presence work that enables the formation of weak ties, so that individual and societal benefits related to remote work can be preserved without impacting the generation of new ideas and innovation in general. 
While previous studies have documented the effects of reduced in-person collaboration in the short term\cite{yang2021the}, we show that the shift to remote communication produces a long-lasting impact on the formation of local bridges in collaborative networks, with effects accumulating over time. Expanding on the existing methodology, we demonstrate that paired testing estimates, conventionally used to evaluate the short-term effects of stay-at-home restrictions, do not accurately represent the long-term changes in the social network. We provide an alternative estimation strategy that produces more robust inference about the long-term effects.   

We also would like to highlight several important considerations on our causal inference strategy. We use the Bayesian structural time series method to estimate the long-term effects of the mandatory shift to remote work by constructing a counterfactual prediction of weekday email exchanges using weekend emails data. The validity of our methodology is supported by three main arguments. First, at a weekly level, weekend email exchanges are sufficiently predictive of the weekday emails due to common unobserved seasonality and the fact that work communication tends to spillover into weekends. Second, even if weekend email exchanges are affected by the shift to mandatory remote work, this is not due to changes in co-location of researchers since people typically do not come to the office on weekends. Finally, while there are other potential pandemic-related confounding factors such as changes in childcare which could \textit{a priori} contribute to changes in the MIT email network, it is reasonable to expect such confounding factors to be distributed independently of the  distance between researcher offices. Thus by stratifying connections by the distance between researcher offices, we directly attribute the observed loss in new weak tie formation to a lack of co-location. Still, our approach is not without its limitations. In particular, the predictive power of the weekend email exchanges is limited by the fact that we only observe the network for three months prior to intervention and could benefit from adding data for prior years.

Our results suggest that the loss of social connections that otherwise spontaneously emerge in shared spaces can not be immediately restored by simply returning back to offices. When designing work-from-home policies, firms and organizations should consider ways to promote serendipitous interactions across organizational units if they want to retain efficient discovery and transmission of novel information. Still, our initial findings on the hybrid work model that followed the reintroduction of partial in-person collaboration at MIT show a slight recovery in the number of weak ties -- especially between researchers who are once again co-located. This hints at the possibility of establishing a work balance trade-off by combining in-person and remote interactions among colleagues, which could inform the transition to a hybrid, post-COVID-19 ‘new normal’.

\section{Methods}
\subsection{Ethical Review}
This research was reviewed and classifed as exempt by the
Massachusetts institute of Technology (MIT) Committee on the Use of Humans as Experimental Subjects (MIT’s Institutional Review Board), because the research was secondary use research involving the use of de-identifed data.

\subsection{Data Preprocessing}

Data was analyzed using python 3.7.9, numpy 1.21.5, pandas 1.1.5, statsmodels 0.12.1, osmnx 1.0.1, python-flint 0.3.0, scipy 1.6.0, and geopandas 0.8.1. 

We start with a fixed set of anonymized researchers $\mc R$ (research staff, faculty, and postdocs) from 112 different research units. As required by MIT staff for privacy reasons, these researchers are grouped via 10 random partitions (no researchers are shared between the groups in a partition) of $\mc R$, $\mc R_i = \coprod_j G_j^i$, $1 \leq i \leq 10$. We describe below how to estimate individual-level data from aggregated data which was necessary in order to ensure that the studied users were active through the entire time period of interest. Each group $G_j^i$ contains at least 5 researchers, and all researchers in $G_j$ belong to the same research unit. Let $g_{ij} \in \mc R$ denote the collection of researchers in $G^i_j$. For a pair of groups $G^i_j$,$G^i_k$, and a day $d$ our data contains the sum $W^i_{jk}$ of all emails sent between groups $j$ and $k$ for randomization $i$:
\begin{equation}\label{eq:dioph}
W^i_{jk} = \sum_{g_{ij} \in G^i_j} \sum_{g_{ik} \in G^i_k} \Emails_d(g_{ij},g_{ik})
\end{equation}
For each research unit $U$, let $\mc R^U$ denote the collection of researchers in the research unit. Because each group contains only researchers from a single research unit, each equation \eqref{eq:dioph} is a sum over researchers from at most two research units. Grouping the equations \eqref{eq:dioph} by pairs of research units $U,V$ on each day $d$ we obtain a collection of constrained linear systems of Diophantine equations 
\begin{align}
W^U_V =& A^U_V x^U_V \\
 x^U_V \geq& 0
\end{align}
where $x$ is the column vector whose entries are $\Emails_d(g_{ij},g_{ik})$, $1 \leq i \leq 10$, $g_{ij} \in \mc R^U$, $g_{jk} \in \mc R^V$ and $A^U_V$ is the matrix of coefficients of the equations \eqref{eq:dioph}.
Each of these linear systems is guaranteed to have at least one solution (the actual number of emails sent), but may be underdetermined. If there is a unique solution, we use the Hermite normal form \cite{Bradley1971AlgorithmsFH} of $A^U_V$ to find it. If the system is underdetermined, we use non-negative matrix factorization \cite{Lee1999LearningTP} with an $\ell^1$ penalty in order to quickly estimate a sparse non-negative solution, as the algorithms which compute exact sparse integer solutions are slow. This procedure yields an estimate for each day $d$ and each pair $u,v \in \mc R$ of $\Emails_d(u,v)$. In the Supplementary Information we show that our approximate solutions are very close to true solutions.

\subsection{Network formation}
To rule out changes in the data due to departures from the university or new hires, we ensure that each user sent at least one email over the university network before the end of the 2020 spring semester \emph{and} at least one email after May 20, 2021. Denote this set of active users by $\mc A$.

We are missing data from December 23, 2020 and January 19,20, and 21 of 2021;  because these days are during the winter holiday at MIT this does not significantly affect our analysis. For each of the remaining 562 days from December 26, 2019 through July 15, 2021, we obtain a weighted, undirected network whose nodes represent (anonymized) individuals. Fix a day $d$, for each user $u$ in $\mc A$, let $\Nb_d(u)$ denote the number of people whom $u$ emailed on day $d$. For a pair of users $u,v \in \mc A$, let $\Emails_d(u,v)$ denote the (estimated as in the previous section) number of emails sent from $u$ to $v$ on day $d$.  To rule out massmails, let $\mc A_d \subseteq \mc A$ be the subset
\[
\mc A_d = \{u \in \mc A \mid 1 \leq \Nb(u) < 100 \}.
\]
Define a weighted, undirected network $N_d$ with nodes $\mc A_d$. For two nodes $u,v$, there is an edge $(u,v)$ if $\Emails(u,v) \geq 1$ and $\Emails(v,u) \geq 1$. The weight of the edge is defined to be $\min(\Emails_d(u,v),\Emails_d(v,u))$. Although the email data is partially estimated due to randomization and aggregation, for $> 66$ \% of the edges with non-zero weight in the estimated network $N_d$ we were able to recover the true number of emails sent. If we include all edges between users contributing to a non-zero weight edge in at least one random aggregation of the network (but which may have weight zero in the estimated network), $> 99.9$\% of edges have the ground truth number of emails. When building the undirected network $N_d$, we consider four possible time-windows during which emails can be reciprocated: the same day (daily), within 5 business days (weekly), within 10 business days (bi-weekly), or within 21 business days (monthly). For results on weekly, bi-weekly, and monthly networks see Supplementary Information. Previous studies have found that more than 90\% of emails are replied to the same day that they are sent, with more than half being replied to within 47 minutes \cite{EmailReplyTimes}. Requiring emails to be reciprocated the same day hides interactions between users who typically respond slowly to emails; however, it is useful for filtering out massmails, observing sharp discontinuities in the data, and for increasing the power of hypothesis tests. Allowing longer periods of reciprocation captures weaker ties missed in the daily reciprocated email network, but we are forced to sacrifice some statistical power either to autocorrelation or lower sample size; additionally the networks become more saturated, destroying some topological features of interest. Examples of daily, weekly, bi-weekly, and monthly email networks are reported in the Supplementary Information.


To examine whether this hybrid mode of work returned tie formation to pre-pandemic levels, we first restrict to a collection of 2,206 researchers who sent at least 5 emails after September 2021 and before May 2020, then proceed as above to form networks from December 23, 2020 to October 31, 2021. We choose a stricter requirement for inclusion in the network than previously as we observe many users becoming inactive starting in summer 2021. 

When comparing February 2020 to February 2021, we pair the days in the two months as follows: the first Tuesday of the MIT semester in February 2020 is paired with the first Tuesday of the MIT semester in February 2021, etc. For each pair of days $(d_{2020}^i,d_{2021}^i)$, we consider the set of users $\mc A_{d_{2020}^i} \cap \mc A_{d_{2021}^i}$ who were active on both days and, as above, form a pair of undirected networks $(G_{2020}^i,G_{2021}^i)$ whose edges $(u,v)$ correspond to reciprocated emails on $d_{2020}^i$ (for $G_{2020}^i$) or $d_{2021}^i$ (for $G_{2021}^i$). Directly comparing these networks allows us to completely remove the effects of seasonality or any difference in makeup of active users in February 2020 and February 2021. 

\subsection{Link-centric preferential attachment}

The goal of our model is not to serve as a tool for prediction, but to understand the mechanism via which distance impacts link formation. Through experimentation, we found that using link-centric preferential attachment alone to propagate a dynamic network produced daily networks which had too many local bridges. Thus we use a two step approach which first produces an intermediate network using link-centric preferential attachment, then adds edges in a way which increases the clustering coefficient of the network. This is analogous to the reverse of the Watts-Strogatz method \cite{Watts1998CollectiveDO}, where the intermediate network has high clustering coefficient and local bridges are added afterwards. 


Our link formation model has the following parameters:
\begin{itemize}
    \item $P$, the {\bf p}eriodicity of the model. $P$ controls how much the graph on day $d$ looks like the graphs on days $d-7$. In other words, the higher the value of $P$, the closer the dynamic network is to being 7-periodic (or 5-periodic if weekends are removed).
    \item $O$, the tendency to connect with {\bf o}ld links. The higher the value of $O$, the more likely it is that a given link will connect with a previous partner rather than someone new. This is one of the standard parameters from a vanilla link-centric preferential attachment model, and this parameter decays exponentially in the number of days between contact: $O = ce^{-d}$ where $c$ is a constant and $d$ is the number of days since the link last appeared.
    \item $N$, the tendency to reach out to {\bf n}ew people. This is typically the complement of $O$ in vanilla link-centric attachment models (we have more parameters than the standard two).
    \item $D$, the tendency to connect with people in the same {\bf d}epartment.
    \item $F$, the tendency to be introduced to a mutual {\bf f}riend.
\end{itemize}

The parameters $O,N$, and $P$ rely on a memory dictionary which stores the days on which a given edge has appeared. For all of the above parameters $\{P,O,N,D,F\}$, we include interaction terms 
\[
\{C_P,C_O,C_N,C_D,C_F\}
\]
controlling the extent to which co-location amplifies or dampens the effect. For example, from the empirical data we conclude that co-location should dampen periodicity while amplifying the probability of reaching out to new partners.

Let $e = (u,v) \in \mc A \times \mc A$ be a pair of nodes. For each parameter $Q \in \{P,O,N,D,F\}$ let $\bbm 1_Q$ denote the associated indicator variable. Specifically, 
\begin{align*}
\bbm 1_P(e) =& \begin{cases}
1 & \text{if $u,v$ connected 7 days in the past}\\
0 & \text{otherwise}
\end{cases}\\
\bbm 1_N(e) =& \begin{cases}
1 & \text{if $u,v$ have never connected}\\
0 & \text{otherwise}
\end{cases}\\
\bbm 1_O(e,d_c) =& \begin{cases}
\sum_{d_p} e^{-(d_c-d_p)/562} & \text{if $u,v$ have previously connected}\\
0 & \text{otherwise}
\end{cases}\\
\bbm 1_D(e) =& \begin{cases}
1 & \text{if $u,v$ are in the same research unit}\\
0 & \text{otherwise}
\end{cases}\\
\tau(e) =& \begin{cases}
1 & \text{if $u,v$ have offices within 150 meters}\\
0 & \text{otherwise}
\end{cases}
\end{align*}
where $d_c$ is the current day, and $d_p$ are the past days on which the tie $e$ was present in the network. The 150 meter cutoff is chosen based on the empirical results.

Consider the set $E$ of all edges which appear on at least one day in the empirical data. Note that the use of $E$ rather than the set of all possible edges makes this model unsuitable for prediction tasks. On each day $d$, we start by adding an edge $e \in E$ to the random network $G^d_1$ with probability proportional to
\begin{align*}
    p_{e} \propto &\tau(e)\sum_{Q \in \{P,O,N,D\}} C_Q  \bbm 1_Q(e) Q\\ 
    + & (1-\tau(e))\sum_{Q \in \{P,O,N,D\}} \frac{1}{C_Q} \bbm 1_Q(e) Q
\end{align*}
If $C_Q > 1$ then co-location amplifies the effect of parameter $Q$, while if $C_Q < 1$ it dampens the effect. In total, in the first step we add $\eps_1$ edges to $G^d_1$, where $\eps_1 \sim \mc N(1000,10)$.

In the second step, we add the parameter $F$,
\[
\bbm 1_F(e) = \begin{cases}
1 & \text{if $d_{G_1}(u,v) = 2$ }\\
0 & \text{otherwise}
\end{cases}
\]
where $G^d_1$ is the random network constructed in step 1, and $d_G$ is the usual (unweighted) shortest path distance in the graph $G$. In words, $\bbm 1_F(e) = 1$ if adding $e$ will close a triangle in the network. A new edge $e$ is added to the network in step two with probability
\begin{align*}
    p_{e} \propto &\tau(e)\sum_{Q \in \{P,O,N,D,F\}} C_Q \bbm 1_Q(e) Q\\ 
    + & (1-\tau(e))\sum_{Q \in \{P,O,N,D,F\}} \frac{1}{C_Q}\bbm 1_Q(e) Q
\end{align*}
In total we add $\eps_2$ edges to $G^d_1$ in the second step, where $\eps_2 \sim \mc N(500,10)$. Including the parameter $F$ has the effect of raising the expected number of triangles in the random graph, and hence lowering the percentage of edges which are local bridges.

With parameters fixed, we proceed to form networks one day at a time, adding the edges from the current network to a memory dictionary after formation. To model the effect of remote work, for each day $d$ after March 23, 2020, we set the distance between researcher offices to be a fixed constant larger than 650 meters (all other parameters remain fixed). For parameter values we set:

\begin{tabular}{lllll}
    $P = 80000$ & $N = 2000$ & $O = 1$ & $D = 90000$ & $F = 200000$\\
    $C_P = \frac{2}{3}$ & $C_N = 2$ & $C_O = 1$ & $C_D = \frac{10}{11}$ & $C_F = \frac{2}{3}$
\end{tabular}

\subsection{Regression Discontinuity}
Figure \ref{fig:num_weak_bridges} (panels a and b) (respectively Figure \ref{fig:combined_ties_fall2021} panel c) show the drop in weak tie and new weak tie formation (respectively increase in weak ties) due to the policy change on March 23rd, 2020. We used Regression Discontinuity Designs (RDD) \cite{thistlethwaite1960regression, hahn2001identification, lee2010regression} to estimate the causal impact of the policy change. RDDs are a classic, quasi-experimental procedure for estimating treatment effects in observational studies. In an RDD, treatment assignment is determined by an assignment variable rather than through randomization.

For an RDD to be valid, we need only assume that the response is continuous with the assignment variable near the cutoff and that subjects cannot precisely manipulate the assignment variable \cite{hahn2001identification, lee2010regression}. Panels a and b in Figure \ref{fig:num_weak_bridges} and panel c in Figure \ref{fig:combined_ties_fall2021} show that the responses (weak ties and new weak ties) are continuous with the assignment variable time, albeit observed with noise. The assignment variable, time, is not precisely manipulable by subjects since the announced policy was not known far in advance. Furthermore, there would be little reason to manipulate assignment since subjects are free to send emails at the same rate before and after the policy change. 

In  Figure \ref{fig:num_weak_bridges} panel a and Figure \ref{fig:combined_ties_fall2021} panel c, we model the weekly mean number of weak ties with the discontinuous linear regression 
\begin{equation} \label{eqn:rdd_weakties}
    Y = \alpha + \eta D + \beta_2 D (X - c) + \epsilon
\end{equation}
where $c$ is the cutoff date (either March 23rd, 2020 or September 8, 2021) and $D$ is the binary variable
\begin{equation*}
    D = \begin{cases} 
      1 & X \geq c \\
      0 & X < c
   \end{cases}
\end{equation*}
that indicates if the date $X$ is before or after the policy change date $c$. The error term $\epsilon$ is assumed to be heteroskedastic white noise. The coefficient $\eta$ is the impact of the policy, and measures the gap between the two sides of the regression. We estimate $\eta$ and the other coefficients with generalized least squares with AR(n) structured covariance matrix with $n=5$ and report the value of $\hat{\eta}$ and its p-value in Figure \ref{fig:num_weak_bridges}. We use heteroskedasticity-robust estimators for standard errors, so that in total standard errors are robust to autocorrelation (from the AR(5) GLS) and heteroskedasticity \cite{NBERt0055}.

In panel a, we assumed a discontinuous order one polynomial trend line because the data did not display any apparent higher-order non-linear behavior. The data were subset to January 3rd, 2020 to October 1st, 2020, to semi-localize our regression around the discontinuity, which reduces bias in $\hat{\eta}$ \cite{lee2010regression}, and to avoid influence from the two outlying regions (before January 3rd, 2020 and during December 2020). These outliers correspond to winter break at MIT and represent a natural and expected decrease in weak ties not due to the policy change. 

In Figure 2 panel b, we similarly model the rate of new weak tie formation over time. We assumed a second-order discontinuous polynomial trend (Equation \ref{eqn:rdd_weaktierate}) due to the observed parabolic behavior before the cutoff point. Using the same notation as in Equation \ref{eqn:rdd_weakties}, our linear regression is given by
\begin{equation} \label{eqn:rdd_weaktierate}
    Y = \alpha + \eta D + \beta_1(X - c) + \beta_2(X - c)^2 + \beta_3 D (X - c) + \beta_4 D (X - c)^2 + \epsilon.
\end{equation}
The coefficient $\eta$ is, again, the causal impact of the policy change. We report the value of $\widehat{\eta}$ and its p-value in Figure \ref{fig:num_weak_bridges}.

\subsection{Bayesian Structural Time Series}

We stress that when using Bayesian methods, reported CIs are credible intervals of the predicted dependent variable (also called posterior predictive intervals), and $p$-values are posterior tail probabilities. Bayesian structural time-series (BSTS) combines a state-space model for time-series data and Bayesian model averaging for parameter selection and estimation \cite{bstsBrodersen}.  

As a state-space model, BSTS combines three components of state: a local linear trend:
\begin{align*}
    \mu_{t+1} &= \mu_t + \delta_t + \eta_{\mu,t}\\
    \delta_{t+1} &= \delta_t + \eta_{\delta,t},
\end{align*}
with $\eta_{\mu,t} \sim \mc N(0,\sigma_\mu^2)$, $\eta_{\delta,t} \sim \mc N(0,\sigma_\delta^2)$; a seasonality component:
\[
\gamma_{t+1} = - \sum_{s=0}^{S-2} \gamma_{t-s} + \eta_{\gamma,t}
\]
with $S$ the number of seasons and $\eta_{\gamma,t}$ again an independent error; and (static) covariates which are predictive of the time series in question before the intervention:
\[
Z_t = \beta^T \mathbf{x}_t.
\]

For the local linear trend and seasonality components, we use the default priors of the CausalImpact library: 
\[
\frac{1}{\sigma} \sim \mc G(10^{-2},10^{-2}s_y^2) \qquad s_y^2 = \sum_t \frac{(y_t - \overline y)^2}{n-1}
\]
where $\mc G(-,-)$ denotes a Gamma distribution. For the covariates (the weekend data), in general a spike-and-slab prior is used with the spike defined by
\[
p(\xi) = \prod_{i=1}^J \pi_j^{\xi_j}(1-\pi_j)^{1-\xi_j)}
\]
with $\pi_j$ initialized to $\frac{M}{J}$ where $M$ is the expected model size. The slab part of the spike-and-slab prior is 
\begin{align*}
    \beta_\xi \mid \sigma_\epsilon^2 &\sim \mathcal N(0, \sigma^2_\epsilon (\Sigma_\xi)^{-1})^{-1}\\
    \frac{1}{\sigma^2_\epsilon} &\sim \mc G\left(\frac{\nu_\epsilon}{2},\frac{s_\epsilon}{2} \right)\\
    \Sigma^{-1} &= \frac{1}{n}\left\{\frac{1}{2}X^TX + \frac{1}{2}\mathrm{diag}(X^TX)\right\}
\end{align*}
where $X$ is the covariate data. Because we include only one covariate (the weekly minimum) the spike-and-slab prior collapses to just a normal-inverse Gamma distribution. We use 1000 iterations of Markov Chain Monte Carlo (MCMC) to compute posterior predictive distributions.

Consider the binary variable $X(r,d)$ defined by
\[
X(r,d) = \begin{cases}
1 & \text{researcher $r$ is in their office on day $d$}\\
0 & \text{otherwise}.
\end{cases}
\]
For us, ``treatment" consists of setting $X(r,d)$ to zero for $d$ greater than March 23, 2020 by not permitting researchers to enter their campus office. The time series whose counterfactual we want to estimate is the weekday maximum of the network measure while the covariate is the weekend minimum. 
As most employees are not physically present in their office on weekends, $X(r,d) = 0$ for most $r$ when $d$ is a weekend so that the treatment has little effect. We verify this assumption by looking at the number of distinct MAC addresses connected to routers in on-campus research labs on the weekday and weekend (see Supplementary Information).

When studying the effect of hybrid work, we construct a counterfactual using weekday email data spanning July 22 2020 through October 14, 2020 as a covariate for email data spanning July 28, 2021 through October 20, 2021, aligning so that the start of the Fall 2020 and Fall 2021 semesters coincide. We also remove Memorial Day (a university holiday) from both the 2020 and 2021 data. 

\section{Data Availability}
A subset of the data containing 1000 (anonymized) users is available on Zenodo\cite{carmody_daniel_2022_6809296}. Because of privacy concerns for MIT employees, the entire dataset of emails and pairwise distances cannot be publicly released. Source data for Figures 1,2,3,4,6 is available with this manuscript.

\section{Code Availability}
The code used to analyze the data can be obtained from Code Ocean\cite{51f79280-6c58-4e9a-a5ed-3aafdc05cea4}.

\section{Acknowledgments}
P.S. and C.R. thank FAE Technology, MipMap, Samoo Architects \& Engineers, GoAigua, DAR Group, Ordinance Survey, RATP, Anas S.p.A., ENEL Foundation and all of the members of the MIT Senseable City Laboratory Consortium for supporting this research. S.L. was supported by the Carlsberg Foundation (CF20-0044) and the Villum Foundation (Nation-scale Social Networks).

\section{Author Contributions}

 D.C., M.M., C.R., and P.S. designed the research. M.M. provided data access. D.C. and T.A.H. processed and analysed the data. D.C., M.M., T.A.H., T.A., and P.S. performed the interpretation and writing. S.L., T.A., and R.D provided theoretical expertise. P.S. and C.R. continuously advised the project. C.R. framed initial hypothesis.

\section{Competing Interests}
The authors declare no competing financial interests.

\section{Research Units}
Here we list the research units used in the study. Only a subset of researchers from each unit appear in the email networks. Note that in addition to standard denominations corresponding to departments (e.g. Chemistry), there are also subgroups corresponding to department heads and institute professors (distinguished professors).

Abdul Latif Jameel Poverty Action Lab, Aeronautics and Astronautics, Anthropology Program, Archaeology, Architecture, Architecture and Planning - Depart Heads, Biology, Brain and Cognitive Sciences, Center for Advanced Urbanism, Center for Biomedical Innovation, Center for Collective Intelligence, Center for Environmental Health Sciences, Center for Global Change Science, Center for Information Systems Research, Center for International Studies, Center for Real Estate, Center for Transportation and Logistics, "Chancellors Office", Chemical Engineering, Chemistry, Civil and Environmental Engineering, Comp Sci and Artificial Intel Lab HQ, Comp Sci and Artificial Intelligence Lab, Comparative Media Studies/Writing, Cybersecurity at MIT Sloan, D-Lab, DAPER Administration, DAPER Intercollegiate Sports, DLC Heads Science, Dean for Student Life - Dept Heads, Dean of Humanities, Arts, and Social Sci, Dean of Science, Department of Biological Engineering, Dept Administrators and Lab Directors, Dept Heads Vice President for Research, Division of Comparative Medicine, Earth, Atmospheric and  Planetary Sciences, Economics, Electrical Engineering-Computer Science, Global Studies and Languages, Haystack Observatory, Health Sciences and Technology Program, History Section, Industrial Performance Center, Inst for Data Systems and Society Profs, Inst for Medical Eng. and Science Prof, Institute Professors, Institute for Data, Systems, and Society, Institute for Medical Eng. and Science, Institute for Soldier Nanotechnologies, J-WAFS, Jameel Clinic for ML in Health, Kavli Inst for Astrophysics and Space Rsrh, Koch Inst - Integrative Cancer Research, Lab for Information and Decision Systems, Laboratory for Nuclear Science, Leaders for Global Operations Program, Libraries, Linguistics and Philosophy, Literature Section, MIT Energy Initiative, MIT Environmental Solutions Initiative, MIT Open Learning, "MIT Program in Womens and Gender Studies", MIT Quest for Intelligence, MIT Sea Grant College Program, MIT Sloan Human Resources, MIT-SUTD Collaboration, MIT.nano, MITii, Materials Research Laboratory, Materials Science and Engineering, Mathematics, McGovern Institute for Brain Research, Mechanical Engineering, Media Lab, Microsystems Technology Laboratories, Music and Theater Arts Section, Nuclear Reactor Laboratory, Nuclear Science and Engineering, OSATT-Technology Licensing Office, Office of the President, Office of the Provost, Open Learning, J-WEL, Open Learning, J-WEL Research, Projects, Open Learning, J-WEL, Higher Ed, Open Learning, Playful Journey Lab, Physics, Picower Institute for Learning and Memory, Plasma Science and Fusion Center, Political Science, Prog in Science, Technology, and Society, Program in Art, Culture and Technology, Program in Media Arts and Sciences, ROTC Air Force, ROTC Army, ROTC Navy, Research Laboratory of Electronics, SCC Dept/Lab/Center/ Director Org, SHASS Department Heads, School of Architecture and Planning, School of Engineering, School of Engineering Professors, Schwarzman College of Computing, Sloan School of Management, SoE Dept/Lab/Center/Director Org, Sociotechnical Systems Research Center, Supply Chain Management Program, System Design and Management Program, Urban Studies and Planning, VP for Research, World Wide Web Consortium

\section{Local bridges}
\begin{definition}
Given an undirected network $N$, an edge $(u,v)$ in $N$ is a \emph{local bridge} if there is no node $n$ such that there are edges $(u,n)$ and $(v,n)$. In other words, the endpoints of a local bridge have no common neighbors. 
\end{definition}

As the definition of a local bridge given above requires an undirected network, we will restrict to reciprocated communications when studying local bridges. The link between weak ties and local bridges comes via the following definition from Granovetter \cite{GranWeakTies}.

\begin{definition}
Fix an undirected network $N$ together with a labeling of each edge in $N$ as either ``strong" or ``weak". We say that a node $A$ violates the Strong Triadic Closure Property if it has strong ties to two other nodes $B$ and $C$, and there is no edge between $B$ and $C$. We say a node $A$ satisfies the Strong Triadic Closure Property if it does not violate it.
\end{definition}

From here, a well-known argument by contradiction implies that  local bridges and a weak ties are intimately related in networks which satisfy Strong Triadic Closure.

\begin{theorem}[Granovetter \cite{GranWeakTies}]
If a node $A$ in an undirected network $N$ satisfies the Strong Triadic Closure Property and is involved in at least two strong ties, then any local bridge it is involved in must be a weak tie.
\end{theorem}

There are two distinct ways to lose local bridges in a network: first, an edge which is a local bridge on day $d$ can fail to exist in the network on day $d+k$ for some $k > 0$; second, an edge which is a local bridge on day $d$ can be embedded in a triangle in the network on day $d+k$.  

\section{Cumulative plots and robustness of network formation}

Here we plot more detailed versions of results from the main text which show the evolution of the structural time series model uncertainty over time. Note that if our covariate (weekend data) becomes less predictive of weekday data, then the width of the posterior predictive interval will expand over time to compensate. Specifically, Supplementary Figure \ref{fig:num_weak_bridges} corresponds to Figure 2 in the main text, showing additionally the intersection over union of the network edge sets on days $d$ and $d+7$ each week in panel h. If people are not forming new weak ties, then assuming a constant number of ties over time (supported by the non-significant difference in number of emails sent before/after COVID), people must be holding onto old ties for longer or forming ties which already have a large overlap with their ego network. In both cases there would be increased redundancy in information spread. 
Panel h directly demonstrates a statistically significant increase in week to week redundancy, measure as  periodicity, in email networks due to the shift to remote work. Supplementary Figure \ref{fig:combined_ties_fall2021all} shows the evolution of uncertainty over time for Figure 4 in the main manuscript detailing the change in new weak ties with a partial re-introduction of co-location. Supplementary Figure \ref{fig:num_weak_bridges_nosmall} is identical to Supplementary Figure \ref{fig:num_weak_bridges} except that we do not prune nodes with more than 100 outgoing connections from the network in order to show that our results are robust to this pruning. Similarly Supplementary Figure \ref{fig:spatial_nosmall} is identical to Figure 3 in the main manuscript but computed without first pruning network nodes. We again note that there are no qualitative differences from the results in the main text. 

\begin{figure}
    \centering
    \includegraphics[width=\linewidth]{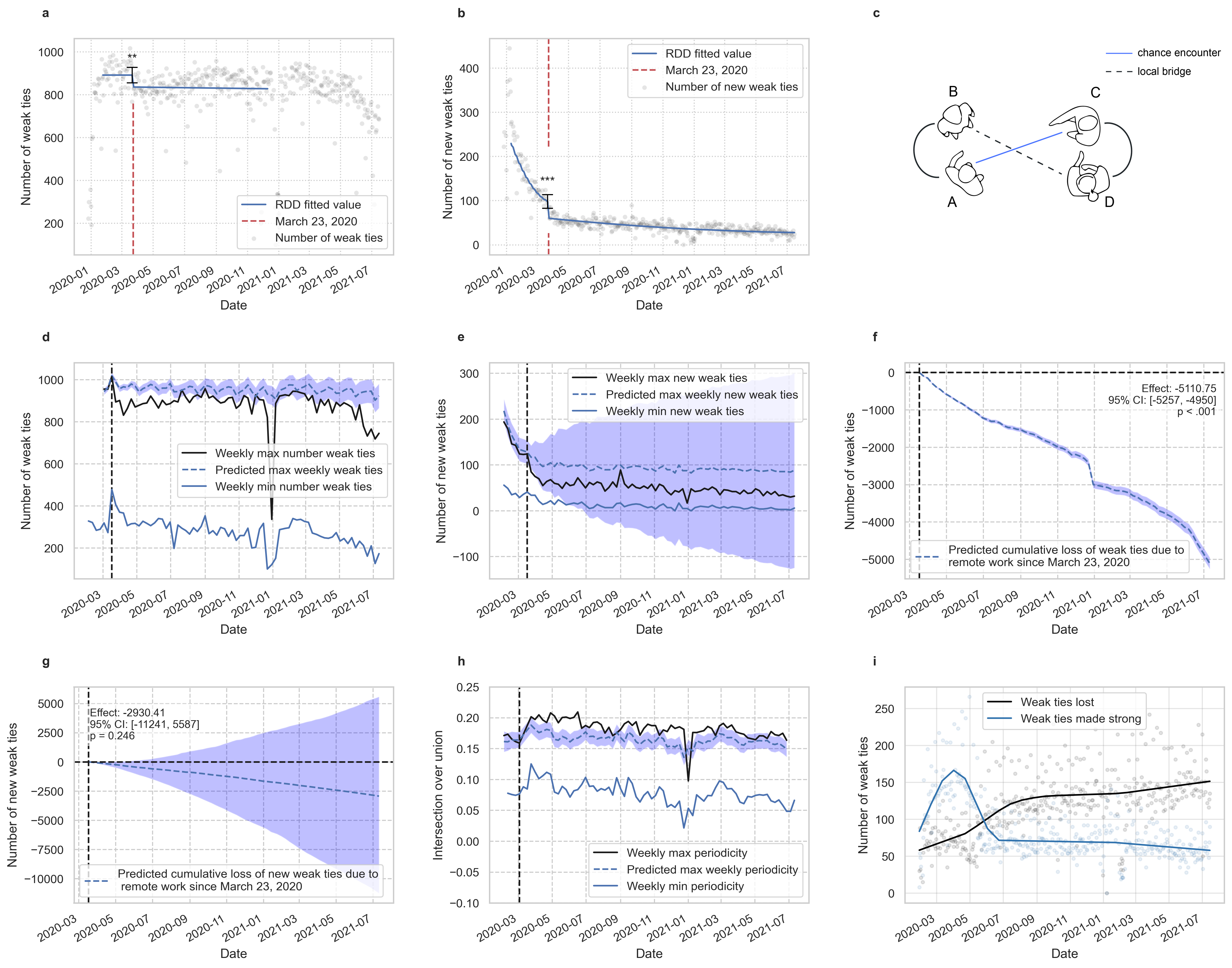}
    \caption{{\bf Changes in the structure of weak ties in the email network}. ***$: p < .001$, **$: .001 < p <.01$, *$: .01 < p < .05$. {\bf a}, There is a drop of 55.70 (6.2 \%) in the mean number of local bridges (weak ties) caused by the switch to fully remote work ($p = .002$, 95\% CI: [-91.627, -19.774]). {\bf b}, A statistically significant drop of -38.03 (38.7\%) in the mean number of new (not previously seen) weak ties appearing each weekday ($p < .001$, 95\% CI: [-53.38,     -22.68]). Statistics represent two-sided $z$ tests. {\bf c} Local bridges in a social network can be formed from chance encounters. {\bf d}, A predicted loss of  75.16 weak ties ($p < .001$, 95\% CI: [-77.44, -72.75]). {\bf e}, A predicted loss of 43.09 new weak ties ($p = 0.23$, 95\% CI: [-171.04, 79.08]). {\bf f}, A cumulative drop of 5110 in the number of weak ties throughout an entire year. {\bf g}, A cumulative drop of 2930 in new weak ties. {\bf h}, A 0.02 (10.81\%) increase in the week to week similarity of network edges ($p < .001$, 95\% CI: [7.75\%,13.9\%]). {\bf i}, The number of weak ties which become strong or are churned in a 30 day rolling window. Shaded regions show a 95\% posterior predictive interval computed using Bayesian structural time series. $n_{\mathrm{pre}} = 42$ days, $n_{\mathrm{post}} = 188$ days for panels {\bf a}, {\bf b}. $n_{\mathrm{pre}} = 8$ weeks, $n_{\mathrm{post}} = 72$ weeks for panels {\bf d}-{\bf h}. Fitted values/intervals use the mean as the measure of central tendency.}
    \label{fig:num_weak_bridges}
\end{figure}

\begin{figure}
    \centering
    \includegraphics[width=0.8\linewidth]{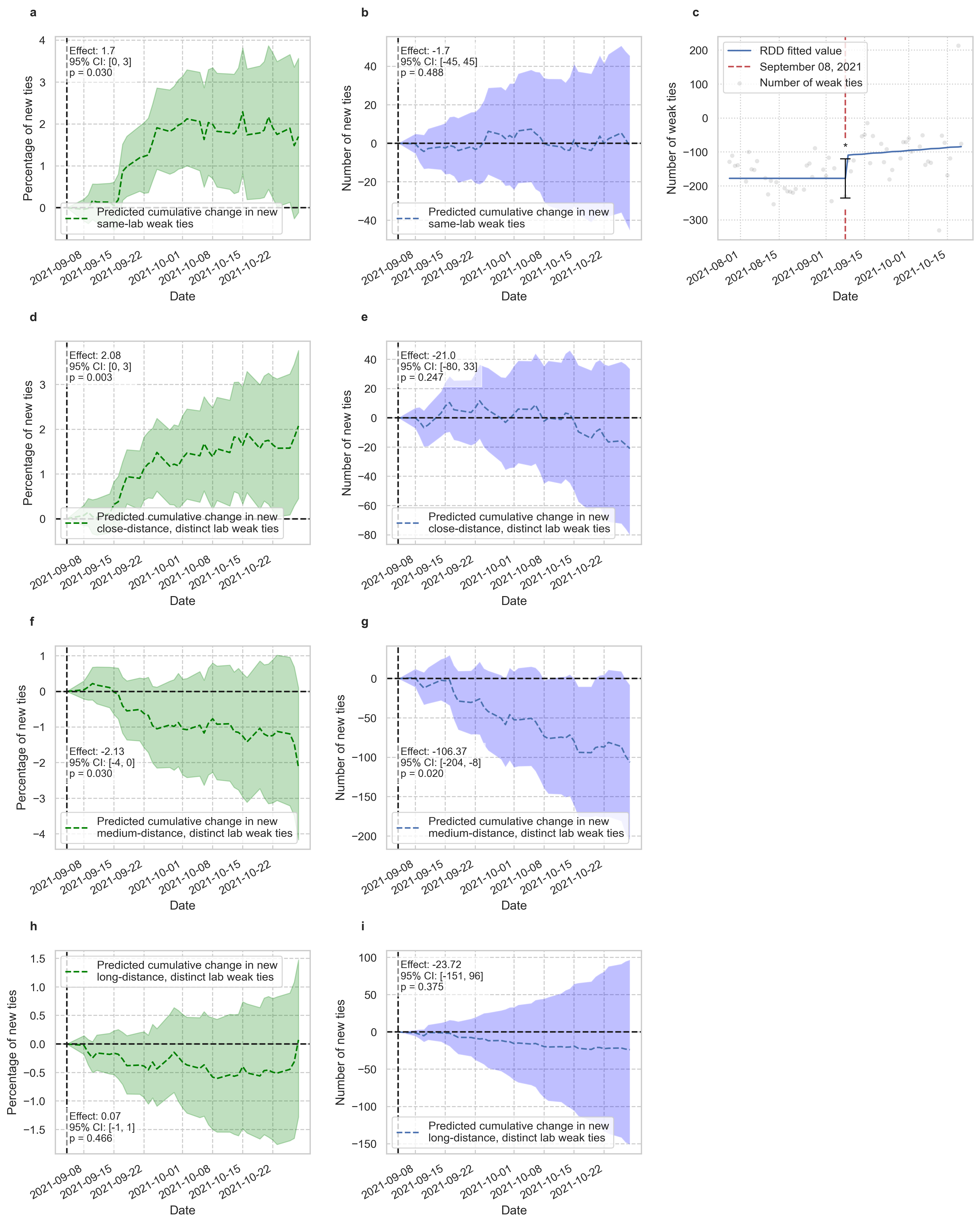}
    \caption{{\bf The effect of re-introducing co-location between researchers.} ***$: p < .001$, **$: .001 \leq p <.01$, *$: .01 \leq p < .05$. {\bf a,d,f,h}, The change due to hybrid work in the \emph{percentage} of new weak ties between researchers in the same lab, researchers in distinct labs at distance less than 150m, researchers in distinct labs at distance between 150 and 650m, and researchers in distinct labs at distance larger than 650m. {\bf b,e,g,i}, The changes in the \emph{total number} of new ties between researchers at the distance thresholds detailed above. $n_{\mathrm{pre}} = 29$ days, $n_{\mathrm{post}} = 35$ days for the panels listed above. {\bf c}, A discontinuity design (two-sided $z$ test) on the \emph{difference} of the number of weak ties in September 2020 and September 2021 $(\mathrm{effect} = 67.5, p = 0.022, 95\%$ CI: $[9.740,	125.266], n_{\mathrm{pre}} = 34$ days, $n_{\mathrm{post}} = 26$ days). Fitted values/intervals use the mean as the measure of central tendency.}
    \label{fig:combined_ties_fall2021all}
\end{figure}

\begin{figure}
    \centering
    \includegraphics[width=\linewidth]{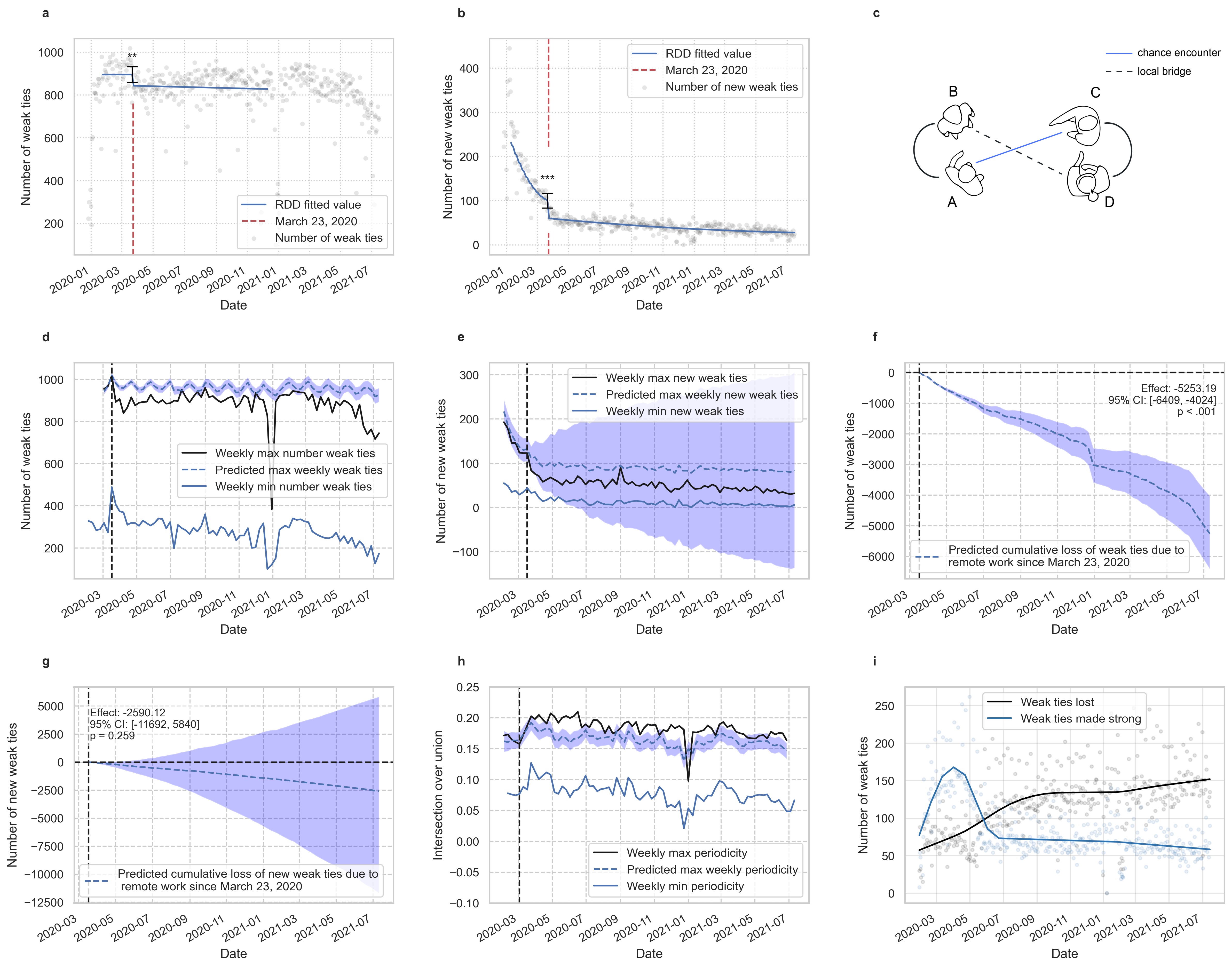}
    \caption{{\bf Changes in weak ties in the MIT email network with no pruned users after the shift to remote work}. ***$: p < .001$, **$: .001 < p <.01$, *$: .01 < p < .05$. {\bf a}, There is a drop of 52.0 in the mean number of local bridges (weak ties) caused by the switch to fully remote work ($p = .005$, 95\% CI: [-88.334, -15.670]). {\bf b}, A statistically significant drop of -39.8 in the mean number of new (not previously seen) weak ties appearing each weekday ($p < .001$, 95\% CI: [-56.602, -22.947]). Statistics represent two-sided $z$ tests. {\bf c} Local bridges in a social network can be formed from chance encounters. {\bf d}, A predicted loss of  77.25 weak ties ($p < .001$, 95\% CI: [-94.44, -60.4]). {\bf e}, A predicted loss of 38.1 new weak ties ($p = 0.29$, 95\% CI: [-166.4, 86.51]). {\bf f}, A cumulative drop of 5253 in the number of weak ties throughout an entire year. {\bf g}, A cumulative drop of 2590 in new weak ties. {\bf h}, A 0.02 (11.9\%) increase in the week to week similarity of network edges ($p < .001$, 95\% CI: [8.9\%,15.0\%]). {\bf i}, The number of weak ties which become strong or are churned in a 30 day rolling window. Shaded regions show a 95\% posterior predictive interval computed using Bayesian structural time series. $n_{\mathrm{pre}} = 42$ days, $n_{\mathrm{post}} = 188$ days for panels {\bf a}, {\bf b}. $n_{\mathrm{pre}} = 8$ weeks, $n_{\mathrm{post}} = 72$ weeks for panels {\bf d}-{\bf h}. Fitted values/intervals use the mean as the measure of central tendency.}
    \label{fig:num_weak_bridges_nosmall}
\end{figure}

\begin{figure}
    \centering
    \includegraphics[width=\linewidth]{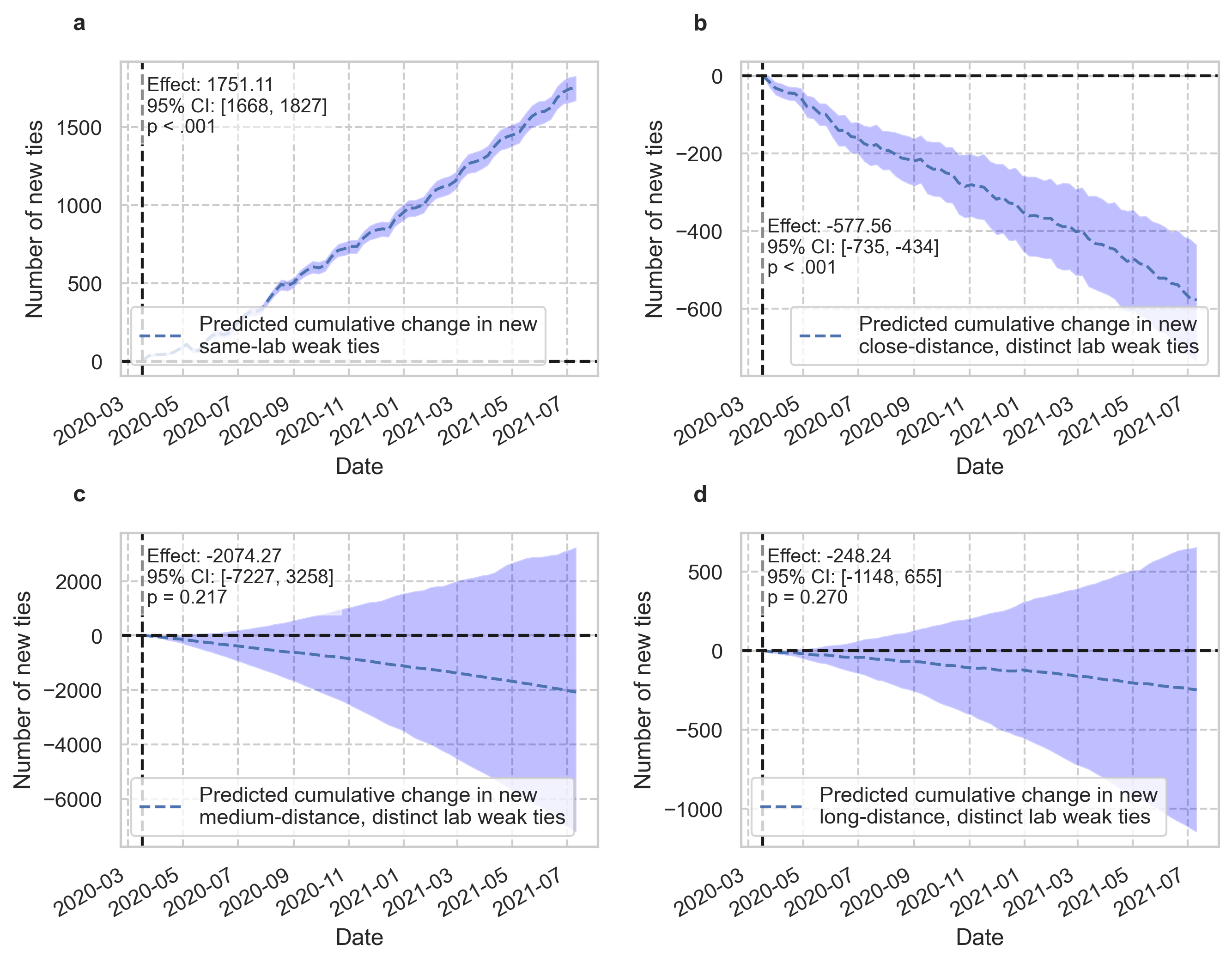}
    \caption{{\bf Formation of new weak ties stratified by distance in a network where no users are pruned.} {\bf a}, The estimated effect  between researchers in the same lab. {\bf b}, The estimated effect between researchers in distinct labs within 150 meters. {\bf c}, The estimated effect between researchers in distinct labs between 150 and 650 meters. {\bf d}, The estimated effect between researchers in distinct labs further than 650 meters. Shaded regions represent 95\% posterior predictive intervals computed using Bayesian structural time series with a synthetic counterfactual constructed from weekend data $n_{\mathrm{pre}} = 8$ weeks, $n_{\mathrm{post}} = 72$ weeks for all panels. Fitted values/intervals use the mean as the measure of central tendency.}
    \label{fig:spatial_nosmall}
\end{figure}

\section{Estimating the number of researchers on campus}

\begin{figure}
    \centering
    \includegraphics[width=\linewidth]{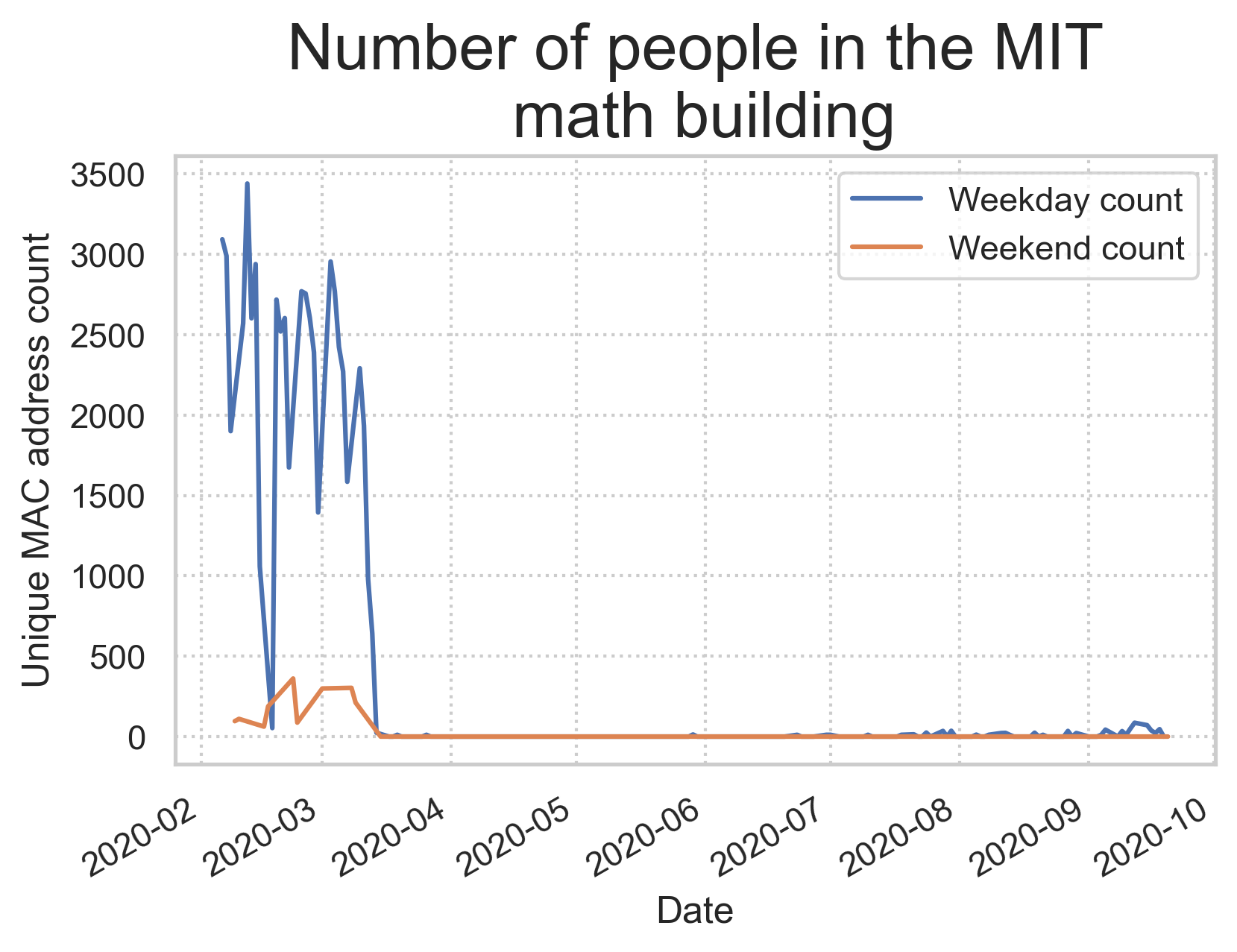}
    \caption{{\bf The number of users in the MIT math building on weekdays and weekends.} Counts are estimated using the number of distinct MAC addresses connected to access points in MIT building 2.}
    \label{fig:wifi_math_count}
\end{figure}

\begin{figure}
    \centering
    \includegraphics[width=\linewidth]{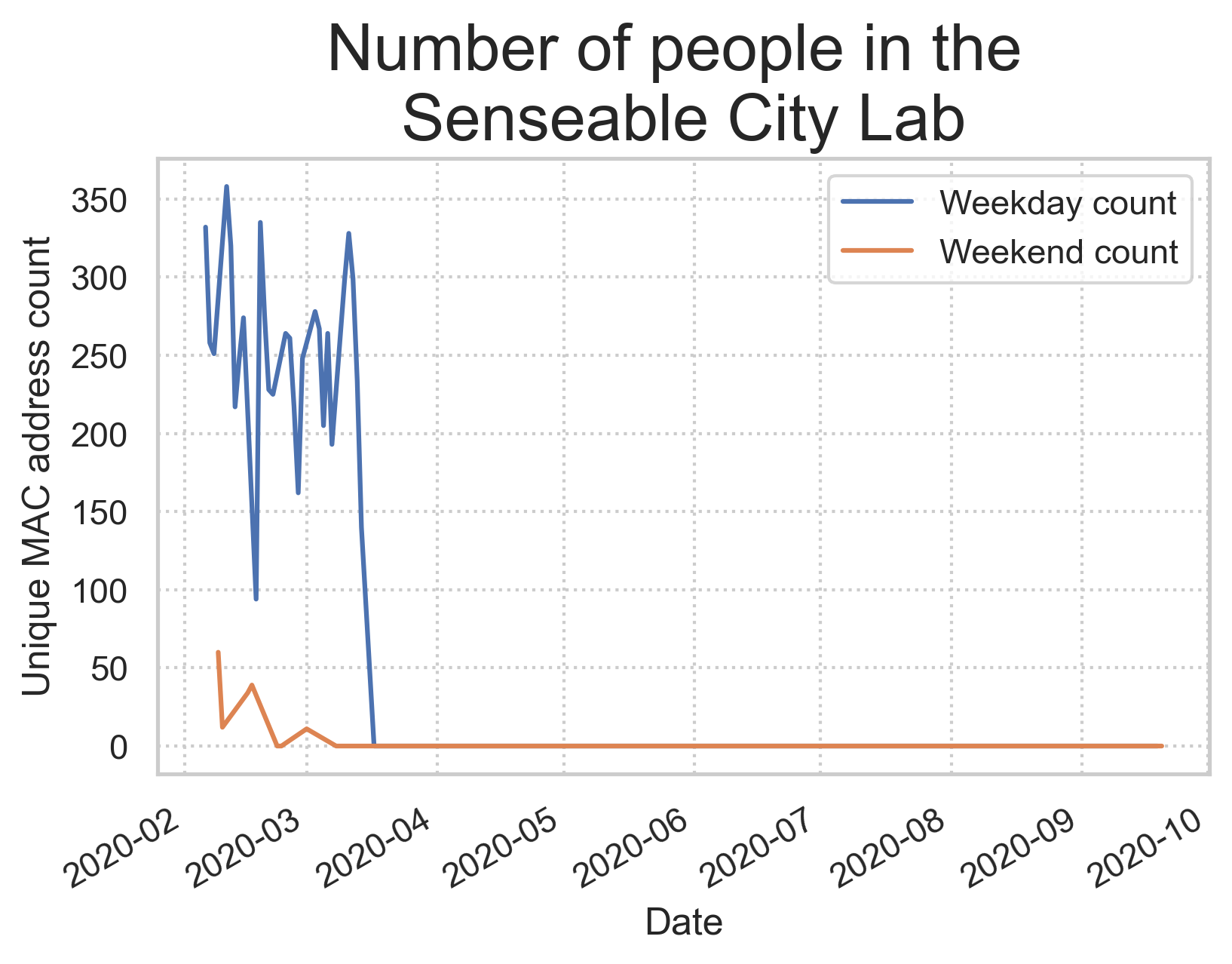}
    \caption{{\bf The number of users in the Senseable City Lab on weekdays and weekends.} Counts are estimated using the number of distinct MAC addresses connected to access points in MIT building 9 room 216.}
    \label{fig:wifi_scl_count}
\end{figure}

To justify the use of weekend email data to generate a synthetic counterfactual, we estimate the number of distinct researchers co-located on weekends and weekdays before and after the pandemic lockdown using WiFi data. Specifically, we count the number of unique MAC addresses connected to a given MIT campus WiFi router each hour from February 3, 2020 until October 10, 2020. Supplementary Figure \ref{fig:wifi_math_count} shows that, as expected, far fewer people are on campus during the weekend. To confirm that this effect is not due to a lack of classes on the weekends, Supplementary Figure \ref{fig:wifi_scl_count} shows the number of unique MAC addresses in the Senseable City Lab on weekends and weekdays -- a room in which no classes are taught.  

\section{Robustness of the tie formation model}

It is well known that standard random graph models like Erd\H{o}s-Renyi, Watts-Strogatz, and Barab\'{a}si-Albert do not fully capture the topological features of real social networks \cite{10.5555/1809753}. For example, the Erd\H{o}s-Renyi and Barab\'{a}si-Albert models have unrealistically low clustering coefficients, while the Watts-Strogatz model has an unrealistic degree distribution.

To verify that a lack of co-location is what causes our model to reproduce the drop in weak ties seen in the empirical data, we check that with no changes in co-location our model produces no changes in weak ties in Supplementary Figure \ref{fig:combined_simu_nomod}. Supplementary Figures \ref{fig:combined_simu_nodpt} through \ref{fig:combined_simu_notri} show the results of ablation studies on our model from removing parameters. For each parameter except $O$, we see changes to the output of the simulation which makes it inconsistent with empirical results. Because the $O$ parameter is small and unaffected by co-location, this is unsurprising.

Finally, Supplementary Figure \ref{fig:office_distances} shows the distribution of distances between researcher offices at MIT in order to demonstrate that our choices of distance bins each contain a non-trivial number of researchers. Almost all researchers' offices are within 2 km of one another. 


\begin{figure}
    \centering
    \includegraphics[width=\linewidth]{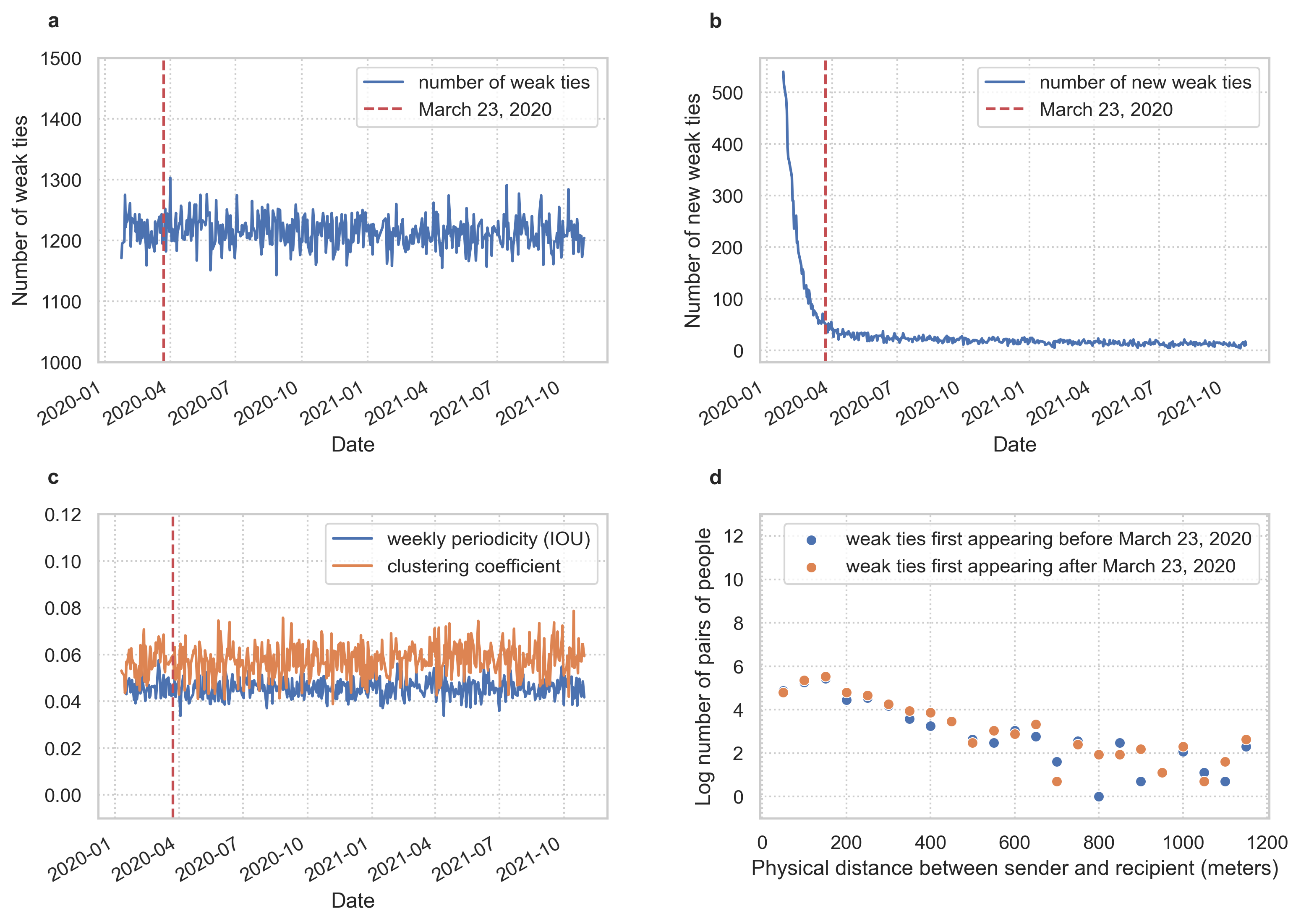}
    \caption{{\bf Simulation results when leaving the distance between researchers unchanged after March 23, 2020}. {\bf a}, The number of weak ties. {\bf b}, The number of new weak ties entering the network. {\bf c}, The weekly periodicity and daily clustering coefficient. {\bf d}, The number of simulated weak ties between users in distinct research units first appearing between February 4 and March 23, 2020 versus between March 23, 2020 and May 22, 2020.}
    \label{fig:combined_simu_nomod}
\end{figure}

\begin{figure}
    \centering
    \includegraphics[width=\linewidth]{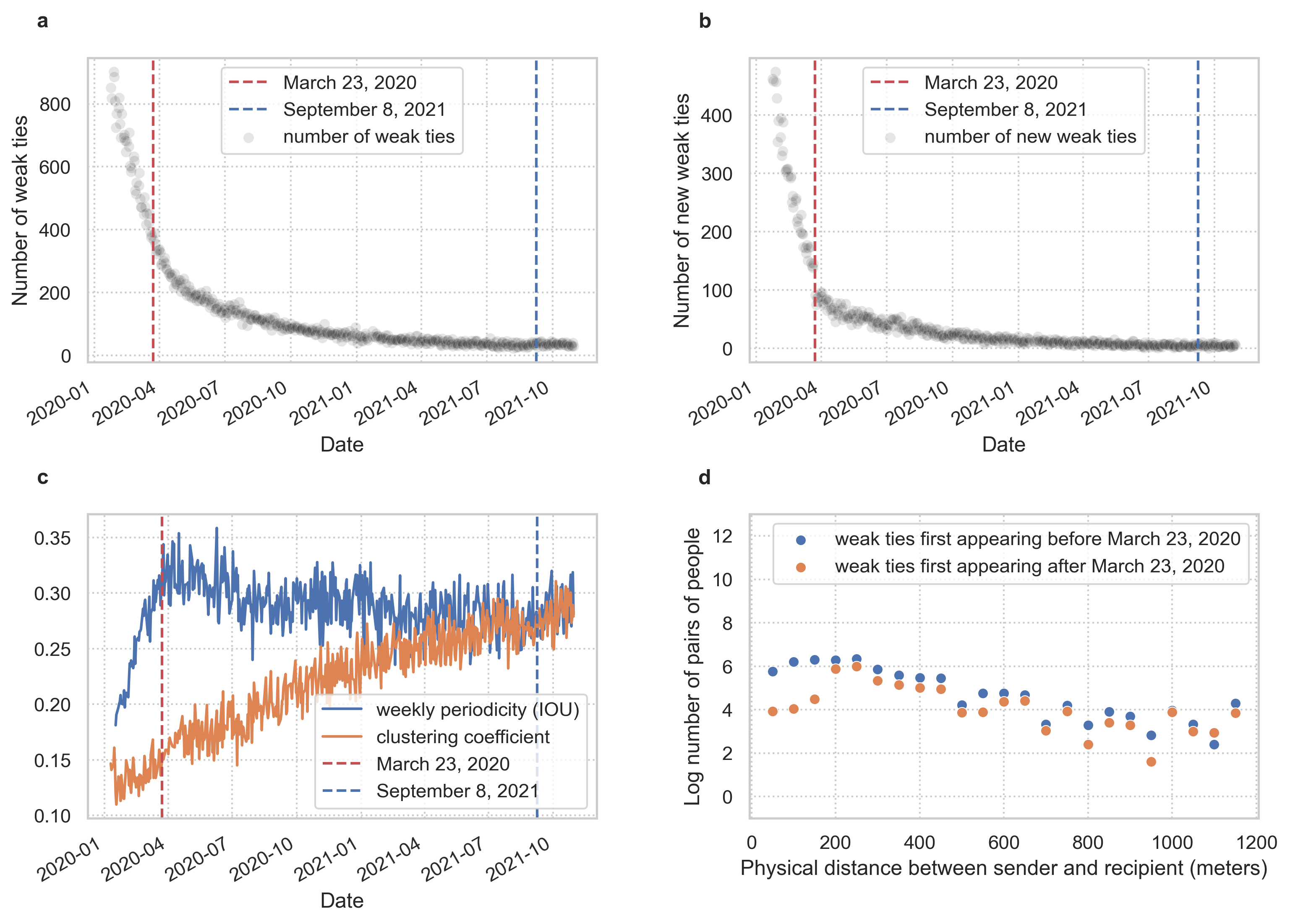}
    \caption{{\bf Simulation results when removing the focal closure (same research unit) parameter}. {\bf a}, The number of weak ties. {\bf b}, The number of new weak ties entering the network. {\bf c}, The weekly periodicity and daily clustering coefficient. {\bf d}, The number of simulated weak ties between users in distinct research units first appearing between February 4 and March 23, 2020 versus between March 23, 2020 and May 22, 2020.}
    \label{fig:combined_simu_nodpt}
\end{figure}

\begin{figure}
    \centering
    \includegraphics[width=\linewidth]{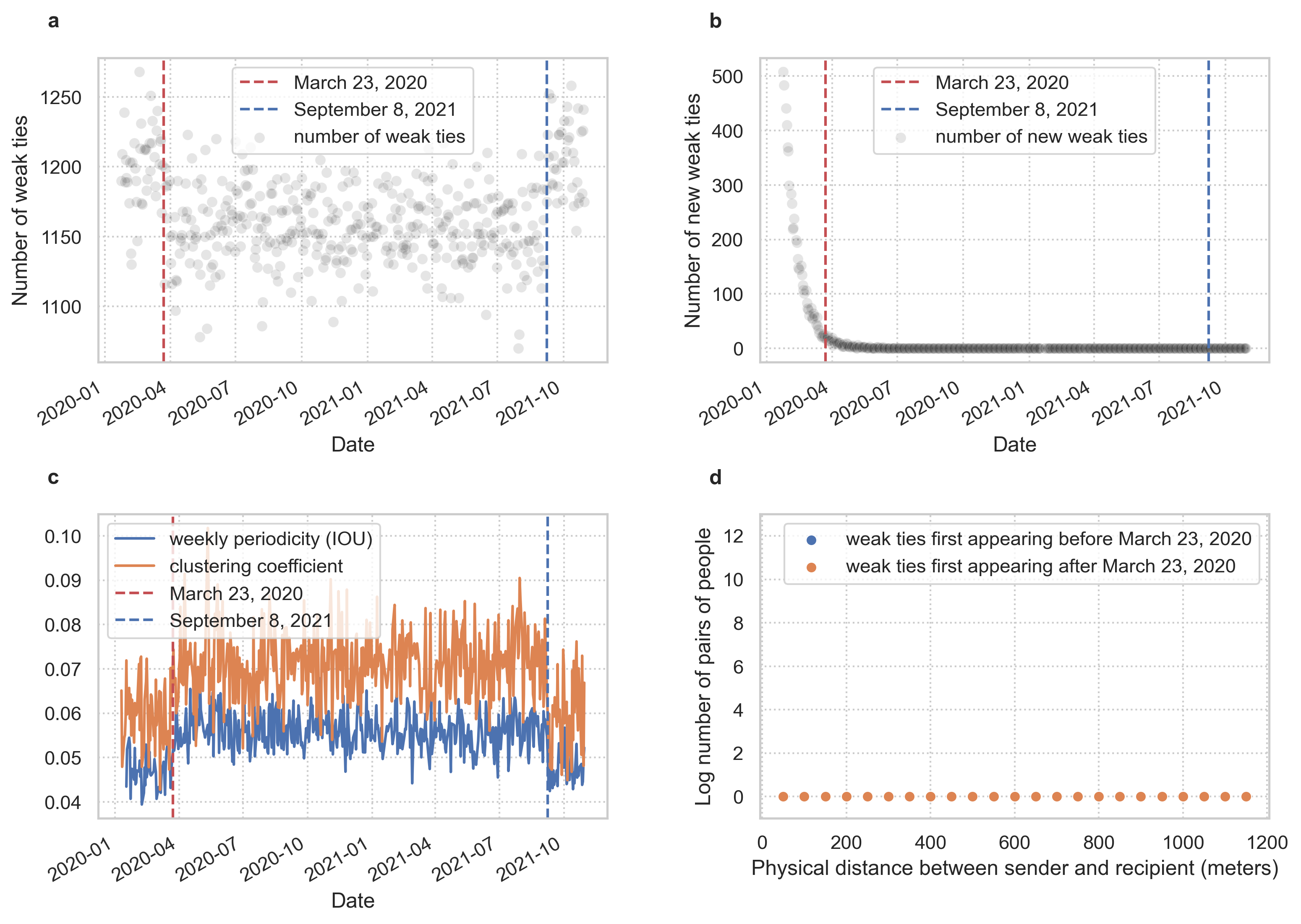}
    \caption{{\bf Simulation results when removing the new tie parameter}. {\bf a}, The number of weak ties. {\bf b}, The number of new weak ties entering the network. {\bf c}, The weekly periodicity and daily clustering coefficient. {\bf d}, The number of simulated weak ties between users in distinct research units first appearing between February 4 and March 23, 2020 versus between March 23, 2020 and May 22, 2020.}
    \label{fig:combined_simu_nonew}
\end{figure}

\begin{figure}
    \centering
    \includegraphics[width=\linewidth]{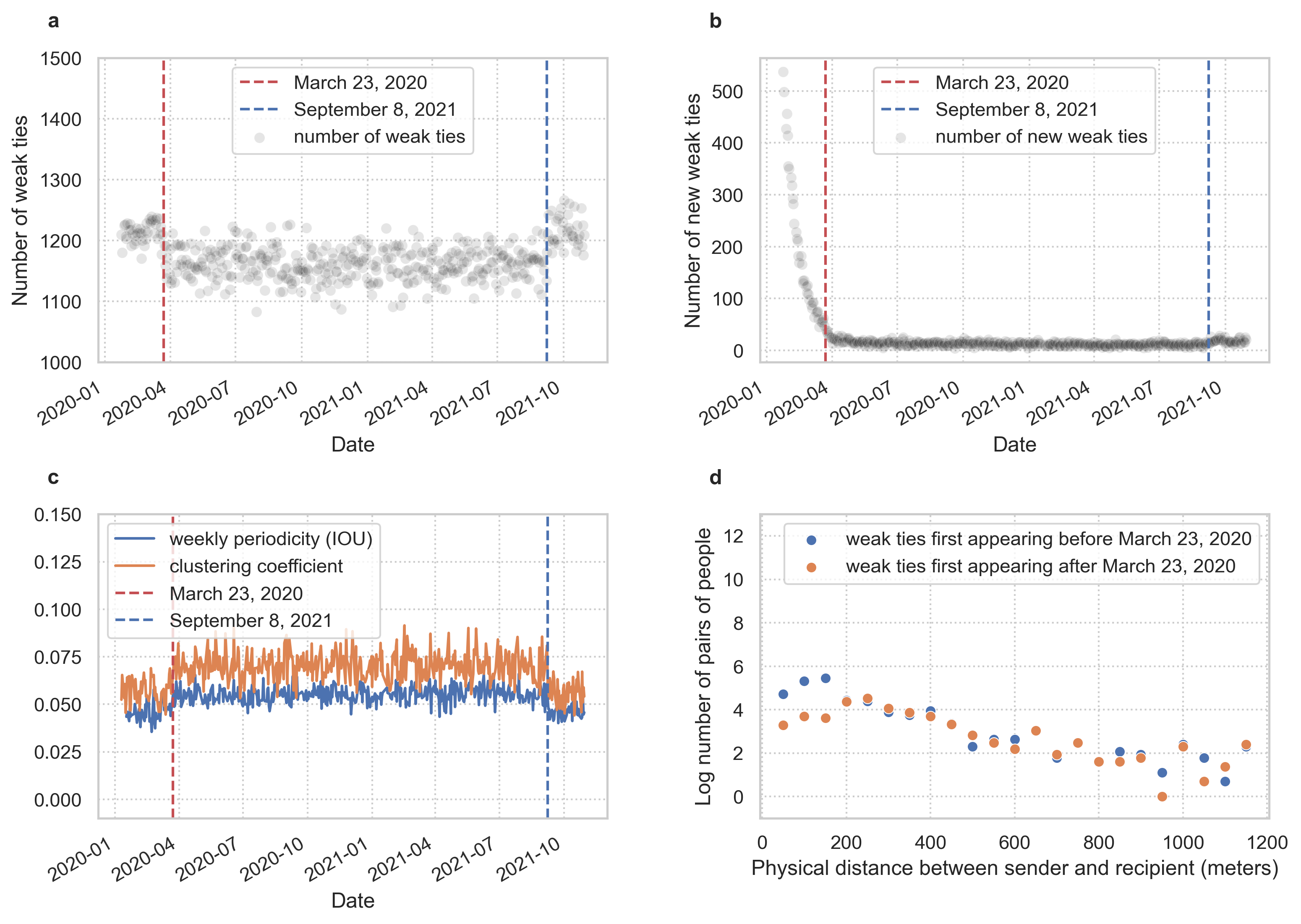}
    \caption{{\bf Simulation results when removing the old tie parameter}. {\bf a}, The number of weak ties. {\bf b}, The number of new weak ties entering the network. {\bf c}, The weekly periodicity and daily clustering coefficient. {\bf d}, The number of simulated weak ties between users in distinct research units first appearing between February 4 and March 23, 2020 versus between March 23, 2020 and May 22, 2020.}
    \label{fig:combined_simu_noold}
\end{figure}

\begin{figure}
    \centering
    \includegraphics[width=\linewidth]{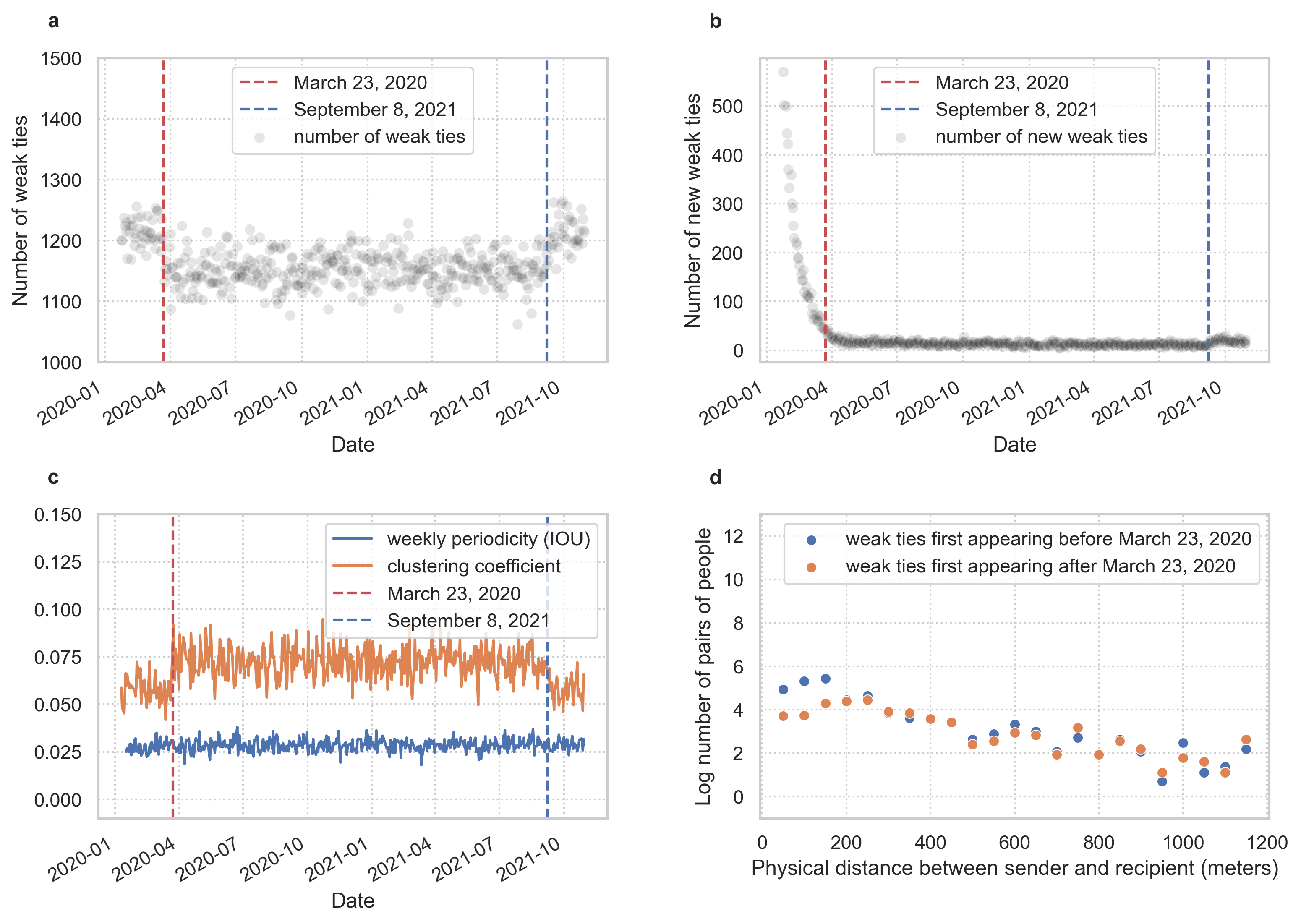}
    \caption{{\bf Simulation results when removing the periodicity parameter}. {\bf a}, The number of weak ties. {\bf b}, The number of new weak ties entering the network. {\bf c}, The weekly periodicity and daily clustering coefficient. {\bf d}, The number of simulated weak ties between users in distinct research units first appearing between February 4 and March 23, 2020 versus between March 23, 2020 and May 22, 2020.}
    \label{fig:combined_simu_noper}
\end{figure}

\begin{figure}
    \centering
    \includegraphics[width=\linewidth]{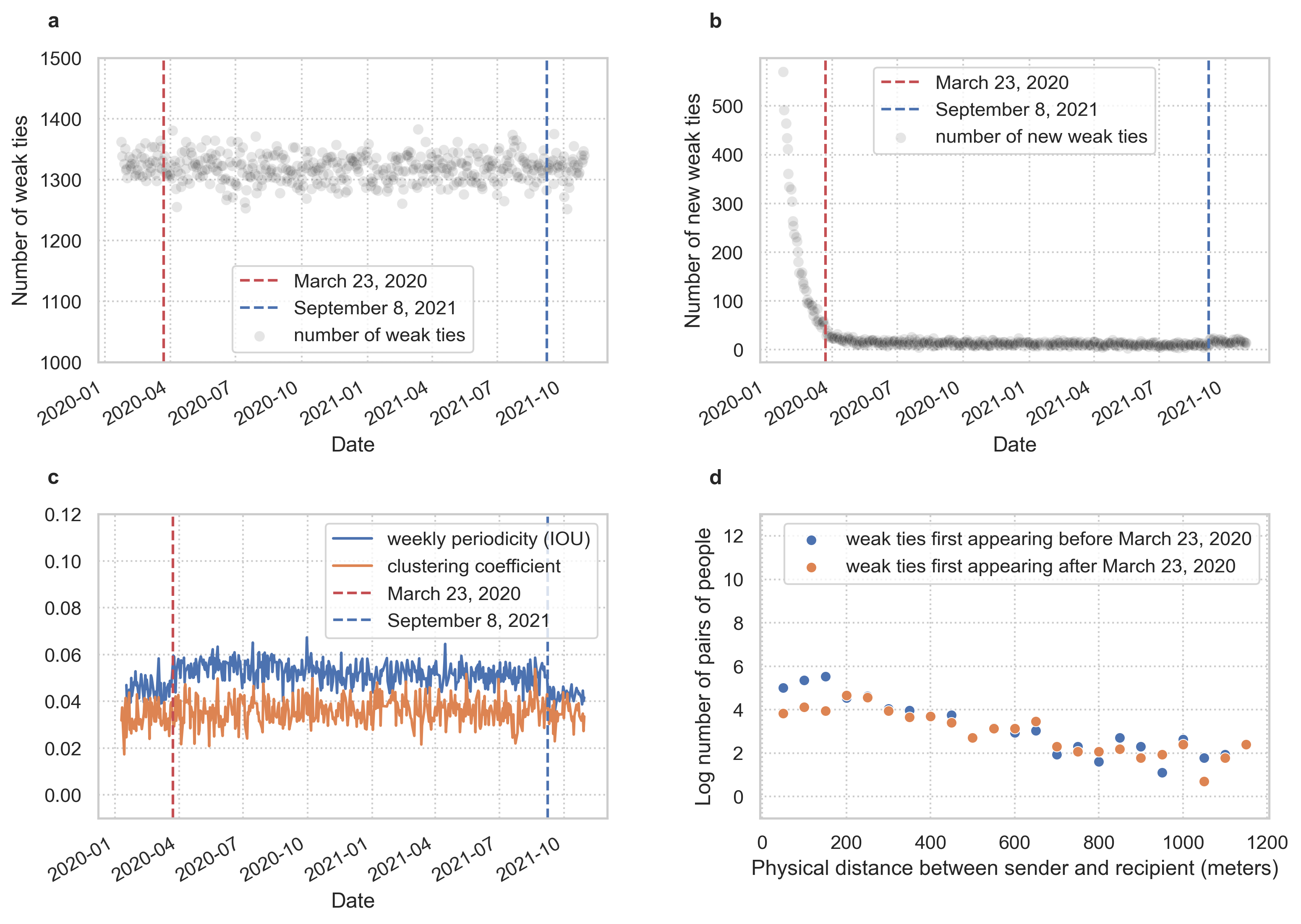}
    \caption{{\bf Simulation results when removing the triadic closure parameter}. {\bf a}, The number of weak ties. {\bf b}, The number of new weak ties entering the network. {\bf c}, The weekly periodicity and daily clustering coefficient. {\bf d}, The number of simulated weak ties between users in distinct research units first appearing between February 4 and March 23, 2020 versus between March 23, 2020 and May 22, 2020.}
    \label{fig:combined_simu_notri}
\end{figure}

\begin{figure}
    \centering
    \includegraphics[width=0.7\linewidth]{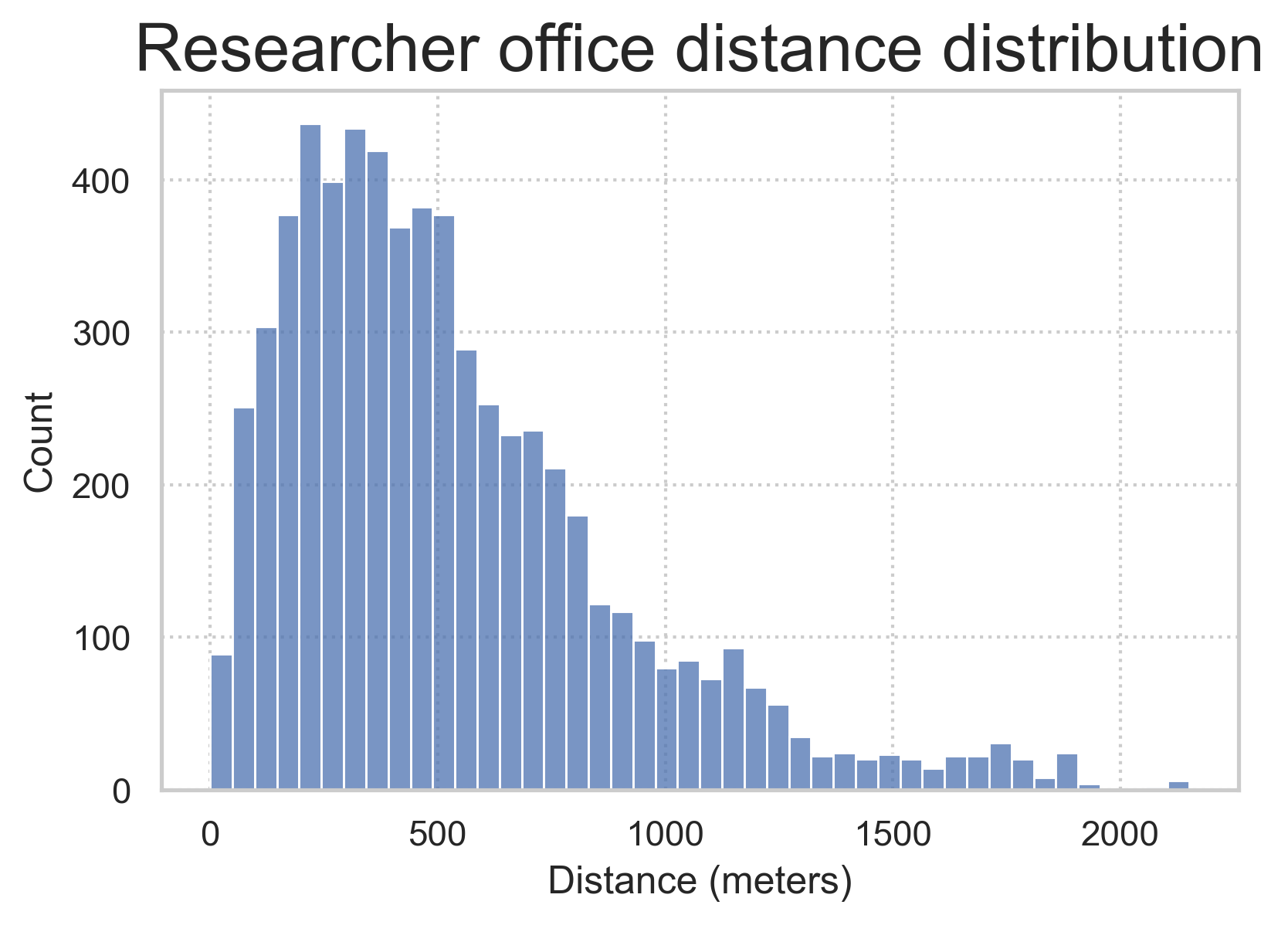}
    \caption{The distribution of distances between researcher offices.}
    \label{fig:office_distances}
\end{figure}

\section{Descriptive statistics for weekly, biweekly, and monthly emails networks}

\begin{figure}
    \centering
    \includegraphics[width=\linewidth]{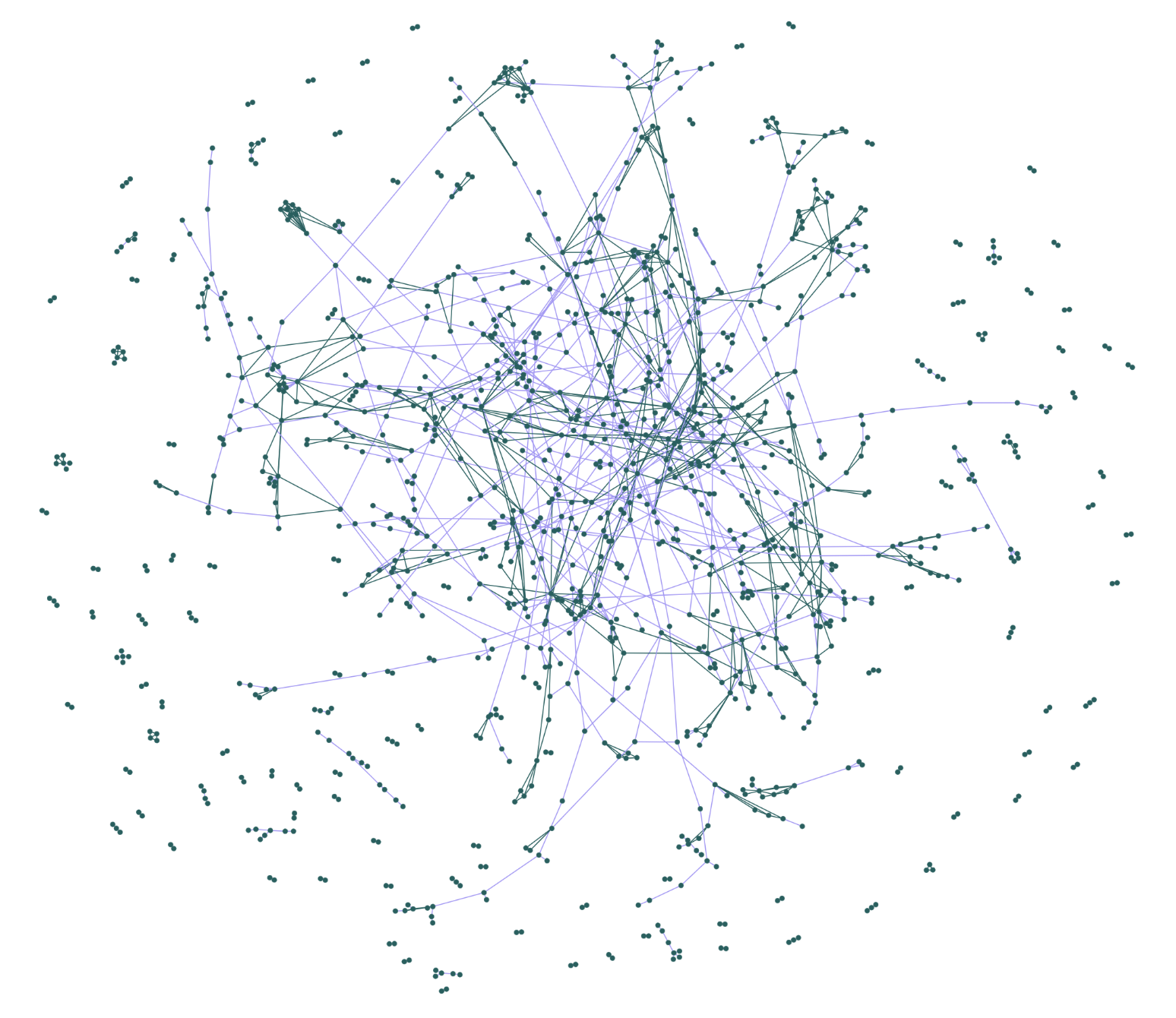}
    \caption{{\bf An example daily email network}. We allow emails to be reciprocated only on the same business day. Local bridges are purple.}
    \label{fig:daily_graph_ex}
\end{figure}

\begin{figure}
    \centering
    \includegraphics[width=\linewidth]{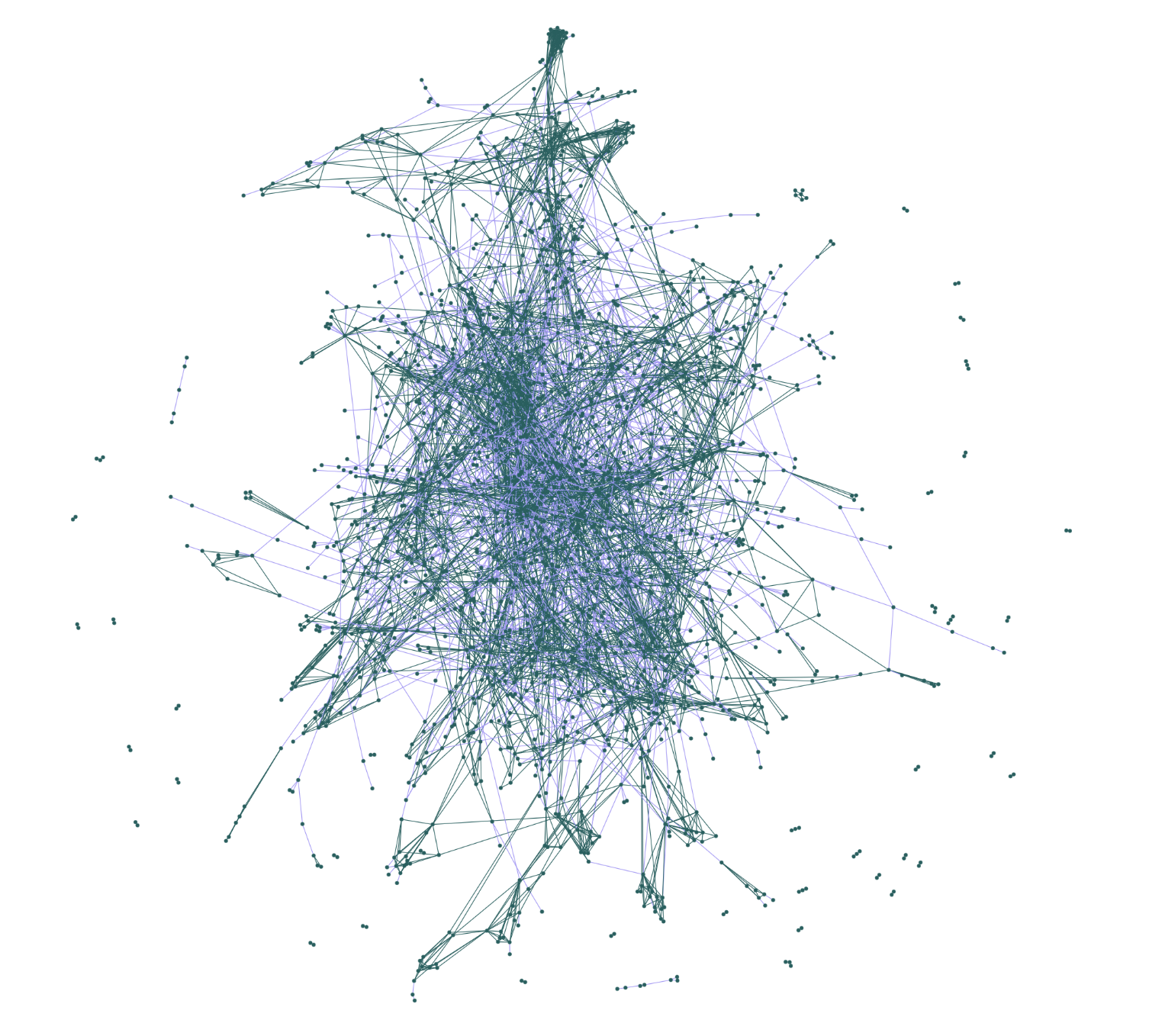}
    \caption{{\bf An example weekly email network}. We allow emails to be reciprocated up to 5 business days later. Local bridges are purple.}
    \label{fig:weekly_graph_ex}
\end{figure}

\begin{figure}
    \centering
    \includegraphics[width=\linewidth]{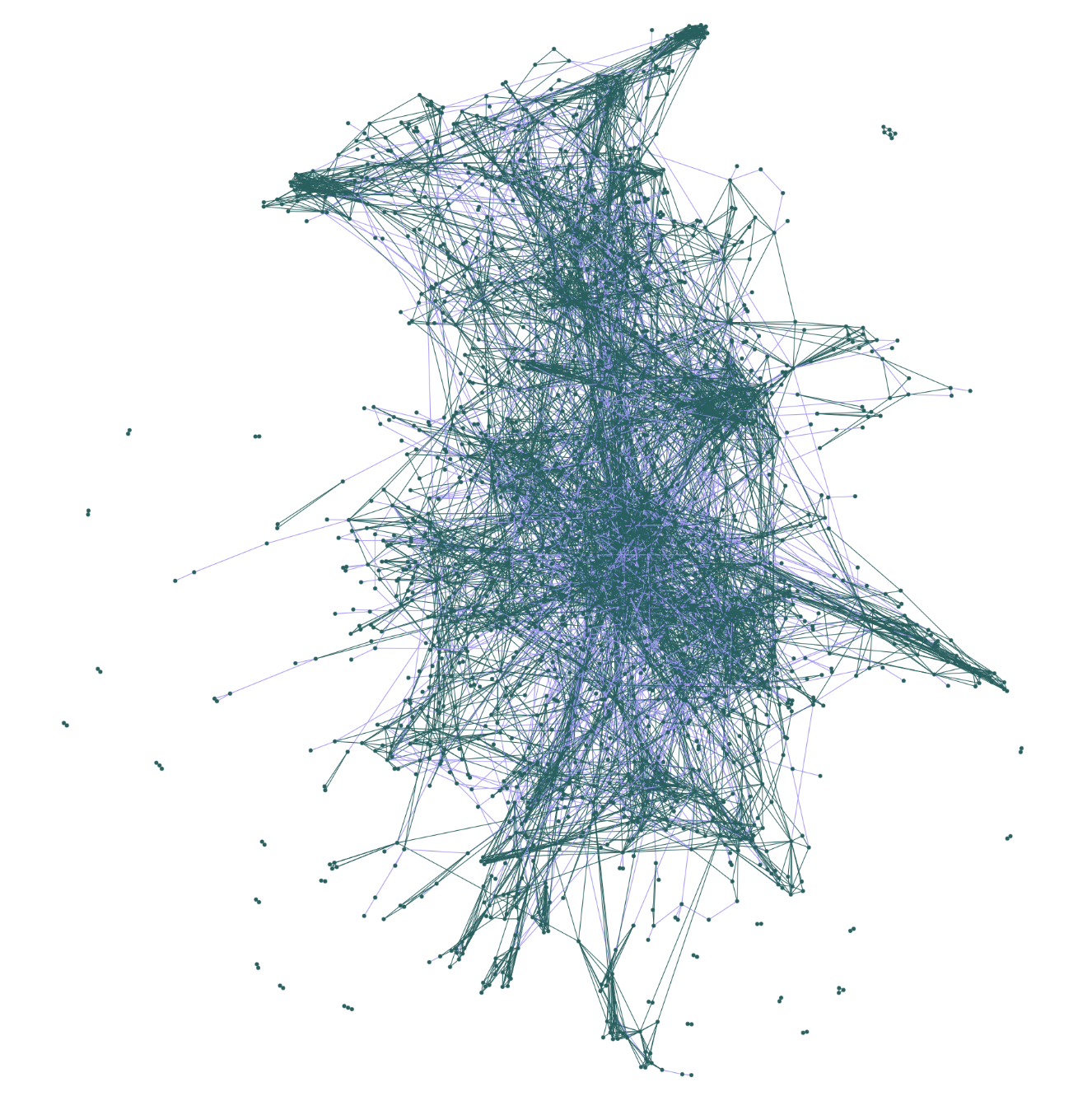}
    \caption{{\bf An example biweekly email network}. We allow emails to be reciprocated up to 10 business days later. Local bridges are purple.}
    \label{fig:biweekly_graph_ex}
\end{figure}

\begin{figure}
    \centering
    \includegraphics[width=\linewidth]{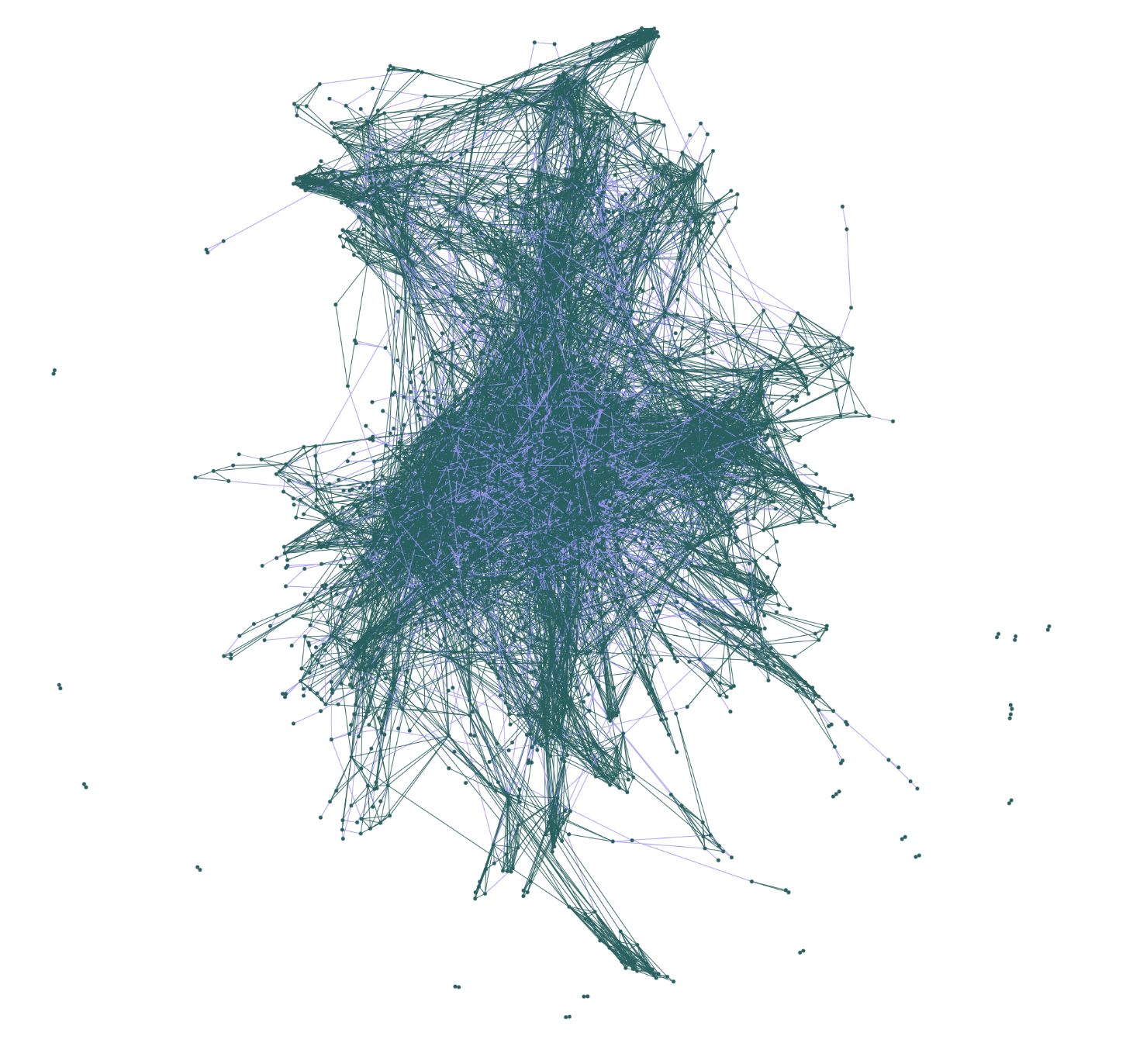}
    \caption{{\bf An example monthly email network}. We allow emails to be reciprocated up to 21 business days later. Local bridges are purple.}
    \label{fig:monthly_graph_ex}
\end{figure}

\begin{figure}
    \centering
    \includegraphics[width=\linewidth]{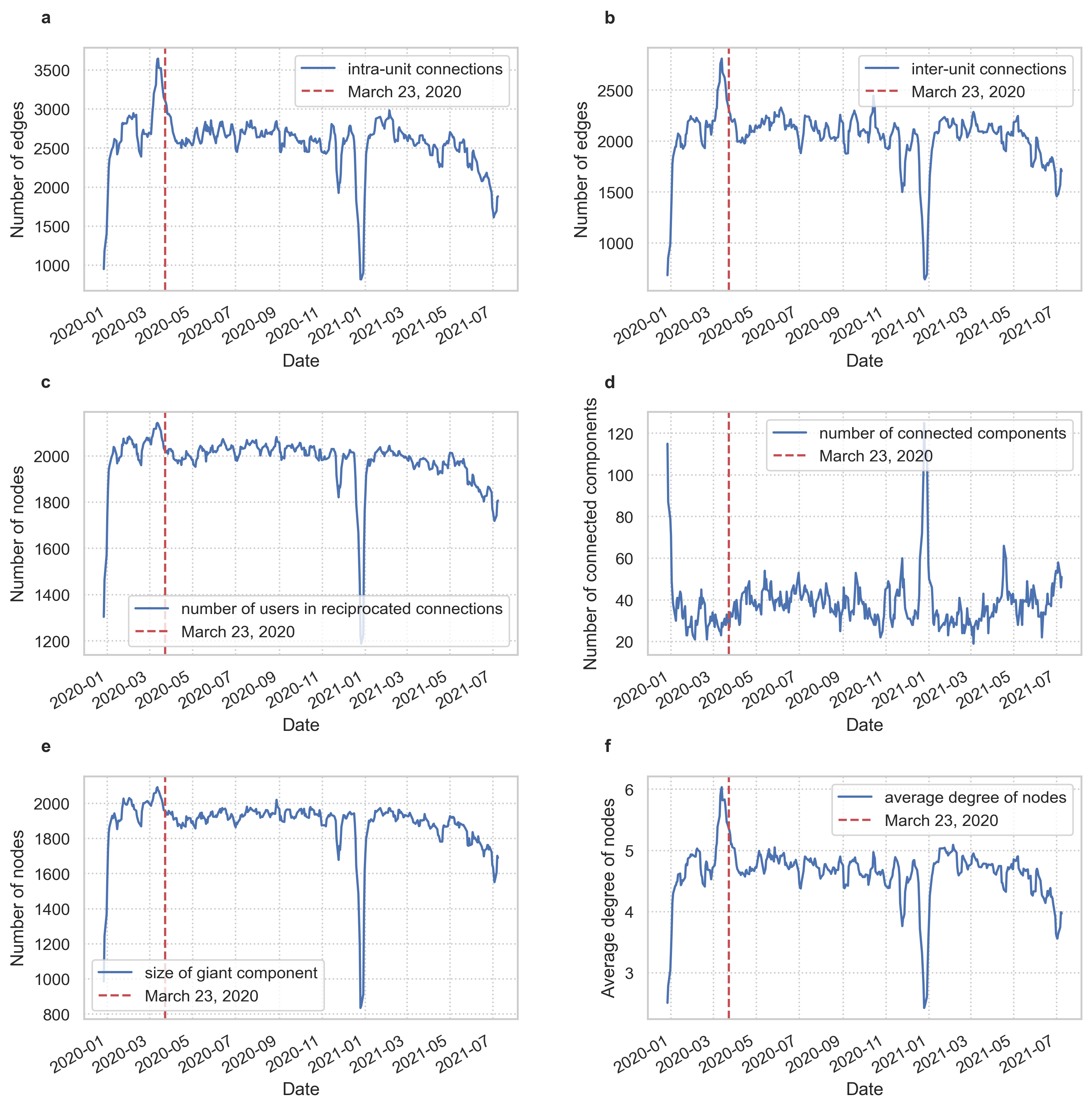}
    \caption{{\bf Summary statistics for weekly email networks.} {\bf a}, The number of connections (edges) between users who are in the same research unit. {\bf b}, The number of connections (edges) between users who are in distinct research units. {\bf c}, The number of nodes in the daily email network who are attached to at least one edge. {\bf d}, The number of connected components. {\bf e} The number of nodes in the largest connected component. {\bf f} The average degree of each node in the weekly email networks.}
    \label{fig:combined_summary_weekly}
\end{figure}

\begin{figure}
    \centering
    \includegraphics[width=\linewidth]{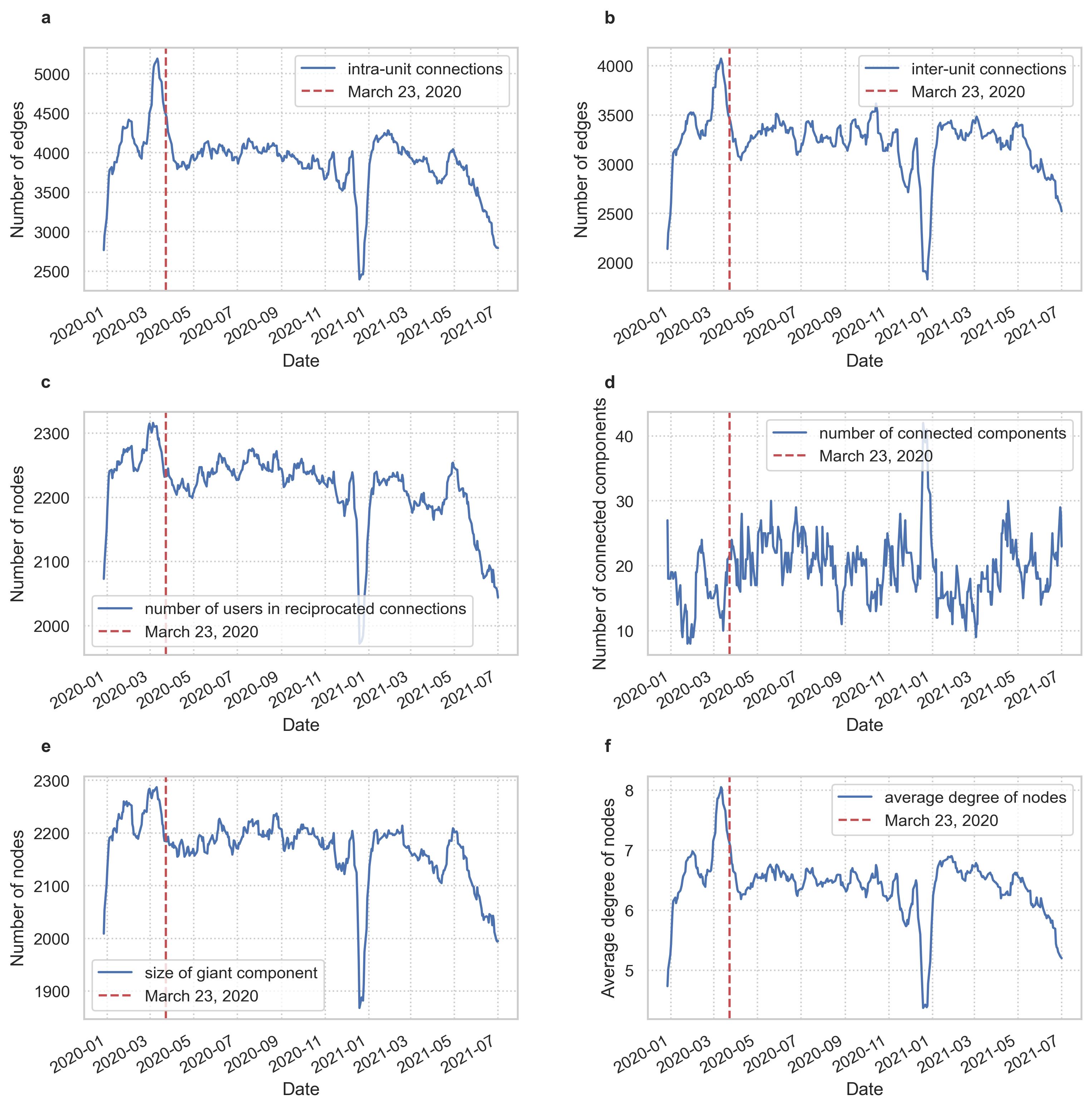}
    \caption{ {\bf Summary statistics for biweekly email networks.} {\bf a}, The number of connections (edges) between users who are in the same research unit. {\bf b}, The number of connections (edges) between users who are in distinct research units. {\bf c}, The number of nodes in the daily email network who are attached to at least one edge. {\bf d}, The number of connected components. {\bf e} The number of nodes in the largest connected component. {\bf f} The average degree of each node in the biweekly email networks.}
    \label{fig:combined_summary_biweekly}
\end{figure}

\begin{figure}
    \centering
    \includegraphics[width=\linewidth]{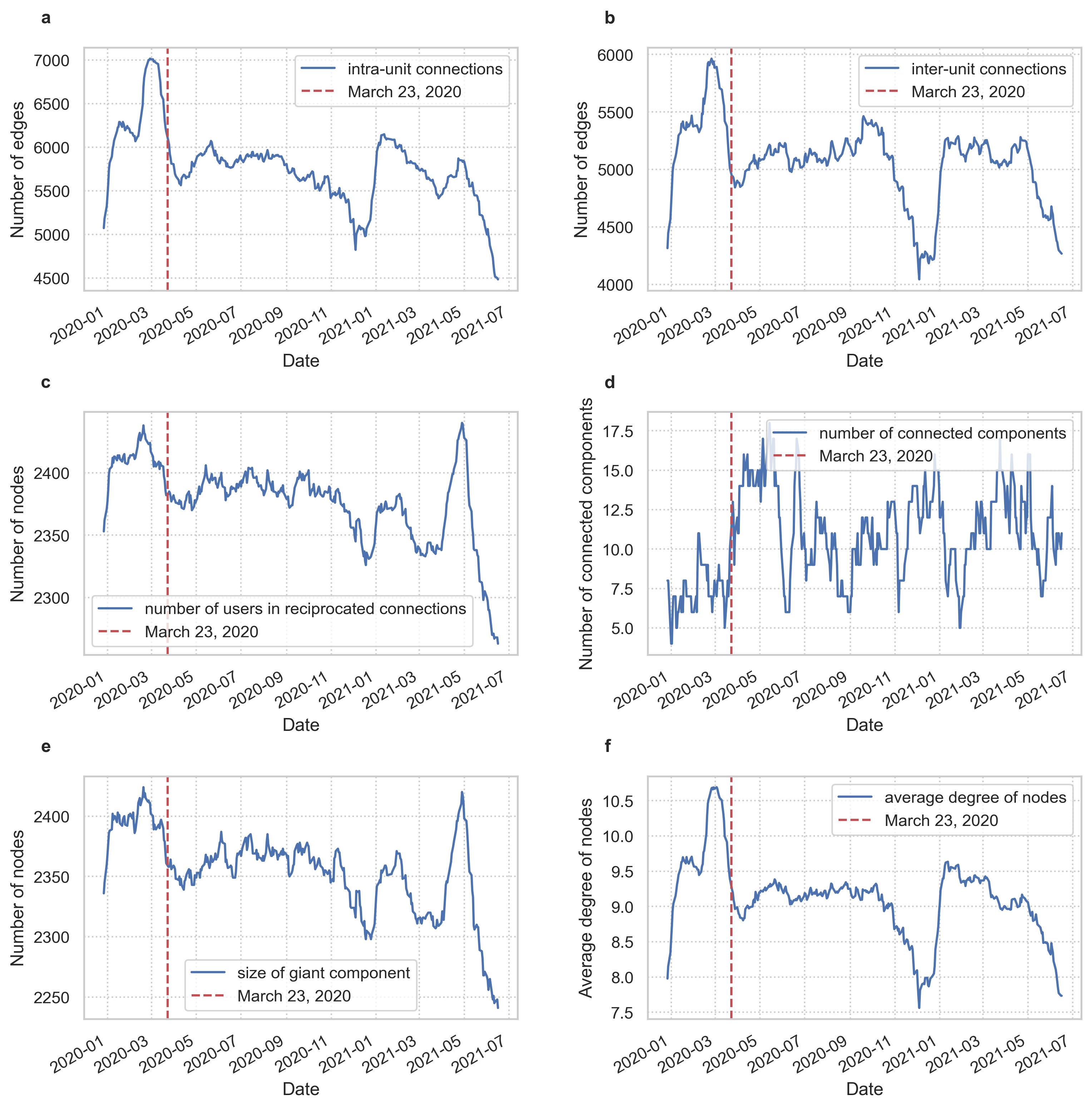}
    \caption{{\bf Summary statistics for monthly email networks.} {\bf a}, The number of connections (edges) between users who are in the same research unit. {\bf b}, The number of connections (edges) between users who are in distinct research units. {\bf c}, The number of nodes in the daily email network who are attached to at least one edge. {\bf d}, The number of connected components. {\bf e} The number of nodes in the largest connected component. {\bf f} The average degree of each node in the monthly email networks.}
    \label{fig:combined_summary_monthly}
\end{figure}

Requiring emails to be reciprocated the same day is useful to increase the number of independent pre-lockdown data points for statistical purposes, but we lose email exchanges which are stretched over the course of a week or longer. The mean response time in an email exchange is closely related to the frequency of communication: if the mean response time between two people in the 562 days of data is $d$ days, then the total number of reciprocated exchanges (ignoring the number of emails sent per exchange) is at most $\frac{562}{d}$. Thus by restricting to daily reciprocated emails, we may miss the contributions of some infrequent ties. To expand the breadth of ties under consideration and study how incorporating lagged communications affects our results, we construct email networks whose edges represent the number of reciprocated emails exchanged within 5  business days, 10 business days, and 21 business days. Supplementary Figures \ref{fig:daily_graph_ex}-\ref{fig:monthly_graph_ex} show sample weekly, biweekly, and monthly reciprocated email networks from April 4, 2020. Supplementary Figures \ref{fig:combined_summary_weekly} through \ref{fig:combined_summary_monthly} show standard network measures for emails networks whose edges represent emails reciprocated within one week, two weeks, or one month. In contrast to the daily email networks, there are seemingly long-lasting changes in the size of the giant component in the networks; we are unable to quantify this statistically as our construction of a synthetic counterfactual from weekend email data breaks down when we allow emails to be reciprocated over any period longer than 2 days.

\begin{figure}
    \centering
    \includegraphics[width=\linewidth]{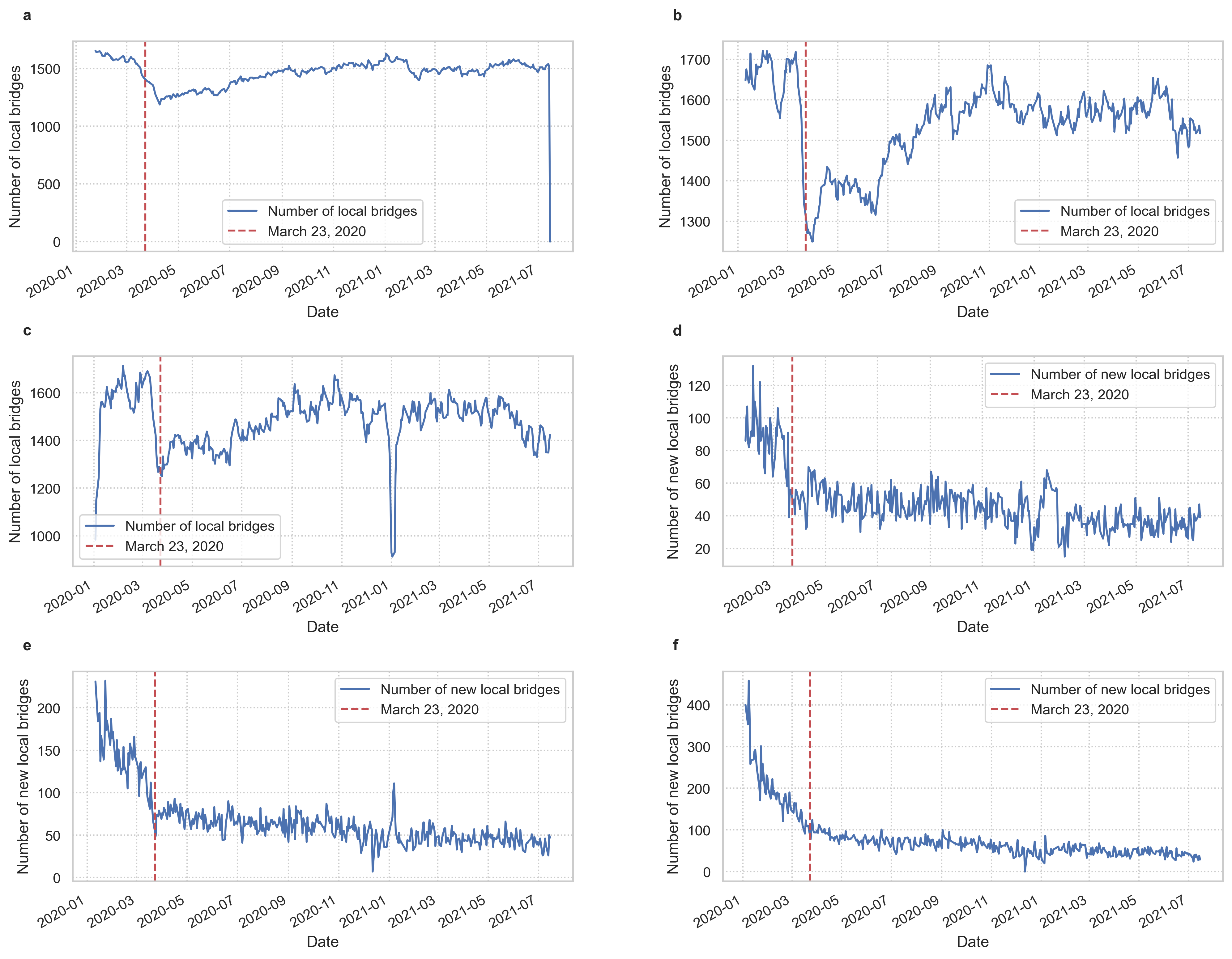}
    \caption{ {\bf a}, The number of local bridges each day in monthly reciprocated email networks. {\bf b}, The number of local bridges each day in biweekly reciprocated email networks. {\bf c}, The number of local bridges each day in weekly reciprocated email networks. {\bf d}, The number of new (not previously seen) local bridges in monthly reciprocated email networks. {\bf e}, The number of new (not previously seen) local bridges in biweekly reciprocated email networks. {\bf f}, The number of new (not previously seen) local bridges in weekly reciprocated email networks.}
    \label{fig:weekly_tie_info}
\end{figure}

Supplementary Figure \ref{fig:weekly_tie_info} shows the number of local bridges and new local bridges in weekly, biweekly, and monthly reciprocated email networks. The number of new weak ties in each case seems to stabilize between 40-60 new weak ties in agreement with the number of new weak ties in the daily email networks. This can be partially explained by the fact that we use a sliding window approach with a stride of one to construct each network. Curiously, the rate of weak tie formation (number of new ties per day) seems to approach a constant around 45 regardless of the reciprocation window.

\section{Alternate definitions of weak tie}

\begin{figure}
    \centering
    \includegraphics[width=\linewidth]{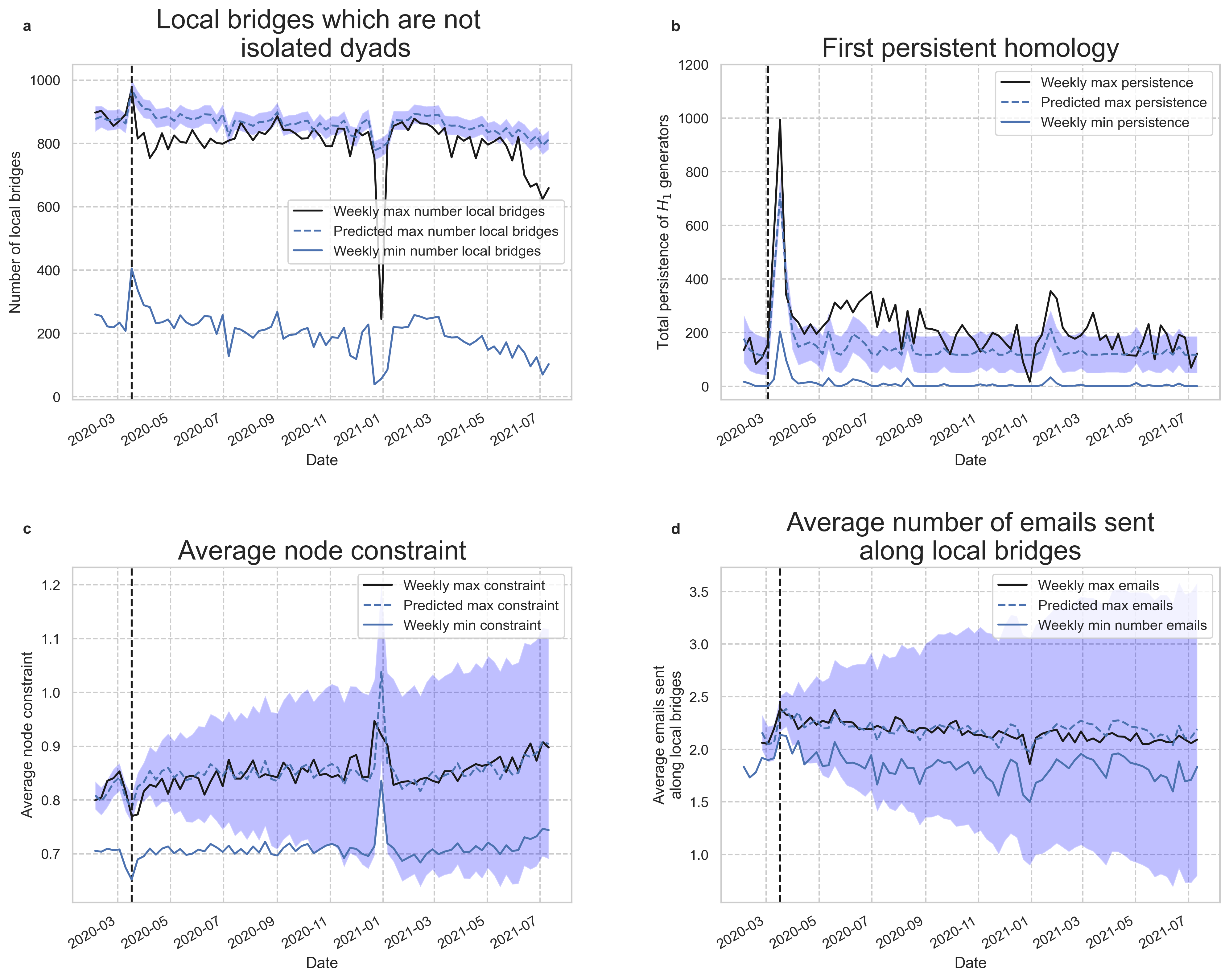}
    \caption{{\bf Other measures of topological holes in networks.} {\bf a}, The number of local bridges each day, excluding those local bridges which are isolated dyads (effect: -60.46, $p < .001$, 95\% CI: [-70.78, -50.3]). {\bf b}, The total persistence of all generators of the first persistent homology with finite death (effect: 71.98, $p < .001$, 95\% CI: [44.07, 99.4]). {\bf c}, The average node constraingt, in the sense of Burt (effect: 0.0, $p = .5$, 95\% CI: [-0.13, 0.12]). {\bf d}, The average width of each local bridge each day measured as the total number of emails sent along the bridge (effect: -0.03, $p = 0.5$, 95\% CI: [-0.85, 0.75]). Posterior predictive intervals computed using Bayesian structural time series. $n_{\mathrm{pre}} = 8$ weeks, $n_{\mathrm{post}} = 72$ weeks for all panels. Fitted values/intervals use the mean as the measure of central tendency.}
    \label{fig:alternate_bridge}
\end{figure}

From the definition of local bridge, one can see that an isolated dyad -- an edge whose source and target have degree one -- is a local bridge. However, such edges do not correspond to the intuitive notion of a ``bridge" as a link between distinct communities. Supplementary Figure \ref{fig:alternate_bridge} panel a shows that the observed drop in local bridges persists even when we ignore local bridges which are isolated dyads. 

A previous study on the effect of remote work on tech employee collaboration patterns defined bridging connections as those connections in monthly communication networks with a low local constraint \cite{BurtHoles}, and weak ties as ties with below-median communication time \cite{yang2021the}. Our definition of weak ties (local bridges) is more similar to their definition of bridging tie than their definition of weak tie. Supplementary Figure \ref{fig:alternate_bridge} panels b and c show Burt's measure of structural holes and a measure of structural holes commonly used in topological data analysis, respectively. In both cases, there are no significant long-term changes in the structure of holes in the network. Panel d confirms that there is no significant change in the number of emails sent along local bridges. Though this may seem to be in contrast to the results found by Microsoft, it's important to note that they primarily studied communication media other than email.

\begin{figure}
    \centering
    \includegraphics[width=\linewidth]{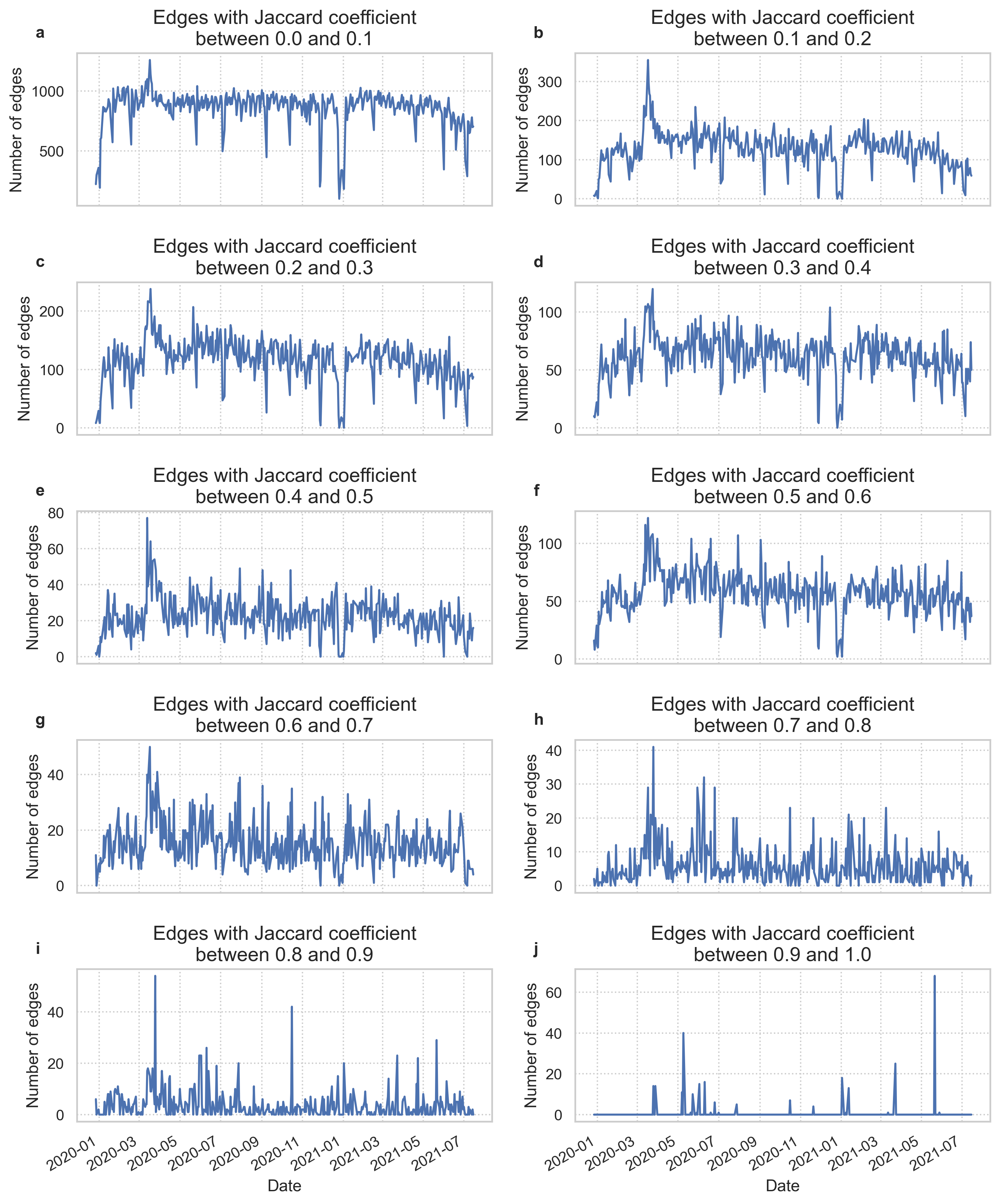}
    \caption{{\bf Number of edges with fixed Jaccard coefficients in daily email networks}. The Jaccard coefficient of an edge is defined as the intersection over union of the neighbor sets of each edge. Data reported as edge counts within fixed bins.}
    \label{fig:all_jaccard}
\end{figure}

\begin{figure}
    \centering
    \includegraphics[width=\linewidth]{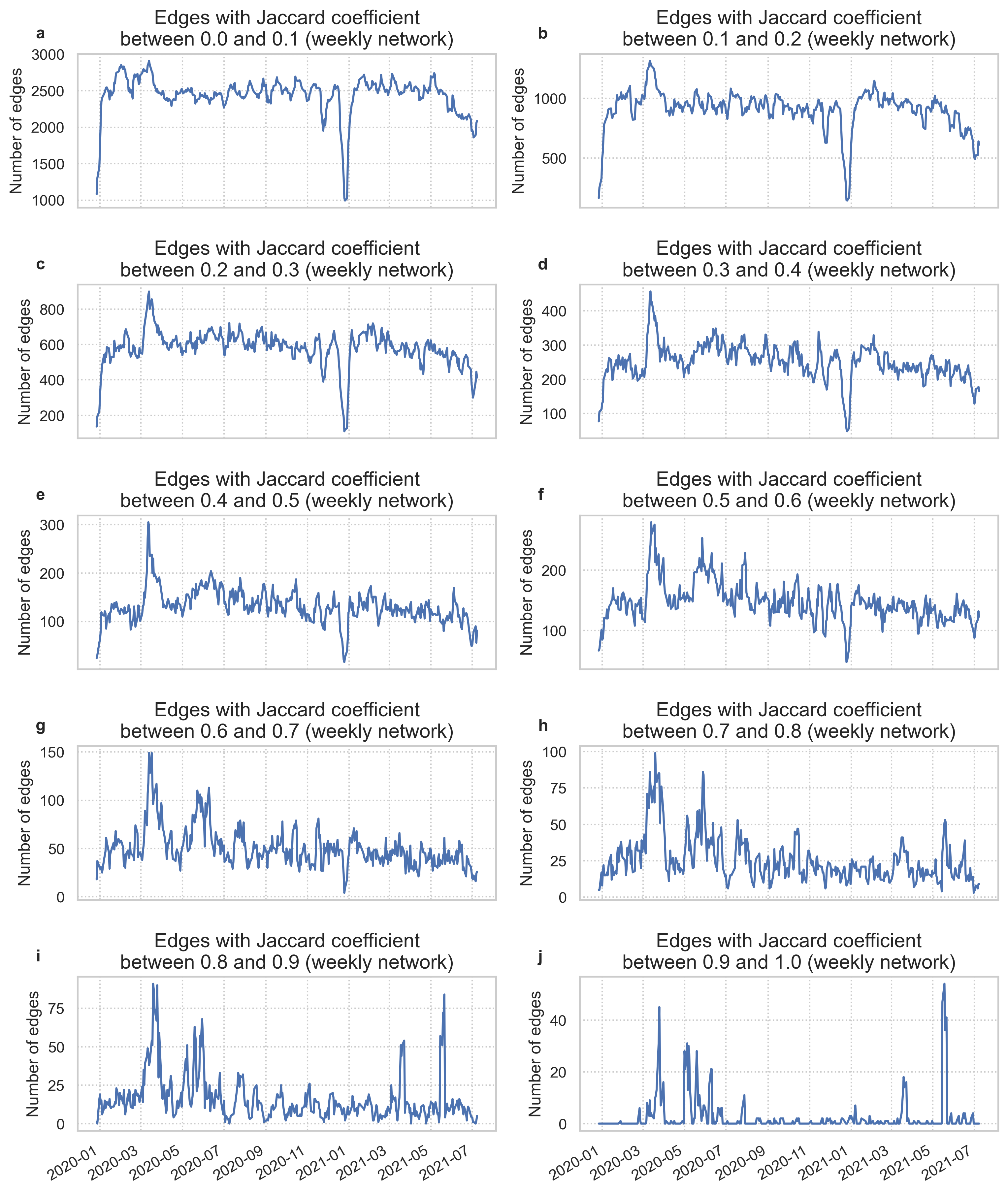}
    \caption{{\bf Number of edges with fixed Jaccard coefficients in weekly email networks}. The Jaccard coefficient of an edge is defined as the intersection over union of the neighbor sets of each edge. Data reported as edge counts within fixed bins. }
    \label{fig:all_jaccard_weekly}
\end{figure}

\begin{figure}
    \centering
    \includegraphics[width=\linewidth]{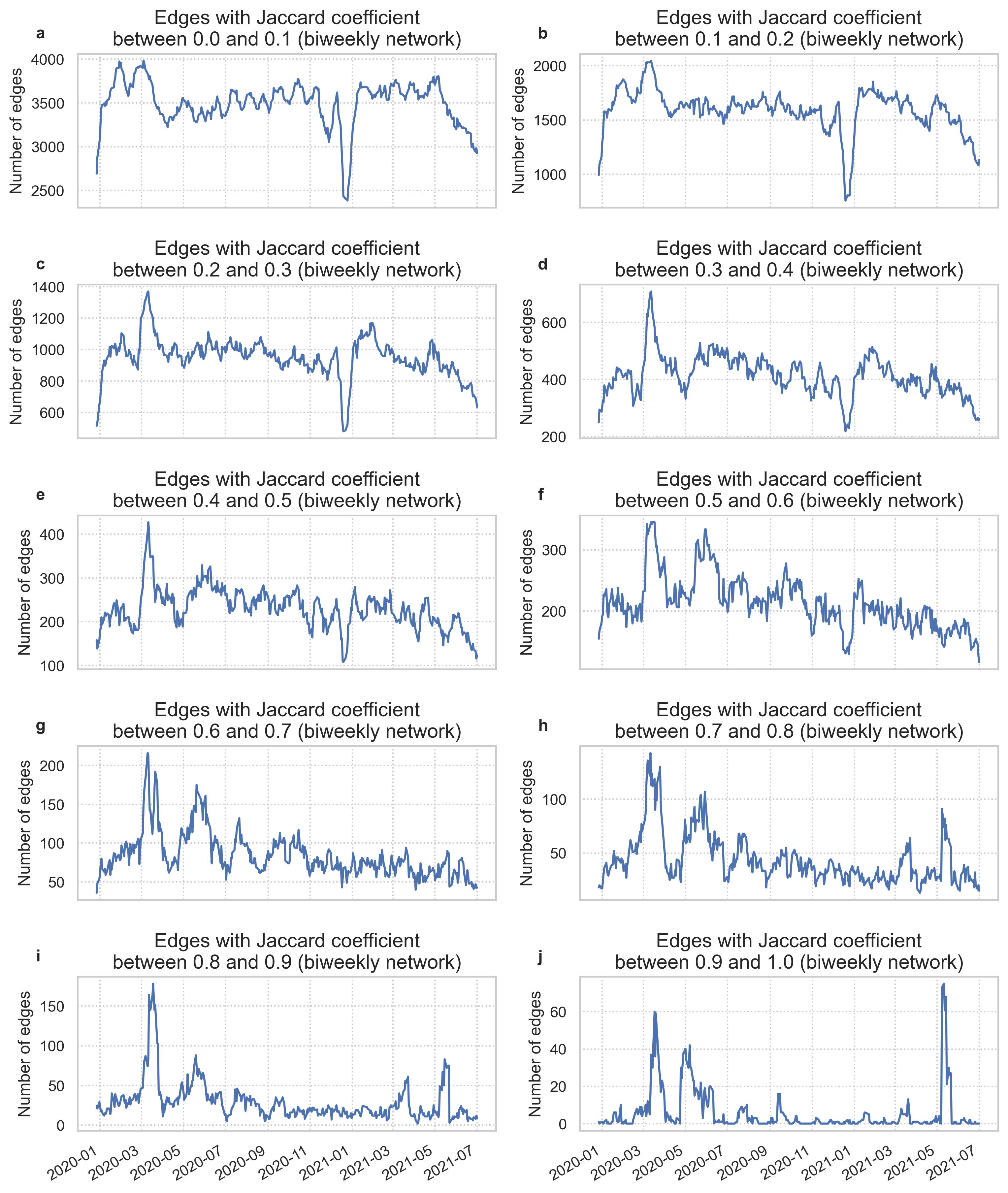}
    \caption{{\bf Number of edges with fixed Jaccard coefficients in biweekly email networks}. The Jaccard coefficient of an edge is defined as the intersection over union of the neighbor sets of each edge. Data reported as edge counts within fixed bins.}
    \label{fig:all_jaccard_biweekly}
\end{figure}

\begin{figure}
    \centering
    \includegraphics[width=\linewidth]{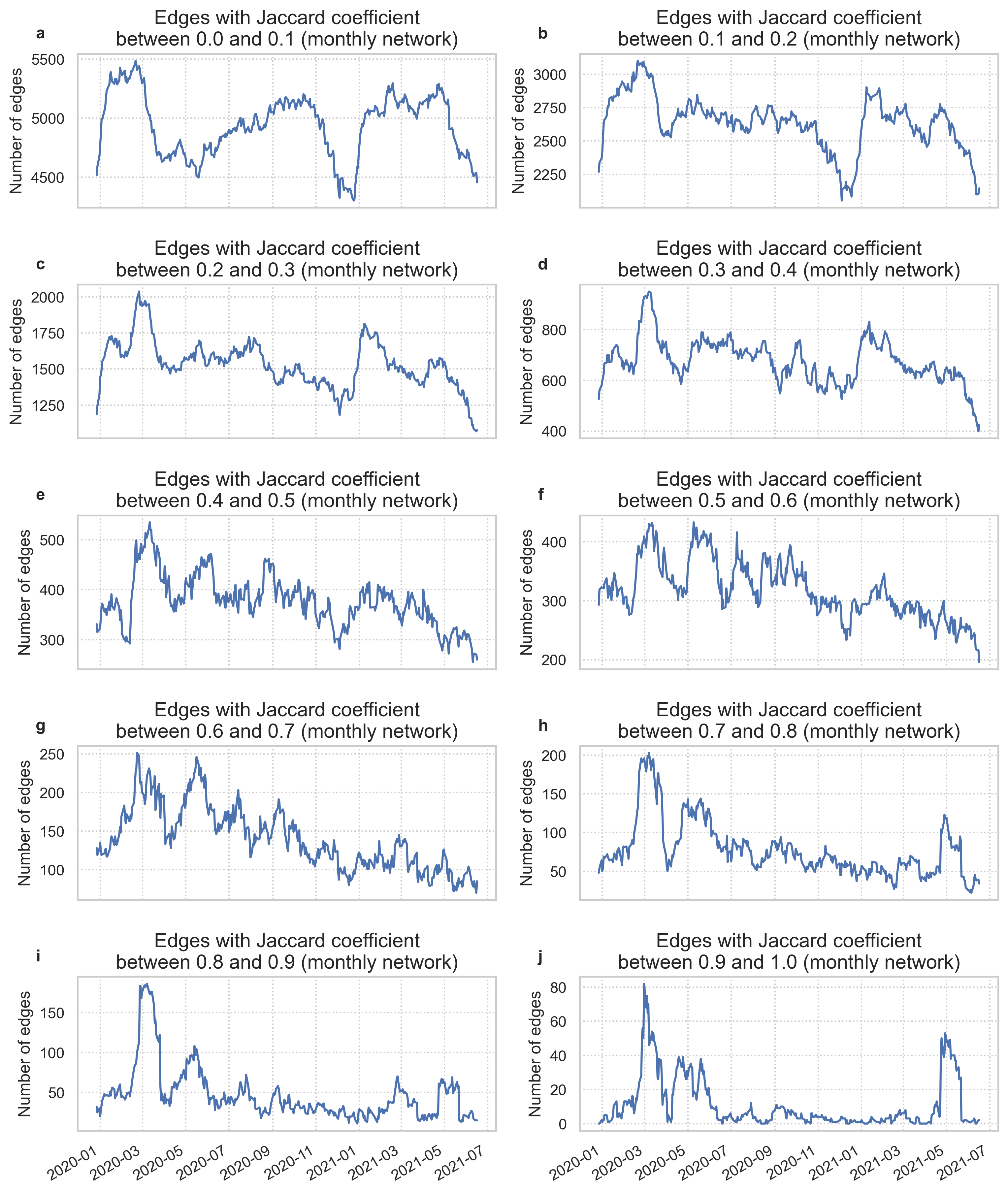}
    \caption{{\bf Number of edges with fixed Jaccard coefficients in monthly email networks}. The Jaccard coefficient of an edge is defined as the intersection over union of the neighbor sets of each edge. Data reported as edge counts within fixed bins.}
    \label{fig:all_jaccard_monthly}
\end{figure}

The Jaccard coefficient of an edge $(u,v)$ is defined as $\frac{\|N(u) \cap N(v)\|}{\|N(u) \cup N(v)\|}$ where $N(u)$ is the set of neighbors of $u$ in the email network, and similarly for $v$. Local bridges are edges with Jaccard coefficient 0. Supplementary Figures \ref{fig:all_jaccard}-\ref{fig:all_jaccard_monthly} show the number of edges with different Jaccard coefficients in daily, weekly, biweekly, and monthly email networks. The most noticeable prolonged changes in number of edges occurs in sets of edges with low Jaccard coefficient. 

\begin{figure}
    \centering
    \includegraphics[width=\linewidth]{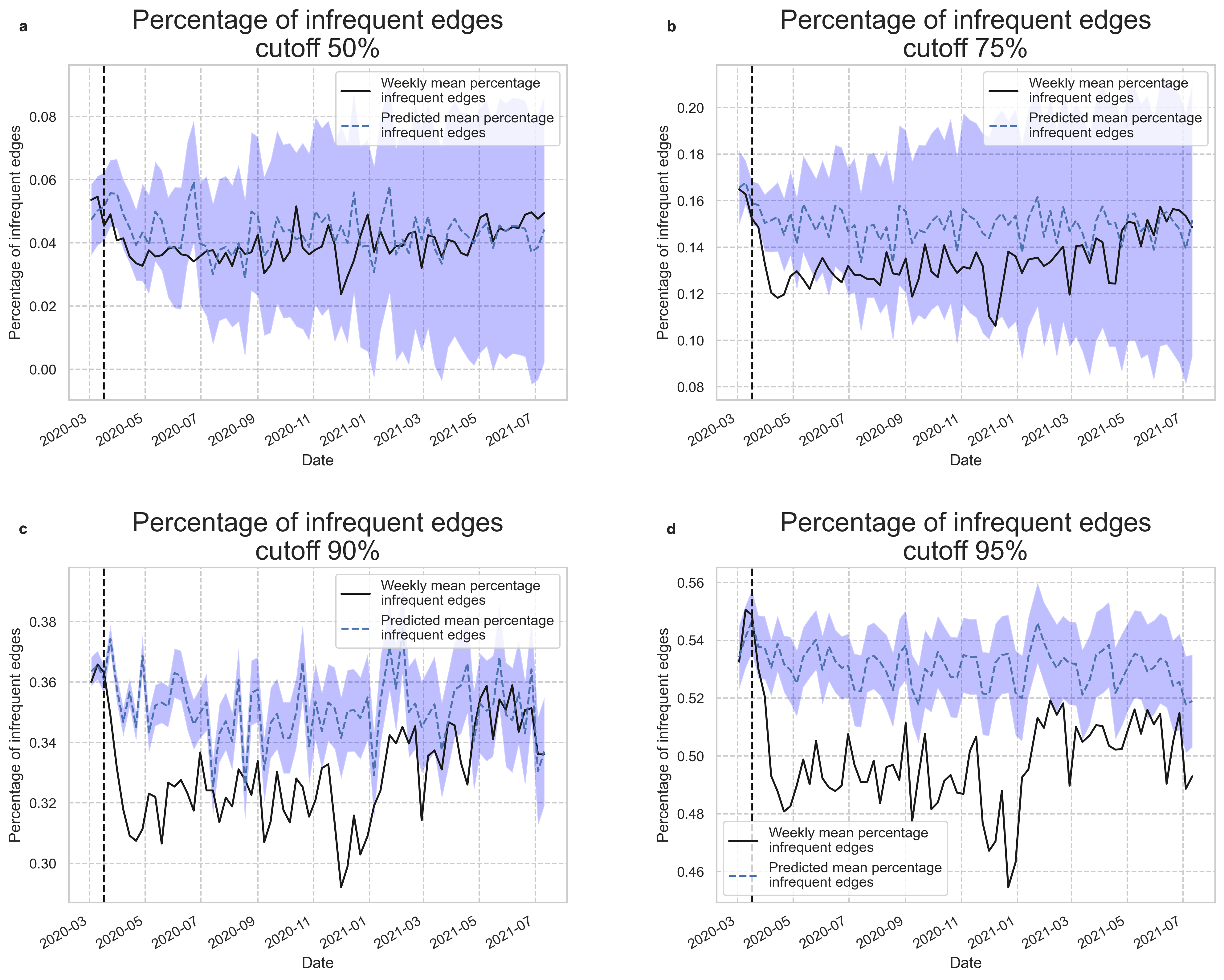}
    \caption{ {\bf Infrequent edge counts measured for different frequency thresholds.} {\bf a}, The number of infrequent edges at threshold 0.5 each day (effect: -0.0036, $p < .01$, 95\% PPI: [-0.0053, -0.0019]). {\bf b}, The number of infrequent edges at threshold 0.75 each day (effect: -0.016, $p = .18$, 95\% PPI: [-0.049, 0.018]). {\bf c}, The number of infrequent edges at threshold 0.9 each day (effect: -0.02, $p < .001$, 95\% PPI: [-0.0230, -0.0216]). {\bf d}, The number of infrequent edges at threshold 0.95 each day (effect: -0.034, $p = 0.5$, 95\% PPI: [-0.041, -0.025]). Posterior predictive intervals computed using Bayesian structural time series. $n_{\mathrm{pre}} = 8$ weeks, $n_{\mathrm{post}} = 72$ weeks for all panels. Fitted values/intervals use the mean as the measure of central tendency.}
    \label{fig:freq_def_ties}
\end{figure}

\begin{figure}
    \centering
    \includegraphics[width=\linewidth]{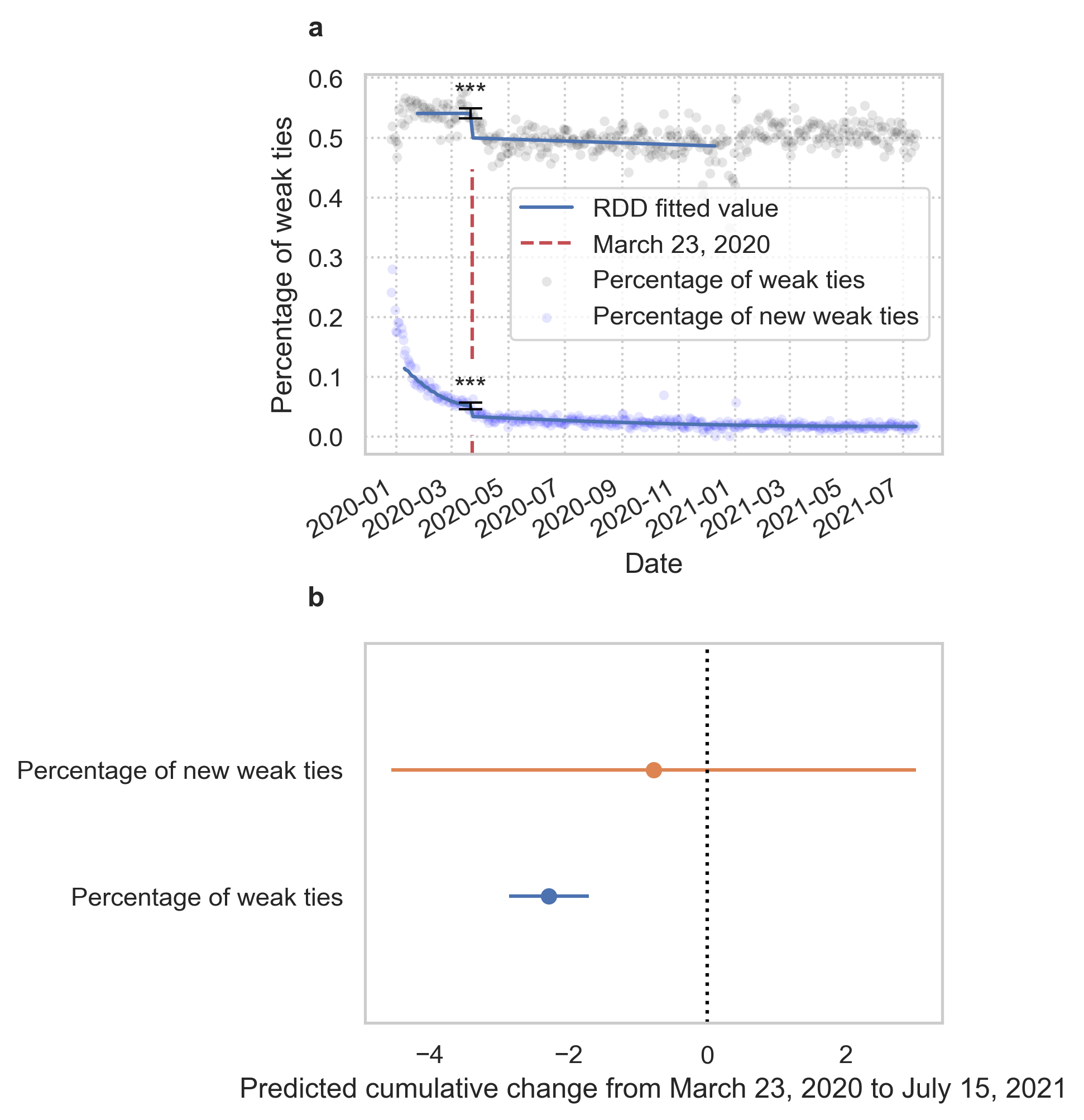}
    \caption{ {\bf Changes in infrequent edges in the MIT email network after the shift to remote work}. ***$: p < .001$, **$: .001 \leq p <.01$, *$: .01 \leq p < .05$. {\bf a}, A drop of 0.0175 in the percentage of infrequent edges after March 23, 2020 ($p < .001$, 95\% CI: [-0.049	-0.033]). There is a drop of 0.041 in the mean percentage of new (not previously seen) infrequent edges appearing each weekday after March 23, 2020 ($p < .001$, 95\% CI: [-0.023	-0.012]). Statistics represent the results of a two-sided $z$ test corresponding to a local polynomial regression discontinuity design ($n_{\mathrm{pre}} = 42$ days, $n_{\mathrm{post}} = 188$ days). There is a cumulative drop of .034 in the percentage of infrequent edges throughout an entire year ($p < .001$, 95\% PPI: [-.042, -.026]) and a non-significant drop of 2930 new weak ties ($p = .36$, 95\% PPI: [-.064, .048]). Posterior predictive intervals computed using Bayesian structural time series ($n_{\mathrm{pre}} = 8$ weeks, $n_{\mathrm{post}} = 72$ weeks). Fitted values/intervals use the mean as the measure of central tendency.}
    \label{fig:combined_freq}
\end{figure}

\begin{figure}
    \centering
    \includegraphics[width=\linewidth]{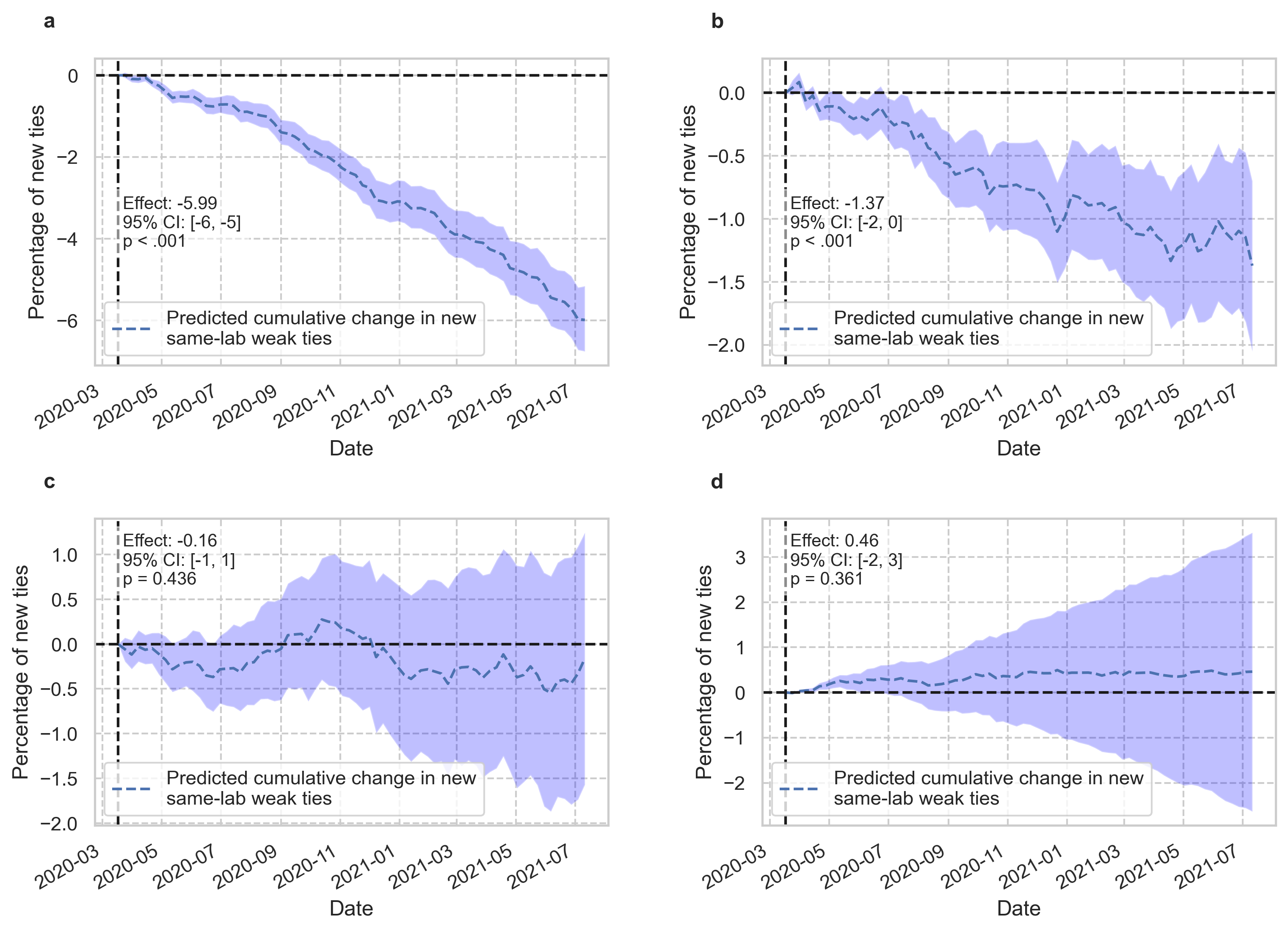}
    \caption{{\bf Formation of new infrequent edges stratified by distance in the MIT email network.} {\bf a} The estimated effect  between researchers in the same lab. {\bf b} The estimated effect between researchers in distinct labs within 150 meters. {\bf c} The estimated effect between researchers in distinct labs between 150 and 650 meters. {\bf d} The estimated effect between researchers in distinct labs further than 650 meters. Shaded regions represent 95\% posterior predictive intervals computed using Bayesian structural time series with a synthetic counterfactual constructed from weekend data $n_{\mathrm{pre}} = 8$ weeks, $n_{\mathrm{post}} = 72$ weeks for all panels. Fitted values/intervals use the mean as the measure of central tendency.}
    \label{fig:freq_spatial}
\end{figure}

Finally, another commonly considered dimension of tie strength is frequency of contact. After computing the number of days on which each tie appears in the network and obtaining a distribution of tie frequency, we choose four different frequency cutoff thresholds to distinguish between ``strong" and ``weak" ties: the 50th, 75th, 90th, and 95th percentile of the frequency distribution. Supplementary Figure \ref{fig:freq_def_ties} shows the predicted percentage of weak ties in the email network on each day for these four cutoffs, and we see that by defining ``weak ties" to be ties which occur with frequency below the 95th percentile (ties which occur on less than 12.6\% of the dates in our data) of the frequency distribution, we recover a long-lasting drop similar to the drop in local bridges described in the main manuscript. Supplementary Figures \ref{fig:combined_freq} and \ref{fig:freq_spatial} show that infrequent ties and new infrequent ties respond the same way to a lack of co-location as local bridges, making our results robust to the definition of weak tie.

\section{Preprocessing error}

\begin{figure}
    \centering
    \includegraphics[width=\linewidth]{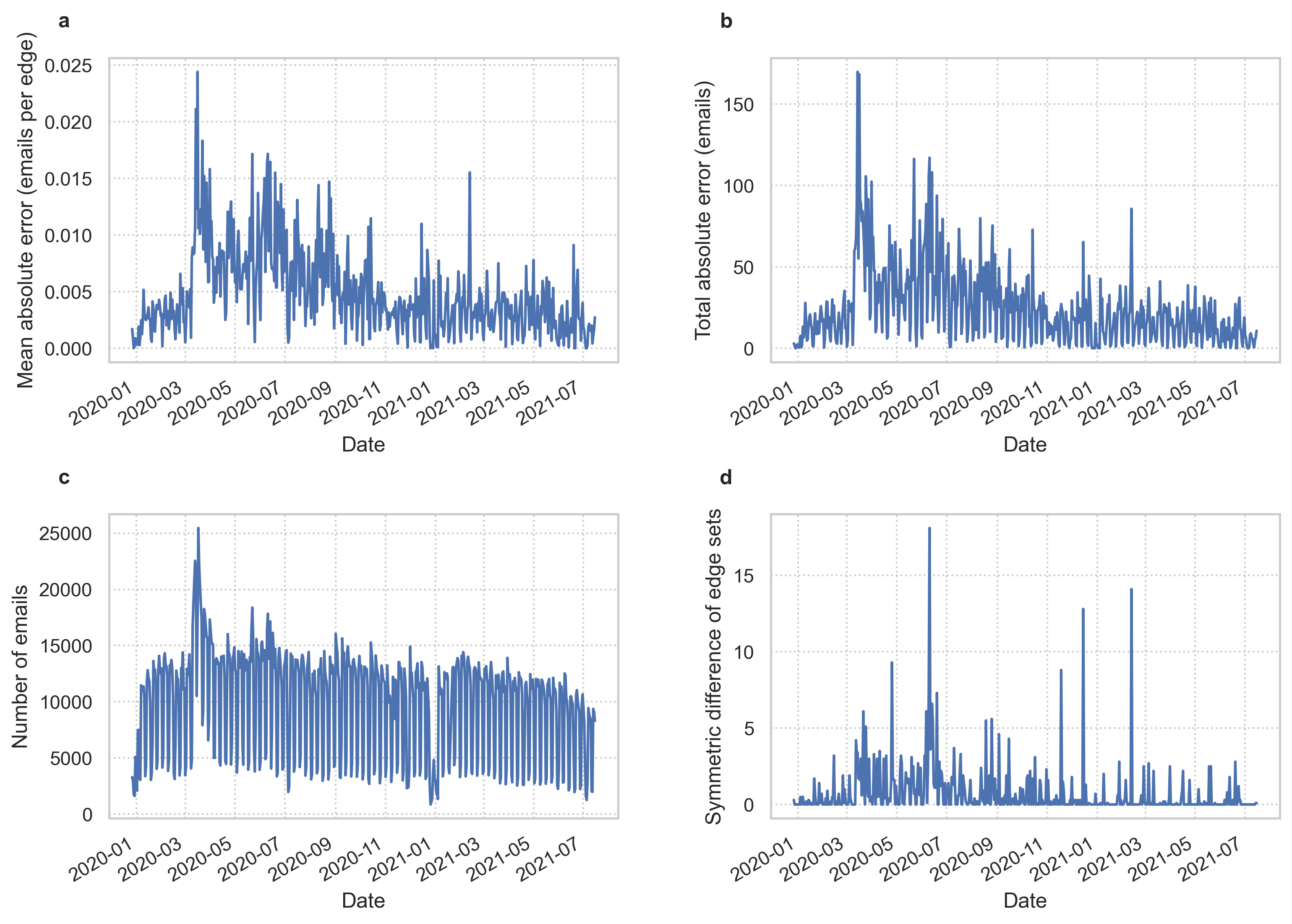}
    \caption{ {\bf Error quantification for estimation of email edge weights using non-negative matrix factorization.} {\bf a}, After re-aggregating the data, the mean error in emails sent between each group of people averaged across the 10 randomizations. {\bf b}, After re-aggregating the data, the total error in emails sent per edge. {\bf c} The total number of emails sent in the data on each day before any preprocessing. {\bf d} After re-aggregating the data, the cardinality of the symmetric difference of estimated and actual edge sets.}
    \label{fig:combined_errors}
\end{figure}

Recall from the Methods section that we estimate the number of emails between each pair of users on each day from randomized, aggregated data. In order to quantify the quality of our estimations, we re-aggregate our individual estimates and compare to each randomization of the aggregated data. The error observed after aggregation is due to our use of non-negative matrix factorization (NMF) in order to obtain approximate sparse solutions to the number of emails sent per user when there is not a unique integer solution. Of those pairs of people who possibly sent a non-zero number of emails on a given day, we can and do solve for the number of emails exactly in approximately 66\% of cases; if we include the pairs which send no emails to one another in at least one random aggregation (and hence send zero emails), we exactly solve for the number of emails between 99.9\% of users. When a system is underdetermined, NMF is substantially faster than solving the constrained system of linear Diophantine equations necessary to produce a sparse, non-negative integer solution. Furthermore, any exact integer solution to an underdetermined system of linear equations would itself be an approximation to the true number of emails sent between users. This is a source of unmeasurable error which is a consequence of working with aggregated data. Supplementary Figure \ref{fig:combined_errors}, panels b and c, show that of the tens of thousands of emails sent each day, our errors are at most in the hundreds. Panel a shows that, per edge in the aggregated network, we never exceed more than 3\% error.  

\section{Statistics tables}
Here we report statistics tables containing test results for each generalized least squares regression. All null hypothesis tests are two-sided.

\begin{table}
\begin{center}
\begin{tabular}{lclc}
\toprule
\textbf{Dep. Variable:}    &        Intra-unit connections         & \textbf{  R-squared:         } &   -0.000  \\
\textbf{Model:}            &      GLSAR       & \textbf{  Adj. R-squared:    } &   -0.000  \\
\textbf{Method:}           &  Least Squares   & \textbf{  F-statistic:       } &    1.898  \\
\textbf{No. Observations:} &          16      & \textbf{  AIC:               } &   -6.145  \\
\textbf{Df Residuals:}     &          15      & \textbf{  BIC:               } &   -5.372  \\
\textbf{Df Model:}         &           0      & \textbf{                     } &           \\
\textbf{Covariance Type:}  &       HC3        & \textbf{                     } &           \\
\bottomrule
\end{tabular}
\end{center}\begin{center}
\begin{tabular}{lcccccc}
\toprule
                   & \textbf{coef} & \textbf{std err} & \textbf{z} & \textbf{P$> \|$z$\|$} & \textbf{[0.025} & \textbf{0.975]}  \\
\midrule
\textbf{Intercept} &       0.0345  &        0.025     &     1.378  &         0.168        &       -0.015    &        0.083     \\
\textbf{c}         &       0.0345  &        0.025     &     1.378  &         0.168        &       -0.015    &        0.083     \\
\bottomrule
\end{tabular}
\caption{{\bf Statistics for intra-unit connections in Figure 1 panel a.} The difference in each pair of days is represented by the variable c.}
    \label{fig:my_label}
\end{center}
\end{table}

\begin{table}
    \begin{center}
\begin{tabular}{lclc}
\toprule
\textbf{Dep. Variable:}    &        Inter-unit connections         & \textbf{  R-squared:         } &    0.000  \\
\textbf{Model:}            &      GLSAR       & \textbf{  Adj. R-squared:    } &    0.000  \\
\textbf{Method:}           &  Least Squares   & \textbf{  F-statistic:       } &    2.877  \\
\textbf{No. Observations:} &          16      & \textbf{  AIC:               } &   -10.90  \\
\textbf{Df Residuals:}     &          15      & \textbf{  BIC:               } &   -10.13  \\
\textbf{Df Model:}         &           0      & \textbf{                     } &           \\
\textbf{Covariance Type:}  &       HC3        & \textbf{                     } &           \\
\bottomrule
\end{tabular}
\end{center}\begin{center}
\begin{tabular}{lcccccc}
\toprule
                   & \textbf{coef} & \textbf{std err} & \textbf{z} & \textbf{P$> \|$z$\|$} & \textbf{[0.025} & \textbf{0.975]}  \\
\midrule
\textbf{Intercept} &       0.0366  &        0.022     &     1.696  &         0.090        &       -0.006    &        0.079     \\
\textbf{c}         &       0.0366  &        0.022     &     1.696  &         0.090        &       -0.006    &        0.079     \\
\bottomrule
\end{tabular}
\end{center}
    \caption{{\bf Statistics for inter-unit connections in Figure 1 panel a.} The difference in each pair of days is represented by the variable c.}
    \label{fig:my_label}
\end{table}

\begin{table}
    \begin{center}
\begin{tabular}{lclc}
\toprule
\textbf{Dep. Variable:}    &        Num connected comps.         & \textbf{  R-squared:         } &    0.000  \\
\textbf{Model:}            &      GLSAR       & \textbf{  Adj. R-squared:    } &    0.000  \\
\textbf{Method:}           &  Least Squares   & \textbf{  F-statistic:       } &    8.333  \\
\textbf{No. Observations:} &          16      & \textbf{  AIC:               } &   -24.69  \\
\textbf{Df Residuals:}     &          15      & \textbf{  BIC:               } &   -23.92  \\
\textbf{Df Model:}         &           0      & \textbf{                     } &           \\
\textbf{Covariance Type:}  &       HC3        & \textbf{                     } &           \\
\bottomrule
\end{tabular}
\end{center}\begin{center}
\begin{tabular}{lcccccc}
\toprule
                   & \textbf{coef} & \textbf{std err} & \textbf{z} & \textbf{P$> \|$z$\|$} & \textbf{[0.025} & \textbf{0.975]}  \\
\midrule
\textbf{Intercept} &      -0.0404  &        0.014     &    -2.887  &         0.004        &       -0.068    &       -0.013     \\
\textbf{c}         &      -0.0404  &        0.014     &    -2.887  &         0.004        &       -0.068    &       -0.013     \\
\bottomrule
\end{tabular}
\end{center}

    \caption{{\bf Statistics for the number of connected components in Figure 1 panel a.} The difference in each pair of days is represented by the variable c.}
    \label{fig:my_label}
\end{table}

\begin{table}
    \begin{center}
\begin{tabular}{lclc}
\toprule
\textbf{Dep. Variable:}    &        Giant comp. size         & \textbf{  R-squared:         } &    0.000  \\
\textbf{Model:}            &      GLSAR       & \textbf{  Adj. R-squared:    } &    0.000  \\
\textbf{Method:}           &  Least Squares   & \textbf{  F-statistic:       } &    2.387  \\
\textbf{No. Observations:} &          16      & \textbf{  AIC:               } &   -4.477  \\
\textbf{Df Residuals:}     &          15      & \textbf{  BIC:               } &   -3.704  \\
\textbf{Df Model:}         &           0      & \textbf{                     } &           \\
\textbf{Covariance Type:}  &       HC3        & \textbf{                     } &           \\
\bottomrule
\end{tabular}
\end{center}\begin{center}
\begin{tabular}{lcccccc}
\toprule
                   & \textbf{coef} & \textbf{std err} & \textbf{z} & \textbf{P$> \|$z$\|$} & \textbf{[0.025} & \textbf{0.975]}  \\
\midrule
\textbf{Intercept} &       0.0407  &        0.026     &     1.545  &         0.122        &       -0.011    &        0.092     \\
\textbf{c}         &       0.0407  &        0.026     &     1.545  &         0.122        &       -0.011    &        0.092     \\
\bottomrule
\end{tabular}
\end{center}

    \caption{{\bf Statistics for giant component size in Figure 1 panel a.} The difference in each pair of days is represented by the variable c.}
    \label{fig:my_label}
\end{table}

\begin{table}
\begin{center}
\begin{tabular}{lclc}
\toprule
\textbf{Dep. Variable:}    &        Num. weak ties         & \textbf{  R-squared:         } &    0.054  \\
\textbf{Model:}            &      GLSAR       & \textbf{  Adj. R-squared:    } &    0.045  \\
\textbf{Method:}           &  Least Squares   & \textbf{  F-statistic:       } &    7.481  \\
\textbf{No. Observations:} &         225      & \textbf{  AIC:               } &    2685.  \\
\textbf{Df Residuals:}     &         222      & \textbf{  BIC:               } &    2696.  \\
\textbf{Df Model:}         &           2      & \textbf{                     } &           \\
\textbf{Covariance Type:}  &       HC3        & \textbf{                     } &           \\
\bottomrule
\end{tabular}
\end{center}\begin{center}
\begin{tabular}{lcccccc}
\toprule
                     & \textbf{coef} & \textbf{std err} & \textbf{z} & \textbf{P$> \|$z$\|$} & \textbf{[0.025} & \textbf{0.975]}  \\
\midrule
\textbf{Intercept}   &     891.5135  &       14.142     &    63.038  &         0.000        &      863.795    &      919.232     \\
\textbf{policy}      &     -55.7007  &       18.330     &    -3.039  &         0.002        &      -91.627    &      -19.774     \\
\textbf{policy:time} &      -0.0424  &        0.155     &    -0.273  &         0.785        &       -0.347    &        0.262     \\
\bottomrule
\end{tabular}
\end{center}
\caption{{\bf Statistics for Figure 2 panel a.} This table represents the output of a local polynomial regression discontinuity design on the number of weak ties with cutoff on March 23, 2020. The treatment effect is measured by the policy variable.}
\label{tab:wt_rdd}
\end{table}
 
\begin{table}
    \begin{center}
\begin{tabular}{lclc}
\toprule
\textbf{Dep. Variable:}    &        Num. new weak ties         & \textbf{  R-squared:         } &     0.897  \\
\textbf{Model:}            &      GLSAR       & \textbf{  Adj. R-squared:    } &     0.895  \\
\textbf{Method:}           &  Least Squares   & \textbf{  F-statistic:       } &     321.2  \\
\textbf{No. Observations:} &         385      & \textbf{  AIC:               } &     3195.  \\
\textbf{Df Residuals:}     &         379      & \textbf{  BIC:               } &     3219.  \\
\textbf{Df Model:}         &           5      & \textbf{                     } &            \\
\textbf{Covariance Type:}  &       HC3        & \textbf{                     } &            \\
\bottomrule
\end{tabular}
\end{center}\begin{center}
\begin{tabular}{lcccccc}
\toprule
                      & \textbf{coef} & \textbf{std err} & \textbf{z} & \textbf{P$> \|$z$\|$} & \textbf{[0.025} & \textbf{0.975]}  \\
\midrule
\textbf{Intercept}    &     128.8843  &        8.421     &    15.306  &         0.000        &      112.380    &      145.389     \\
\textbf{time}         &      -0.3442  &        1.374     &    -0.251  &         0.802        &       -3.037    &        2.349     \\
\textbf{time2}        &       0.0550  &        0.036     &     1.538  &         0.124        &       -0.015    &        0.125     \\
\textbf{policy}       &     -46.3961  &        8.661     &    -5.357  &         0.000        &      -63.371    &      -29.421     \\
\textbf{policy:time}  &       0.0975  &        1.374     &     0.071  &         0.943        &       -2.596    &        2.791     \\
\textbf{policy:time2} &      -0.0547  &        0.036     &    -1.529  &         0.126        &       -0.125    &        0.015     \\
\bottomrule
\end{tabular}
\caption{{\bf Statistics for Figure 2 panel b.} This table represents the output of a local polynomial regression discontinuity design on the number of new weak ties with cutoff on March 23, 2020. The treatment effect is measured by the policy variable.}
\end{center}
\label{tab:new_wt_rdd}
\end{table}

\begin{table}
    \begin{center}
\begin{tabular}{lclc}
\toprule
\textbf{Dep. Variable:}    &        Num weak ties         & \textbf{  R-squared:         } &    0.268  \\
\textbf{Model:}            &      GLSAR       & \textbf{  Adj. R-squared:    } &    0.240  \\
\textbf{Method:}           &  Least Squares   & \textbf{  F-statistic:       } &    14.30  \\
\textbf{No. Observations:} &          55      & \textbf{  AIC:               } &    624.9  \\
\textbf{Df Residuals:}     &          52      & \textbf{  BIC:               } &    630.9  \\
\textbf{Df Model:}         &           2      & \textbf{                     } &           \\
\textbf{Covariance Type:}  &       HC3        & \textbf{                     } &           \\
\bottomrule
\end{tabular}
\end{center}\begin{center}
\begin{tabular}{lcccccc}
\toprule
                     & \textbf{coef} & \textbf{std err} & \textbf{z} & \textbf{P$> \|$z$\|$} & \textbf{[0.025} & \textbf{0.975]}  \\
\midrule
\textbf{Intercept}   &    -177.6800  &        9.296     &   -19.114  &         0.000        &     -195.899    &     -159.461     \\
\textbf{policy}      &      67.5030  &       29.471     &     2.290  &         0.022        &        9.740    &      125.266     \\
\textbf{policy:time} &       0.8652  &        2.439     &     0.355  &         0.723        &       -3.915    &        5.645     \\
\bottomrule
\end{tabular}
\caption{{\bf Statistics for Figure 4 panel c.} This table represents the output of a local polynomial regression discontinuity design on the number of weak ties with cutoff on September 8, 2021. The treatment effect is measured by the policy variable.}
\label{fig:my_label}
\end{center}
\end{table}

\begin{table}
\begin{center}
\begin{tabular}{lclc}
\toprule
\textbf{Dep. Variable:}    &        Simulated weak ties         & \textbf{  R-squared:         } &    0.346  \\
\textbf{Model:}            &      GLSAR       & \textbf{  Adj. R-squared:    } &    0.341  \\
\textbf{Method:}           &  Least Squares   & \textbf{  F-statistic:       } &    88.49  \\
\textbf{No. Observations:} &         285      & \textbf{  AIC:               } &    2730.  \\
\textbf{Df Residuals:}     &         282      & \textbf{  BIC:               } &    2741.  \\
\textbf{Df Model:}         &           2      & \textbf{                     } &           \\
\textbf{Covariance Type:}  &       HC3        & \textbf{                     } &           \\
\bottomrule
\end{tabular}
\end{center}\begin{center}
\begin{tabular}{lcccccc}
\toprule
                     & \textbf{coef} & \textbf{std err} & \textbf{z} & \textbf{P$> \|$z$\|$} & \textbf{[0.025} & \textbf{0.975]}  \\
\midrule
\textbf{Intercept}   &    1220.8649  &        4.313     &   283.042  &         0.000        &     1212.411    &     1229.319     \\
\textbf{policy}      &     -65.2128  &        5.560     &   -11.730  &         0.000        &      -76.109    &      -54.316     \\
\textbf{policy:time} &       0.0255  &        0.026     &     0.982  &         0.326        &       -0.025    &        0.076     \\
\bottomrule
\end{tabular}
\end{center}
\caption{{\bf Statistics for Figure 6 panel a.} This table represents the
output of a local polynomial regression discontinuity design on the number of weak ties with cutoff on March 23, 2020. The treatment effect is
measured by the policy variable. }
\label{fig:my_label}
\end{table}
\begin{table}
\begin{center}
\begin{tabular}{lclc}
\toprule
\textbf{Dep. Variable:}    &        Simulated weak ties         & \textbf{  R-squared:         } &    0.392  \\
\textbf{Model:}            &      GLSAR       & \textbf{  Adj. R-squared:    } &    0.388  \\
\textbf{Method:}           &  Least Squares   & \textbf{  F-statistic:       } &    107.7  \\
\textbf{No. Observations:} &         162      & \textbf{  AIC:               } &    1536.  \\
\textbf{Df Residuals:}     &         160      & \textbf{  BIC:               } &    1542.  \\
\textbf{Df Model:}         &           1      & \textbf{                     } &           \\
\textbf{Covariance Type:}  &       HC3        & \textbf{                     } &           \\
\bottomrule
\end{tabular}
\end{center}\begin{center}
\begin{tabular}{lcccccc}
\toprule
                  & \textbf{coef} & \textbf{std err} & \textbf{z} & \textbf{P$> \|$z$\|$} & \textbf{[0.025} & \textbf{0.975]}  \\
\midrule
\textbf{Intercept} &    1160.2661  &        2.519     &   460.608  &         0.000        &     1155.329    &     1165.203     \\
\textbf{policy}    &      51.8918  &        5.000     &    10.377  &         0.000        &       42.091    &       61.693     \\
\bottomrule
\end{tabular}
\end{center}
\caption{{\bf Statistics for Figure 6 panel a.} This table represents the output of a local polynomial regression discontinuity design on the number of weak ties with cutoff on September 8, 2021. The treatment effect is measured by the policy variable.}
\end{table}

\begin{table}\begin{center}
\begin{tabular}{lclc}
\toprule
\textbf{Dep. Variable:}    &        Simulated new weak ties         & \textbf{  R-squared:         } &     0.990  \\
\textbf{Model:}            &      GLSAR       & \textbf{  Adj. R-squared:    } &     0.989  \\
\textbf{Method:}           &  Least Squares   & \textbf{  F-statistic:       } &     716.8  \\
\textbf{No. Observations:} &         275      & \textbf{  AIC:               } &     1760.  \\
\textbf{Df Residuals:}     &         269      & \textbf{  BIC:               } &     1781.  \\
\textbf{Df Model:}         &           5      & \textbf{                     } &            \\
\textbf{Covariance Type:}  &       HC3        & \textbf{                     } &            \\
\bottomrule
\end{tabular}
\end{center}\begin{center}
\begin{tabular}{lcccccc}
\toprule
                      & \textbf{coef} & \textbf{std err} & \textbf{z} & \textbf{P$> \|$z$\|$} & \textbf{[0.025} & \textbf{0.975]}  \\
\midrule
\textbf{Intercept}    &      67.5013  &        6.819     &     9.899  &         0.000        &       54.137    &       80.866     \\
\textbf{time}         &       2.6929  &        0.892     &     3.018  &         0.003        &        0.944    &        4.441     \\
\textbf{time2}        &       0.2962  &        0.025     &    11.849  &         0.000        &        0.247    &        0.345     \\
\textbf{policy}       &     -47.0380  &        6.902     &    -6.815  &         0.000        &      -60.566    &      -33.510     \\
\textbf{policy:time}  &      -2.7958  &        0.892     &    -3.133  &         0.002        &       -4.545    &       -1.047     \\
\textbf{policy:time2} &      -0.2959  &        0.025     &   -11.837  &         0.000        &       -0.345    &       -0.247     \\
\bottomrule
\end{tabular}
\end{center}
\caption{{\bf Statistics for Figure 6 panel b.} This table represents the output of a local polynomial regression discontinuity design on the number of new weak ties with cutoff on March 23, 2020. The treatment effect is measured by the policy variable.}
\end{table}



\end{document}